\documentclass[11pt]{article}
\usepackage{psfrag,epsfig,tabls}
\usepackage{latexsym} 
\usepackage{amsmath} 
\usepackage{amssymb}

\newcommand{\RRi}{\frac{1}{2R}}
\newcommand{\half}{\frac{1}{2}} 
\begin{document}
%%%%%%%%%%%%%%%%%%%%%%%%%%%%%%%%%%%%%%%%%%%%%%%%%%%%%%%%%%%%%%%%%%%%%%%%%%%
\thispagestyle{empty}
\vspace*{-15mm}
\begin{flushright}
%\hspace*{10mm}hep-lat/0307013 \\
%\hspace*{10mm}UCSD-PTH-03-04 
\end{flushright}
\begin{center}
\vspace*{2mm}
{\LARGE Quenched spectroscopy with fixed-point \vskip2mm 
and chirally improved fermions} 
\vskip7mm
Christof Gattringer$^a$,
Meinulf G\"ockeler$^{b,a}$,
Peter Hasenfratz$^c$,
Simon Hauswirth$^c$,
Kieran Holland$^d$,
Thomas J\"org$^e$,
Keisuke~J.~Juge$^c$,
C.B.~Lang$^f$,
Ferenc Niedermayer$^{c,h}$,
P.E.L.~Rakow$^g$, 
Stefan Schaefer$^a$ and 
Andreas Sch\"afer$^a$ 
[BGR (Bern-Graz-Regensburg) collaboration]
\vskip6mm
$^a$ 
Institut f\"ur Theoretische Physik, Universit\"at
Regensburg \\
D-93040 Regensburg, Germany
\vskip1mm
$^b$ 
Institut f\"ur Theoretische Physik, Universit\"at
Leipzig \\
D-04109 Leipzig, Germany
\vskip1mm
$^c$
Institut f\"ur Theoretische Physik,
Universit\"at Bern \\
CH-3012 Bern, Switzerland
\vskip1mm
$^d$
Department of Physics, University of
California at San Diego \\ 
La Jolla, USA
\vskip1mm
$^e$
Dipartimento di Fisica and INFN, \\
Universit\`{a} di Roma "La Sapienza",
I-00185 Roma, Italy
\vskip1mm
$^f$
Institut f\"ur Theoretische Physik, Universit\"at
Graz \\
A-8010 Graz, Austria
\vskip1mm
$^g$
Department of Mathematical Sciences, University of Liverpool \\
Liverpool L69 3BX, UK
\vskip1mm
$^h$
On leave of absence from E\"otv\"os University, HAS Research Group, 
Budapest, Hungary
\vskip6mm
\begin{abstract}
We present results from quenched spectroscopy
calculations with the parametrized fixed-point  
and the chirally improved Dirac operators. Both these 
operators are approximate solutions of the
Ginsparg-Wilson equation and have good chiral properties. 
This allows us to work at small quark masses and we 
explore pseudoscalar-mass to vector-mass ratios down to 0.28. 
We discuss meson and baryon masses, 
their scaling properties, finite volume effects and compare our
results with recent large scale simulations. 
We find that the size of quenching artifacts of the masses
is strongly correlated with their experimentally observed widths
and that the gauge and hadronic scales are consistent.

\end{abstract}
\end{center}

\newpage
\section{Introduction and summary}

QCD spectroscopy with traditional fermion formulations faces some open
problems. For a Dirac operator with poor chiral properties the
fluctuations of the small eigenvalues limit the minimal bare quark
masses one can work at. Wilson fermions have large cut-off effects and
years of experience have demonstrated
\cite{GF11,cppacsquench}
that it is very hard to reach pseudoscalar 
(PS) masses below 300 MeV in quenched QCD. This
problem becomes even worse if the ${\cal O}(a)$-correcting clover term
is included \cite{ukqcd1,flic}. In full QCD the situation does not seem to be
improved. 
State of the art dynamical fermion simulations stay above $m_\mathrm{PS}
\approx $ 450 MeV \cite{ukqcdfull,jlqcdfull,cppacsfull}. Twisted mass Wilson
fermions are an interesting development \cite{twisted,neile} but need detailed
numerical testing.
For staggered fermions - after recent progress with
reducing the cut-off effects and thus flavor symmetry violation 
\cite{MILC09,anna1,qcdsolved}
- the main open question is, whether staggered fermions can describe QCD with
less than 4 degenerate fermions. The solutions \cite{solutions_gw} of
the Ginsparg-Wilson equation \cite{GiWi82} are free of these
problems: Exact chiral symmetry \cite{exact_chi_lu} kills the dangerous
fluctuations which prevent simulations of light pions and they are
automatically ${\cal O}(a)$ improved \cite{ferenc}.
 
There exists a large number of studies exploring Dirac operators
with chiral symmetry in quenched QCD (for recent reviews, see
\cite{chi_works} and references therein). Different
aspects of hadron spectroscopy have also been considered in
\cite{domain1,domain2,over} with domain-wall \cite{kaplan} and
overlap \cite{neuberger} fermions. However, so far no detailed study 
has been attempted on spectroscopy with light pions.
In this work we explore spectroscopy down to $m_\mathrm{PS}/m_\mathrm{V}=0.28$
using two Dirac operators based on the Ginsparg-Wilson equation:
the parametrized fixed-point (FP) and the chirally improved 
(CI) Dirac operators. Actions with exact chiral symmetry are expensive to
simulate. Our approach is a compromise trying to reduce the cost under
the condition of being able to reach pion masses where chiral
perturbation theory can be safely and effectively used for
extrapolation. 
For comparison, at one lattice spacing and lattice size, we simulated
also an approximate overlap Dirac operator obtained from the FP Dirac 
operator, by performing a Legendre expansion of order 3 to approximate 
the overlap projection. 
Finding the right balance between expense and good physical properties 
of an action is an important issue. Significantly simpler and less 
expensive actions were investigated in \cite{tom} and \cite{wolfgang}.

Fixed-point actions result from solving the real space
renormalization group equations in the weak coupling limit \cite{fp1}.
After the first attempts in 2 and 4 dimensional models
\cite{early_FP_fermions}, 
a finite parametrization of FP fermions suitable for numerical calculations
has been developed \cite{fp3,jorg} giving rise to an approximate
solution of the Ginsparg-Wilson equation. The idea behind the CI
operator is a systematic expansion of a solution of the 
Ginsparg-Wilson equation in terms of simple lattice operators  
\cite{ci1}. CI fermions have been shown to give rise to a Dirac
operator with good chiral properties both in 2 and 4 dimensions \cite{ci2}.

The goal of our spectroscopy calculations with FP and CI fermions is 
to explore a new mass region and check whether the promises of chiral
symmetry, i.e.~small pion masses and good scaling properties, become manifest.
The new results also help to outline the effects of finite volume in
quenched QCD with light pions. In particular we compare 
our results for scaling properties with those of recent large-scale simulations
using staggered or Wilson fermions
\cite{cppacsquench,ukqcd1,MILC09,MILC13}. Preliminary results
for FP and CI fermions were reported in \cite{boston1,boston2,hauswirth,fp4}. 
A CI calculation of vector meson couplings to vector and tensor currents was
presented in \cite{braunetal}.

Some of the results of this paper deviate from those presented in our 
preliminary discussion \cite{boston2,hauswirth}. Our data analysis is more
careful and conservative than before. The statistics is also increased
slightly. Most importantly, however, for the FP operator the masses 
in the decuplet baryon sector have changed mainly due to a corrected bug 
in the code constructing the decuplet baryon out of quark propagators. 
The negative parity partner of the nucleon is studied in the present version
also. 

Let us summarize our main results and give conclusions. Our Dirac operators
allow spectroscopy with pseudoscalar masses down to 220 MeV even on coarse
lattices with $a=0.15 \,\mathrm{fm}$. So close to the chiral limit quenched
spectroscopy is heavily contaminated by topological finite size artifacts in
{\it all} hadronic channels at small and intermediate volumes. It is
mandatory to work with hadron operators where these artifacts are cancelled or
at least reduced. In the pseudoscalar channel we clearly see the chiral 
logarithms and tested different methods to determine the chiral log 
parameter $\delta$. In our large ($L=2.5 \,\mathrm{fm}$) volume
spectroscopy we argue for using masses of stable particles/narrow
resonances as input parameters. The results
suggest an intuitive picture where the size of quenching artifacts in the
hadron masses is strongly correlated with their experimentally observed
widths. (Similar observations have been made earlier in \cite{sommer_rho}.) 
We find consistency between the scales of the gauge and hadronic sectors. We
perform different scaling tests changing the resolution in the range 
$a = 0.08 - 0.15\,\mathrm{fm}$. 
No scaling violation is found within our data beyond the statistical errors. 
Comparing our results with other large scale simulations the
conclusion depends on whether the extrapolated CP-PACS data
really describe the continuum limit. This is a
non-trivial question since the extrapolated continuum result is
significantly different from the measured data obtained on the finest
lattice at $a = 0.05\,\mathrm{fm}$ \cite{cppacsquench}. If the answer
is yes, then the FP action results (and also the CI action results,
although within larger statistical errors) at $a=0.15\,\mathrm{fm}$
show cut-off effects, but they 
are closer to the continuum than the Wilson results at $a=0.05\,\mathrm{fm}$
and the improved staggered results at $a=0.13\,\mathrm{fm}$ and
$0.09\,\mathrm{fm}$ and are comparable to the clover improved results
at $0.07\,\mathrm{fm}$. 
We see no finite volume effects above
$L=1.8\,\mathrm{fm}$. We remark finally that the codes are easy to
parallelize and run well on two supercomputers with different architectures.

In a direct comparison of FP and CI fermions we conclude that in the whole
range of parameters we worked at the two operators essentially
perform equally well in light hadron spectroscopy.
Their scaling properties are similar. 
We find it a reassuring aspect that the physical results
obtained with the two operators are well compatible.

Let us finally conclude with some remarks on the bigger picture. The
quoted pion mass of about 220 MeV was reached rather easily for both
operators. We did not need to go to very fine lattices and as a matter of fact
the data for the lightest quarks were obtained on our coarsest lattice of
$a = 0.15$ fm without any sign of exceptional configurations. When comparing
the cost to e.g. the Wilson operator (without clover term) the number of
floating point operations is increased by a factor of $\approx 36$. On the
other hand, our actions have significantly smaller cut-off effects at 
$a = 0.15\,\mathrm{fm}$ than the Wilson action at 
$a = 0.05\,\mathrm{fm}$ and 
the cost of simulation is increasing as $a^{-p}, p \approx 5$. 
Our actions have a rather complicated structure and traditionally used
full QCD algorithms are not applicable. However, recently many
interesting new ideas and test results were presented \cite{anna1,anna2} which
make us cautiously optimistic. 

The article is organized as follows: In Section 2 we describe the technical
details of our calculations, i.e.\ gauge actions, the lattice spacings
and statistics as well as our fitting procedure.  We also discuss the
numerical implementation and performance of the FP and CI operators.  Section
3 is devoted to topological finite size artifacts. We discuss the operators
used for hadron spectroscopy, their topological finite size artifacts and
address strategies to eliminate or alleviate the problem. In Section 4 we
present results for the hadron masses on the coarsest lattice with the largest
physical volume. We suggest here to look at the quenched spectroscopy somewhat
differently from what is done conventionally.  Section 5 is devoted to scaling
and finite volume effects. We compare the FP and CI results with those of
recent large scale simulations. 
In two appendices we compile numerical~data.

\section{Technicalities}
\label{sect:technicalities}

\subsection{Gauge configurations and sources}

One of the goals of the study presented here is to compare our two 
approaches to chiral symmetry, the FP and CI fermions. In order to do
that in a systematic way we kept some basic parameters of our quenched gauge 
configurations similar for the two approaches, in particular the
lattice spacing and the physical volume. We work at three lattice spacings 
of $a = 0.08\,\mathrm{fm}$, $0.10\,\mathrm{fm}$ and $0.15\,\mathrm{fm}$.
For these lattice spacings we generate ensembles on different lattice
sizes of $8^3 \times 24$, $12^3 \times 24$ and $16^3 \times 32$. 
The parameters of our simulations (see Table~\ref{table:1}) allow us to
make a scaling study with both actions in a fixed volume 
$L = 1.2\,\mathrm{fm}$ at three different lattice resolutions and
a finite volume analysis at fixed $a = 0.15\,\mathrm{fm}$ in boxes with
linear extension $L=1.2-2.5\,\mathrm{fm}$. The lattice spacings quoted in
Table~\ref{table:1} were computed \cite{perfgauge,sommerci}
from the Sommer parameter \cite{sommer_r0}. Their statistical errors are
below 1\% for FP and between 1- 2\% for CI, while their systematic
errors are difficult to estimate on coarse lattices. Our statistics is
100 - 200 configurations for the different ensembles. In three cases
(marked with an asterisk in Table~\ref{table:1}) we calculated a second 
quark propagator on each configuration with a source on the $t=0$
time slice shifted by $L$/2.\footnote {They were needed to improve the
$I=2$  $\, \pi \pi$ scattering length analysis in \cite{boston2}.} For one
ensemble (denoted by ``ov3'' in Table~\ref{table:1}) we 
augmented the FP operator, approximating the overlap projection with an order
3 polynomial, to explore even smaller quark masses. 
This construction exploits the fact that an 
approximate solution of the Ginsparg-Wilson
relation converges rapidly towards an exact solution in an overlap
construction (see, for example, \cite{ferenc,fp3,bietenholz}).

\begin{table}[t]
\begin{center}
\renewcommand{\arraystretch}{1.2} % enlarge line spacing
\begin{tabular}{c|c|c|c|c|l|r} \hline\hline
$D$ & $N_s^3\times N_t$ & $\beta$ & $a(r_0)$ & $L$ & \#conf & 
$m_\mathrm{PS}/m_\mathrm{V} $  \\ 
\hline
FP & $16^3\times 32$ & 3.00 & 0.153 fm  & 2.5 fm & \; 200 & 0.28--0.88 \\ 
FP & $12^3\times 24$ & 3.00 & 0.153 fm  & 1.8 fm & \; $200^*$ & 0.3--0.88  \\
ov3& $12^3\times 24$ & 3.00 & 0.153 fm  & 1.8 fm & \; 100 & 0.21--0.88 \\
FP &  $8^3\times 24$ & 3.00 & 0.153 fm  & 1.2 fm & \; $200^*$ & 0.3--0.88  \\
FP & $12^3\times 24$ & 3.40 & 0.102 fm  & 1.2 fm & \; $200^*$ & 0.34--0.89 \\
FP & $16^3\times 32$ & 3.70 & 0.077 fm  & 1.2 fm & \; 100 & 0.36--0.89 \\ 
\hline
CI & $16^3\times 32$ & 7.90 & 0.148 fm  & 2.4 fm & \; 100 & 0.28--0.85 \\
CI & $12^3\times 24$ & 7.90 & 0.148 fm  & 1.8 fm & \; 100 & 0.36--0.85 \\
CI &  $8^3\times 24$ & 7.90 & 0.148 fm  & 1.2 fm & \; 200 & 0.33--0.85 \\
CI & $16^3\times 32$ & 8.35 & 0.102 fm  & 1.6 fm & \; 100 & 0.33--0.92 \\
CI & $12^3\times 24$ & 8.35 & 0.102 fm  & 1.2 fm & \; 100 & 0.32--0.92 \\
CI & $16^3\times 32$ & 8.70 & 0.078 fm  & 1.2 fm & \; 100 & 0.40--0.95 \\ 
\hline
\end{tabular}
\end{center}
\caption{{}Statistics and basic parameters for the runs with the FP and 
CI Dirac operator. We show the lattice size, the lattice spacing $a$
as determined from the Sommer parameter, the spatial extension in fm, 
the statistics and the range of
pseudoscalar to vector mass ratios we worked at. For three cases,
marked with an asterisk in the last but one column, two quark propagators were
calculated on each configuration. 
 \label{table:1}}
\end{table}

Other choices, such as the gauge action 
and the preparation of the sources, are different for the two
calculations. In the FP study we generate the configurations using
the latest parametrization of the FP gauge action \cite{perfgauge}.
Subsequently we treated the configurations with a renormalization
group motivated smearing~\cite{jorg}, which is
related to the ``renormalization group cycle'' studied in \cite{kovacs}. 
The smearing is considered as part of the parametrization of the 
Dirac operator. Furthermore, we fix the gauge to Coulomb gauge.
We use a Gaussian source with a width of about 1 fm and a point-like
sink.

For the runs with the CI operator we use the L\"uscher-Weisz action 
\cite{luweact} with coefficients from tadpole improved perturbation 
theory. One step of hypercubic blocking \cite{anna1} 
was applied to the gauge configurations. For the CI operator we 
use Jacobi smeared sources \cite{jacobi} and therefore no gauge fixing was
necessary. 
In this case, our sources have a width of about 0.7 fm. We
experimented both with smeared and point-like sinks and found no
essential difference. Most of our results presented here refer to
smeared sinks.
 
With the FP gauge action, we use alternating Metropolis and
pseudo over-relaxation sweeps over the lattice, with 2000 sweeps for
thermalization and 500 sweeps to separate between different
configurations. The number of separation sweeps is a worst-case
estimate based on measurements of autocorrelation times for simple
gluonic operators \cite{rufenacht}. We found no correlation between 
the hadron propagators on different configurations either. 

For the L\"uscher-Weisz action, a sweep contained 4 Metropolis steps and 1
pseudo over-relaxation step. On the largest lattices, 3000 sweeps were
used for thermalization and 1000 sweeps to separate the
configurations. It has been checked that this procedure decorrelates
even the topological charge of the configurations quickly \cite{topch}.

We use periodic boundary conditions both for the gauge and for the
fermion fields in the runs with the FP operator. For the CI operator 
the temporal boundary conditions of the fermions are chosen anti-periodic. 

\subsection{The FP and CI lattice Dirac operators}

The structure of the FP and CI Dirac operators is similar. They both have
fermion offsets essentially on the hypercube only. In Dirac space not
only $\gamma_\mu$ and $1$ enter, corresponding to
derivative and Wilson terms, but also all the other elements of 
the Clifford algebra.
The structure of these terms is restricted by the symmetries C, P, 
$\gamma_5$-hermiticity and invariance under $90^\circ$ rotations
\cite{fp3,ci1}.

The FP and CI operators differ in the way the
coefficients of the terms are chosen. For the FP operator the coefficients are
determined from the saddle point approximation of the renormalization group
equation. The parametrized FP operator used here is described by 82 couplings
corresponding to 41 independent terms. 
A detailed description of the FP operator is given in \cite{fp3,jorg}. 
For the CI operator the Ginsparg-Wilson equation is mapped onto a system
of coupled quadratic equations for the expansion coefficients of the
operators. After truncation this system can be solved 
numerically and the resulting coefficients give rise to an approximate
solution of the Ginsparg-Wilson equation. The parametrization of the CI
operator used here has 19 coefficients corresponding to 19 independent terms. 
A detailed description can be found in \cite{ci2,ci3,ciweb}.

Our calculations were mainly done on the Hitachi SR8000 parallel computer 
at the Leibniz Rechenzentrum in Munich. A smaller part was performed on 
the IBM SP4 at CSCS in Manno. In the numerical implementation we
pre-calculate the Dirac operator and store it in memory. The FP operator
contains a large number of paths, but the sum of paths for many
couplings can be factorized, which reduces the build-up time
significantly. This part of the code was repeatedly optimized and, in its
present form, the time needed to construct the FP Dirac operator is
negligible in comparison to that of inverting it on one source.
For the CI operator the number of paths used is smaller and a straightforward 
pre-calculation and storage of all paths is possible. For the quark masses 
considered the overhead for 
the pre-calculation is less than 10\% of an inversion on a complete 
basis of 12 source vectors.

Our codes run on the Hitachi SR8000 machine quite efficiently. The FP
Dirac matrix-vector multiplication runs at 6.3 GFLOPS per 
node\footnote {A node contains 8 CPUs of 0.375 GHz and has a theoretical
peak speed of 12 GFLOPS by counting 2 floating-point multiplications and
2 additions per cycle per processor.} and the overall efficiency in
calculating the quark propagator is around 30\% of the peak performance
\cite{hauswirth,computing}. 
The performance of the CI code is similar to that of the FP. Production runs 
were typically done on 2 - 8 nodes using MPI for the communication 
between the nodes. 

The exact FP Dirac operator satisfies the Ginsparg-Wilson equation
(we often set the lattice spacing $a = 1$ for notational convenience)
\begin{equation}
\label{gw}
\gamma_5 D + D \gamma_5 = D \gamma_5 2R D \,,
\end{equation}
with a non-trivial local matrix $R$ which lives on the hypercube 
and is trivial in Dirac space. Note that the inverse of $R$ is also local.
The quark mass is introduced as
\begin{equation} \label{Dm_mod0}
D(m)  = D +m \left( \RRi - \half D \right) \,.
\end{equation}
which assures ${\cal O}(a)$ improvement in spectroscopy. Since the
inverse of $D(m)$ can be written as
\begin{equation}
\label{Dm_inverse}
D(m)^{-1}=\frac{1}{1-m/2} 2R \left[ D 2R + \frac{m}{1-m/2}\right]^{-1} \,,
\end{equation}
the multi-mass solver can be easily generalized to this case. The
overhead due to the presence of the matrix $R$ is small.

The CI operator corresponds to a Ginsparg-Wilson equation with $2R=1$. The
quark mass term of the CI action corresponds to that on the r.h.s.~of
\eqref{Dm_mod0} with $2R=1$.

We typically worked at 10 different quark masses and
inverted the Dirac matrix using the BiCGstab multimass solver
\cite{multimass}. We remark that the results for
different quark masses come from the same set of gauge configurations
and are thus correlated.

\subsection{Fits and errors}
\label{ssect:fitserrs}

We fitted the zero momentum hadron propagators with a single mass
form with 2 parameters (mass $m$ and amplitude $\alpha$) in a time interval
$ t\in [t_0,t_1]$. The $\chi^2$-function is defined as
\begin{equation} 
\label{chi2}
\chi^2=\sum_{t,t'} \{C(t)- f(t)\}\,w(t,t')\,\{C(t')- f(t')\},
\end{equation}
where $C(t)$ is the measured average of the hadron propagator
on the time slice $t$ and $f(t)$ is the function used for fitting,
e.g.\ $f(t) = \alpha \, \exp(- m t)$.

For any positive definite weight $w$ in \eqref{chi2},
independent of the generated configurations, 
one obtains an unbiased estimator for the parameters $\alpha$ and $m$
if the non-linearities caused by the fluctuation of the non-linear
fit parameter $m$ are negligible (assuming, of course, that the ansatz
describes the physics in  $ t\in [t_0,t_1]$). The values of the hadron
propagator at different time slices are strongly correlated.
The standard deviation of the
parameters is minimized if the weight $w$ is chosen to be the inverse of
the exact covariance matrix of $C$: $\mathrm{Cov}(t,t')^{-1}$. Since the
exact covariance matrix is not known, we tested and compared several
possibilities: the measured covariance matrix 
\begin{equation}
\mathrm{Cov}_\mathrm{meas}(t,t') =
\langle [C(t) - \overline{C}(t)][C(t')- \overline{C}(t')]\rangle/(N-1)\\,
\end{equation}
the diagonal part and a parametrized form of it, 
$\mathrm{Cov}_\mathrm{measd}$ and $ \mathrm{Cov}_\mathrm{measp}$, 
respectively. 
Each of these choices is plagued by its own problems. 
$\mathrm{Cov}_\mathrm{meas}$ is correlated with the hadron propagator which 
leads to a biased estimator.\footnote{This bias goes to zero as the
statistics is increased.} 
Due to fluctuations, $ \mathrm{Cov}_\mathrm{meas}$
can have small artificial eigenvalues which occasionally might 
lead to instabilities in the fit. Both problems are reduced somewhat
by a smooth parametrization $ \mathrm{Cov}_\mathrm{measp}$. Taking 
$\mathrm{Cov}_\mathrm{measd}$ reduces these problems further, but the 
corresponding standard deviation of the fit parameters will be larger and
the value of $\chi^2$ will be artificially small and useless. 

In our large volume ($L = 2.5\,\mathrm{fm}$) spectroscopy analysis
we mainly used $\mathrm{Cov}_\mathrm{meas}$ as the weight 
in \eqref{chi2}, but we compared the results to those obtained with
the alternative weight factors discussed above. In a few cases
(typically for the vector meson and $N^*$, the negative parity
partner of the nucleon) the fit values obtained with
different weight factors were not consistent
within their errors. It happened also that the fit with
$\mathrm{Cov}_\mathrm{meas}$ reacted in an unstable way to the choice of the
fit interval $[t_0,t_1]$. In such cases we have used the safest choice
$\mathrm{Cov}_\mathrm{measd}$ for the weight factor and accepted the
corresponding larger  
statistical error. In general, however, the different fits were fully
consistent. The lower bound $t_0$ was fixed by the requirements that
the $\chi^2$ per degree of freedom ($\chi^2_\mathrm{df}$) 
should be ${\cal O}(1)$, the estimated contribution of 
the first excited state should be negligible for $t \geq t_0$ and $t_0$
should be consistent with a visual check of the effective mass and fitted
mass plots. The upper bound $t_1$ was set by requiring that the hadron
propagator should be larger than its error.
 
The small volume ($L = 1.2\,\mathrm{fm}$) and, to a lesser extent, the
intermediate volume ($L = 1.8\,\mathrm{fm}$) cases were more problematic.
First, the topological finite size artifacts (see Section 3) were larger and 
we were forced to take correlators where these effects were canceled (reduced)
which usually increased the statistical errors. In addition, the statistics
were lower for the small volume simulations.
In these cases we decided to use an uncorrelated fit
with the $\mathrm{Cov}_\mathrm{measd}$ weight and to accept the increased 
statistical errors.  

We invested considerable effort in analyzing the data. 
We found, nevertheless, cases (in the smallest $L=1.2\,\mathrm{fm}$ box, 
in particular on the finest $a= 0.08\,\mathrm{fm}$ lattice) where we were 
not able to give a reliable 
error estimate. For those cases we do not quote numbers. 

The statistical errors were estimated by jackknife resampling.
We analyzed the FP and CI data using comparable criteria. 
In Appendices \ref{FP:data} and \ref{CI:data} we list our hadron mass
results for the FP and CI actions together with the temporal 
fit range and the resulting value of $\chi^2_\mathrm{df}$ for the fit. 
We also collect mass ratios, fit coefficients, etc.\ in these appendices.

\section{Hadron operators and their topological finite size artifacts}

In the chiral region the hadron correlators suffer from a topological finite
size effect specific to the quenched approximation
\cite{domain1,GiHoRe01,DoDrHo02}. These topological finite size artifacts
are different in nature from the physical finite size effects which appear
mainly due to light pseudoscalar mesons traveling around the periodic 
boundary. 
The physical finite size effects show up both in quenched and dynamical 
simulations although they are weakened significantly in the quenched 
case \cite{fin_size}.

The topological finite size artifacts come from a different source. When 
replacing the individual propagators in hadron correlators by their spectral 
sum one can isolate the contributions of the zero modes in the resulting 
expressions. 
At any fixed lattice volume $V$ these terms show a power law behavior    
$(am_q)^{-\tau}$ with some non-negative integer $\tau$
as the quark mass $m_q$ approaches zero.
On the other hand, since the abundance of zero modes scales 
only as $\sqrt{V}$ while the density of the non-zero modes is $\sim V$,
at fixed quark mass these effects
go to zero as $1/\sqrt{V}$ such that the effect is significant only 
in small volumes. In a dynamical simulation the zero modes are suppressed 
by the fermion determinant and no topological finite size artifacts occur.

\begin{figure}[tbp]
\includegraphics[width=6cm]{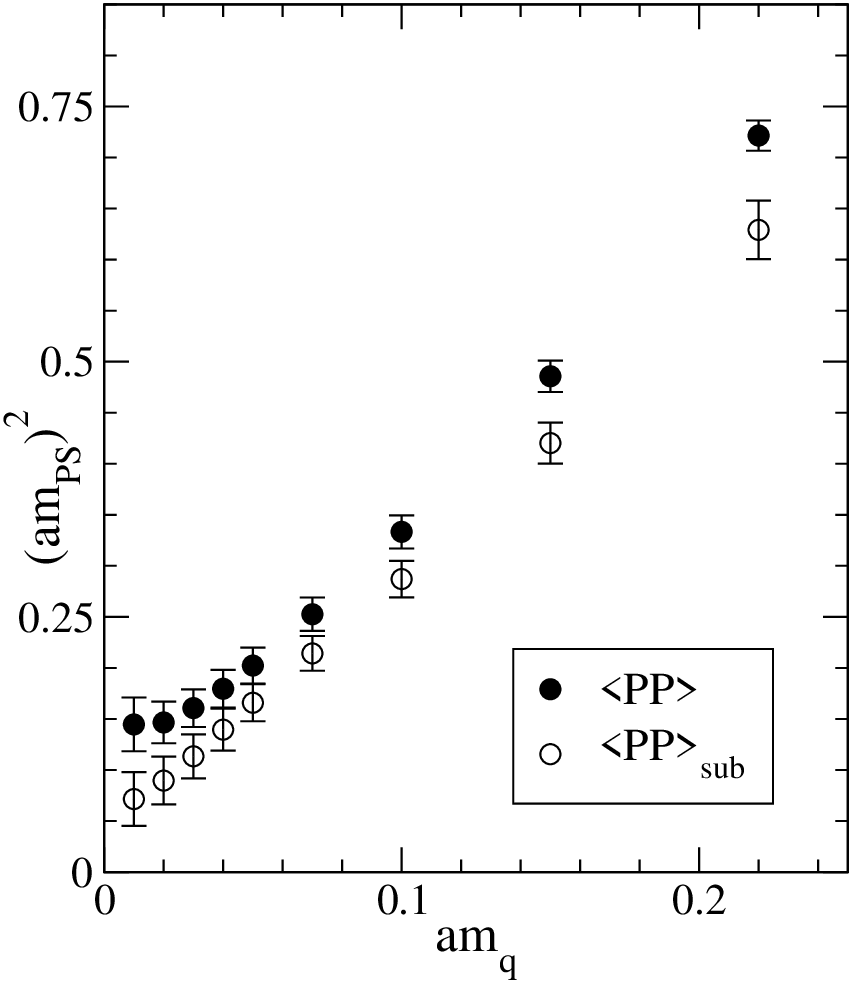}
\includegraphics[width=6cm]{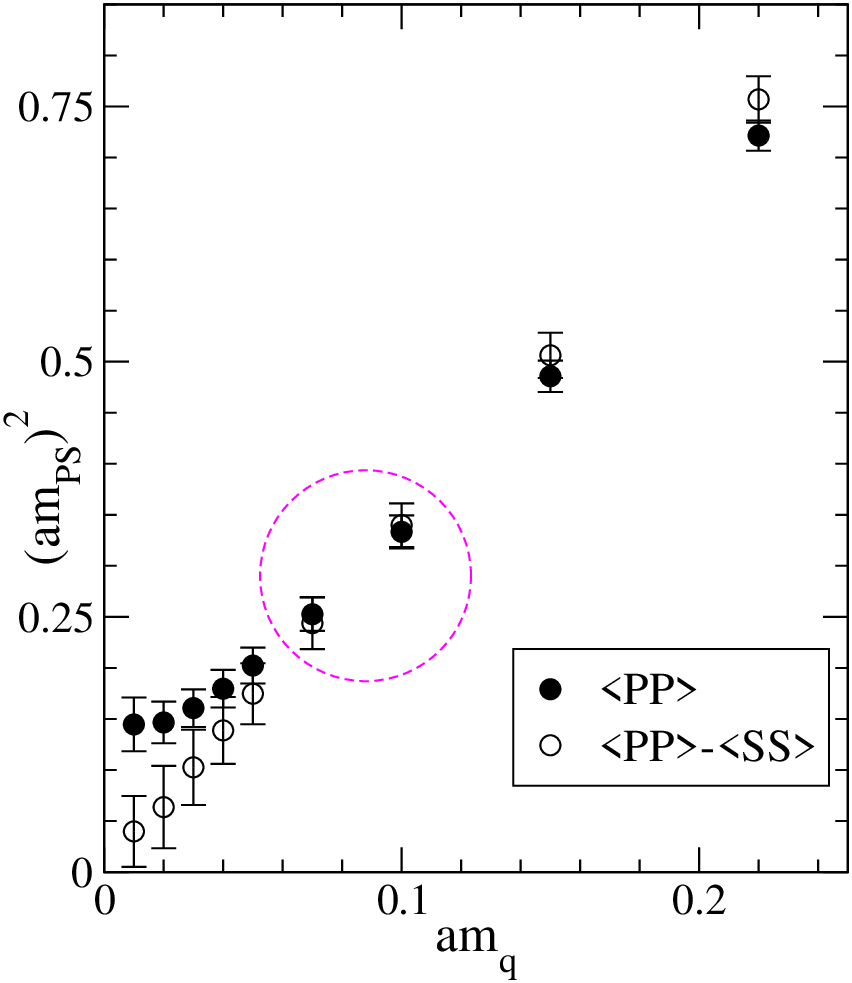}
%\vspace{-7mm}
\caption{{}Effect of the subtraction of the zero modes on the pion mass 
(FP-ov3 operator $6^3\times16$, $a = 0.15~\mathrm{fm}$). 
The mass from $\langle P\,P \rangle$ 
is represented by filled circles, while we use open circles
to represent the results from the explicit removal of the zero modes 
from the quark propagators (l.h.s.\ plot)
and from the correlator $\langle PP \rangle - \langle SS\rangle$ of 
Eq.~(\protect{\ref{subcorr}}) (r.h.s.\ plot).
% zmcorr02.eps zmcorr01.eps
}
\label{fig:1}
\end{figure}

One might consider different strategies to eliminate or reduce these
artifacts in the quenched approximation. The straightforward method of
subtracting the zero modes from the {\it quark} propagator (and thus from all
the hadron propagators) on each configuration is not only expensive, but
dangerous as well. This hand-made procedure might change the hadron propagator
in a non-local manner as illustrated for the pseudoscalar channel in the
l.h.s. of Fig.~\ref{fig:1} \cite{hauswirth}. In this figure filled and open
symbols are used for the 
original and subtracted correlators, respectively. Even at large quark masses,
where the zero mode artifacts cannot play any important role, the predicted
pseudoscalar mass from $\langle P \, P \rangle_\mathrm{sub}$ deviates
from that obtained from $\langle P \, P \rangle$.

In this work we considered a different strategy which, from a
field-theoretical point of view, is safe. There is a considerable
freedom in choosing source/sink operators with given quantum numbers. 
This freedom can be used to cancel or reduce the topological finite size
artifacts. This is the method we applied in the baryon sector, where we
considered the following set of octet and decuplet operators: 
\begin{eqnarray}
N \, & \; = \; &
(d^a {\mathcal C} \gamma_5 u^b) u^c \epsilon_{abc} \; ,
\nonumber
\\
N_4 & \; = \; & 
(d^a {\mathcal C} \gamma_4\gamma_5 u^b) u^c \epsilon_{abc} \; ,
\nonumber
\\
\Delta \, & \; = \; & 
   2(d^a {\mathcal C} \gamma^- u^b) u^c \epsilon_{abc} 
 + (u^a  {\mathcal C} \gamma^- u^b) d^c \epsilon_{abc} \; ,
\nonumber
\\
\Delta_4 & \; = \; & 
   2(d^a {\mathcal C} \gamma_4\gamma^- u^b) u^c \epsilon_{abc} 
 + (u^a  {\mathcal C} \gamma_4\gamma^- u^b) d^c \epsilon_{abc} \;.
\label{nucleonoperators}
\end{eqnarray}
Here ${\mathcal C}$ is the charge conjugation matrix,
$\gamma^-=\gamma_1 - i \gamma_2$, and the 4th direction is the time direction. 

These operators can be combined in different correlators which have different 
quark mass singularities for their topological finite size effect. 
After performing the fermion contractions and isolating the zero mode
contributions one finds the following leading powers for the singularities :

\begin{subequations}
\label{Ncorrs}

\begin{eqnarray}
\langle \overline{N} \, N \rangle & \; \sim \; & (am_q)^{-3} \; ,
\label{Ncorr1}
\\
\langle \overline{N} \, \gamma_4 \, N  \rangle 
& \; \sim \; & (am_q)^{-2} \; ,
\label{Ncorr2}
\\
\langle \overline{N_4} \, N_4  \rangle & \; \sim \; & (am_q)^{-2} \; ,
\label{Ncorr3}
\\
\langle \overline{N_4} \, \gamma_4 \, N_4   \rangle 
& \; \sim \; & (am_q)^{-1} \; .
\label{Ncorr4}
\end{eqnarray}
\end{subequations}
The first and third correlators are even, the second and fourth 
are odd under $t \leftrightarrow T-t$. To improve the signal
we (anti)symmetrized the measured propagators. The separation of the nucleon
and its negative parity partner $N^*$ will be addressed in Section 4.
For the decuplet channel the corresponding
correlators are obtained by replacing the $N$ operators by the $\Delta$ 
operators. 
The mass singularities for the different operators remain the same under
this replacement. 

In the meson sector the standard point-like operators are
\begin{eqnarray}
P & \; = \; & \overline{\psi} \, \gamma_5 \, \psi \; ,
\nonumber
\\
S & \; = \; & \overline{\psi} \, \psi \; ,
\nonumber
\\
A_4 & \; = \; & \overline{\psi} \, \gamma_4 \gamma_5 \, \psi \; ,
\nonumber
\\
V_i & \; = \; & \overline{\psi} \, \gamma_i \, \psi \; .
\label{mesonoperators}
\end{eqnarray}
In these expressions we suppress the flavor content of the operators 
and use $\psi$ for quark fields. All meson correlators we considered are
flavor non-singlet. In these correlators the leading topological finite size
artifacts read:
\begin{eqnarray}
\langle P \, P \rangle & \; \sim \; & (am_q)^{-2} \; ,
\nonumber
\\
\langle S \, S \rangle & \; \sim \; & (am_q)^{-2} \; ,
\nonumber
\\
\langle A_4 \, A_4 \rangle & \; \sim \; & (am_q)^{-1} \; ,
\nonumber
\\
\langle V_i \, V_i \rangle & \; \sim \; & (am_q)^{-1} \; .
\label{mesoncorrelators}
\end{eqnarray}

\begin{figure}[tbp]
\begin{center}
\includegraphics[width=9.5cm,clip]{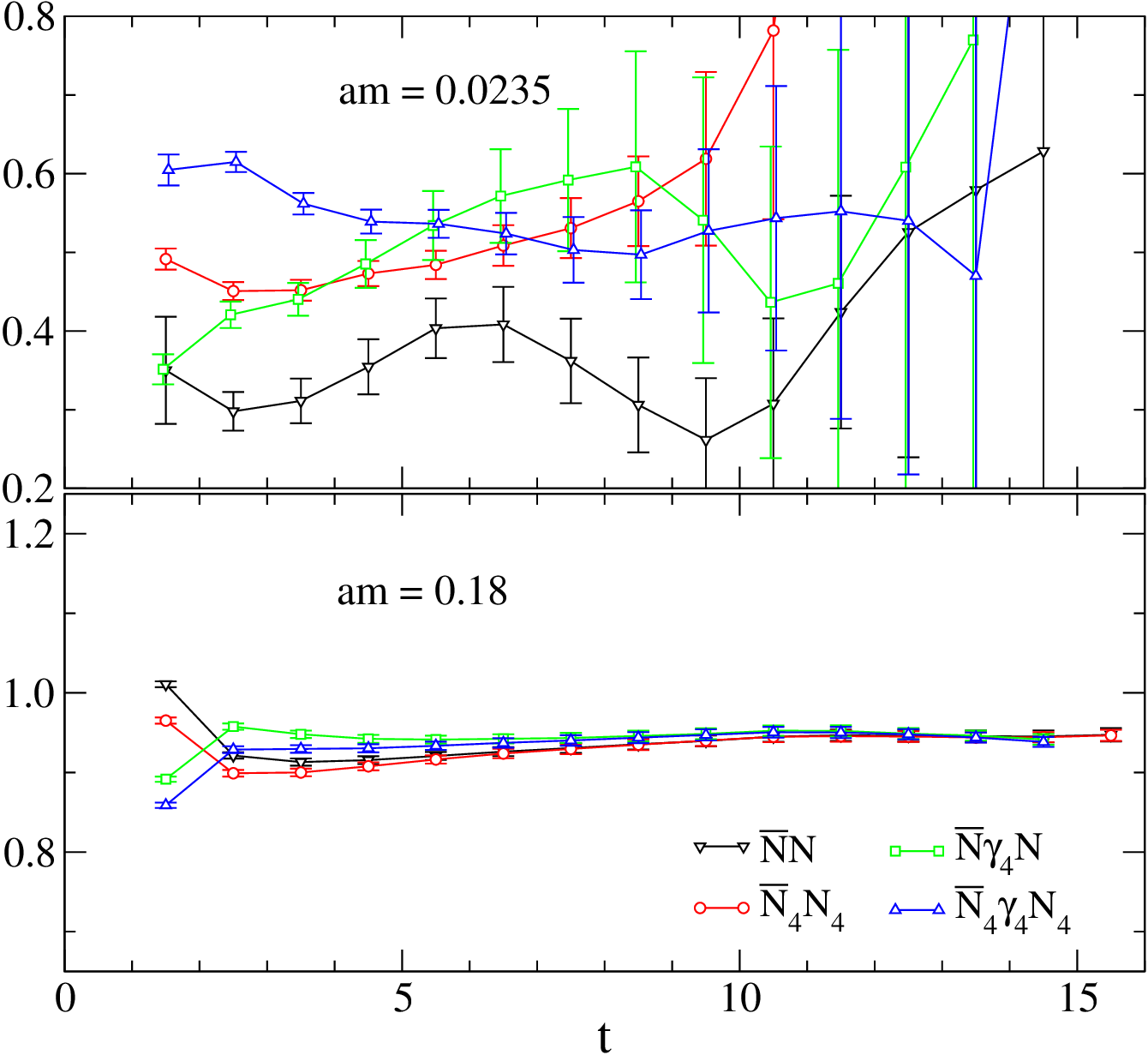}
\end{center}
\vspace{-6mm}
\caption{{}Effective mass plots for the nucleon measured
with four different correlators $\overline{N} N$,
$\overline{N} \gamma_4 N$, $\overline{N_4} N_4 $, 
$\overline{N_4} \gamma_4 N_4$, 
on the $16^3\times 32$ lattice at $a=0.08\,\mathrm{fm}$ using FP.
We show results for a light and a heavy quark  
(corresponding to $m_\mathrm{PS}/m_\mathrm{V}=0.35$ and 0.89).
% meff\_4Nop\_16x32\_b3.70.eps
}
\label{fig:4Nop}
\end{figure}

In Fig.~\ref{fig:4Nop} we show the effective mass plots 
for the 4 different nucleon correlators listed in 
Eq.~\eqref{Ncorrs} for a small (top plot) and a large quark
mass (bottom plot). 
The data are for the $16^3\times 32$ lattice at $a=0.08\,\mathrm{fm}$,
i.e.\ the lattice has a small physical volume of 1.2 fm. 
For the light quark (top plot)
the effective mass curve from the correlator having an 
${\cal O}((am_q)^{-3})$ artifact lies significantly below those of the 
correlators with a reduced ${\cal O}((am_q)^{-1})$ contribution. 
If the quark mass is heavy the artifact is strongly reduced
and all the correlators give consistent results as the bottom plot of
Fig.~\ref{fig:4Nop} illustrates. 

For the pseudoscalar mass we followed the suggestion in
\cite{domain1,GiHoRe01,DoDrHo02} which is based on the fact that for exactly 
chirally symmetric actions the scalar propagator has the 
same topological finite size effect as the pseudoscalar propagator.
Thus we can consider the difference of the pseudoscalar and scalar
2-point functions 
\begin{equation}
\langle PP \rangle \; \; - \; \; \langle SS \rangle \; .
\label{subcorr}
\end{equation}
Since for light quarks
the lightest particle in the scalar channel is much heavier than 
the pseudoscalar, its contribution quickly dies out with time separation. 
For heavy quarks, however, the relative mass difference between 
the pseudoscalar and scalar mass is small and a single-mass fit 
produces a value which is higher than the true pseudoscalar mass.

This observation suggests the following strategy: At small quark
masses, where $\langle PP \rangle$ is strongly distorted, we determine the 
pseudoscalar mass from $\langle PP \rangle - \langle SS \rangle$.
Going towards larger quark masses the
effect of the zero modes is suppressed and we expect a
window where the two correlators lead to consistent mass fits. In
this window and beyond it we use the $\langle PP \rangle$ correlator. 
The r.h.s.\ plot in Fig.~\ref{fig:1} illustrates \cite{hauswirth} these 
expectations on a small lattice for the FP-ov3 operator. 
Filled symbols are used for the 
original $\langle PP \rangle$ correlator and open symbols for the 
difference $\langle PP \rangle - \langle SS \rangle$. The circle indicates 
the window where the zero mode effects are already negligible, but the
scalar mass is much larger than the pseudoscalar and so the correct 
pseudoscalar mass is easily seen in 
the $\langle PP \rangle - \langle SS \rangle$ correlator. 

For the vector meson we used the point-like vector density of
Eq.~\eqref{mesonoperators} 
and we summed over $i=1,2,3$ in the vector meson
propagator Eq.~\eqref{mesoncorrelators}. 
This correlator has a remaining ${\cal O}(m_q^{-1})$ topological
finite size effect. There exist different ways to cancel this
remaining contribution also, but we neither tested nor used them in this
work. 

\begin{figure}[tbp]
\begin{center}
\vspace{-4mm}
\includegraphics[width=9cm,clip]{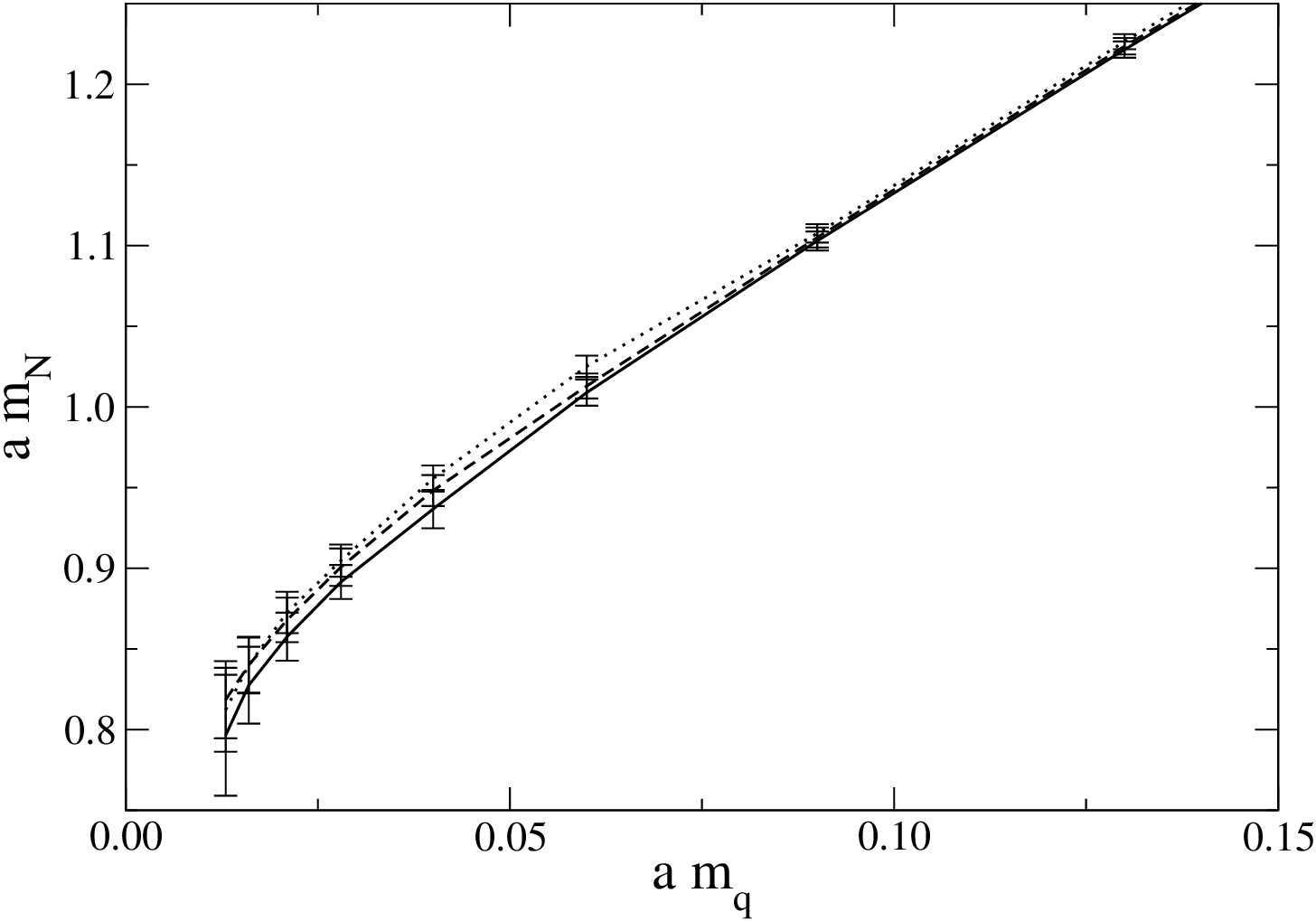}
\end{center}
\vspace{-4mm}
\caption{{}The dependence of the nucleon mass on the bare quark mass was
measured with four different correlators. The lower and upper curves 
correspond to correlators \eqref{Ncorr1}, \eqref{Ncorr2}. 
The correlators  \eqref{Ncorr3}, \eqref{Ncorr4} 
give almost identical nucleon masses corresponding to the curve in
the middle. FP, $16^3 \times 32,~ a=0.153\,\mathrm{fm}$.
% mN\_16x32\_b3.00\_fp.eps
}    
\label{fig:N_curves}
\end{figure}

Having discussed topological finite size artifacts in different operators and
some strategies how to deal with them we now demonstrate that for our 
largest lattice with size $L = 2.5$ fm they are not observed for 
the nucleon. We do not see such artifacts in the vector channel either. 

In Fig.~\ref{fig:N_curves} we show the nucleon mass as a function 
of the quark mass (both in lattice units) 
determined from all 4 correlators listed in 
Eq.~\eqref{Ncorrs}. If the topological finite size artifacts 
were still important for the 2.5~fm lattice then one would expect that 
the curves for the different operators would be different and this 
difference would increase substantially for small quark masses. 
The plots do not show such a behavior and we conclude that for the nucleon 
the topological finite size artifacts are negligible for $L = 2.5$ fm. 
On the other hand, 
as we shall discuss in the next section, we still see a small remaining 
effect in the pseudoscalar sector at small quark masses even in this 
large volume.

\section{Spectroscopy in a large volume}
\label{sect:spectro}

\subsection{Discussion of the raw data}

Here we present our results which were obtained in a box 
$L^3 \times 2L$ with $L = 2.5\,\mathrm{fm}$ for FP, $L = 2.4\,\mathrm{fm}$ 
for CI, on a coarse lattice with
$a(r_0)= 0.153\,\mathrm{fm}$ for FP and $a(r_0)= 0.148\,\mathrm{fm}$ for CI. 
In Section 3 we have shown that for these lattice sizes topological finite
size artifacts are negligible for the baryons in the quark mass range 
we consider. Thus, among the different choices which we have discussed 
for the nucleon correlator,
we here choose the most convenient one. It is a linear combination 
of \eqref{Ncorr1} and \eqref{Ncorr2}
giving rise to a projection onto positive parity. Thus
in forward time direction only the nucleon propagates and only 
its negative parity partner $N^*$ in backward time direction
\cite{nstar}:
\begin{equation}
\langle \overline{N}(0) \, \frac{1}{2} ( 1 + \gamma_4) \, N(t) \rangle
\; \; \sim \; \; A \,\mathrm{e}^{-mt} \; + \; B\, \mathrm{e}^{-m^*(T-t)} \, . 
\label{ncorrlarge}
\end{equation}
For the  CI operator a more detailed study of the nucleon system 
using a basis of three operators and a variational technique to separate 
ground and excited states is presented elsewhere \cite{cinstar}.
\begin{figure}[tbp]
\begin{center}
\vspace{-4mm}
\includegraphics[width=9.9cm,clip]{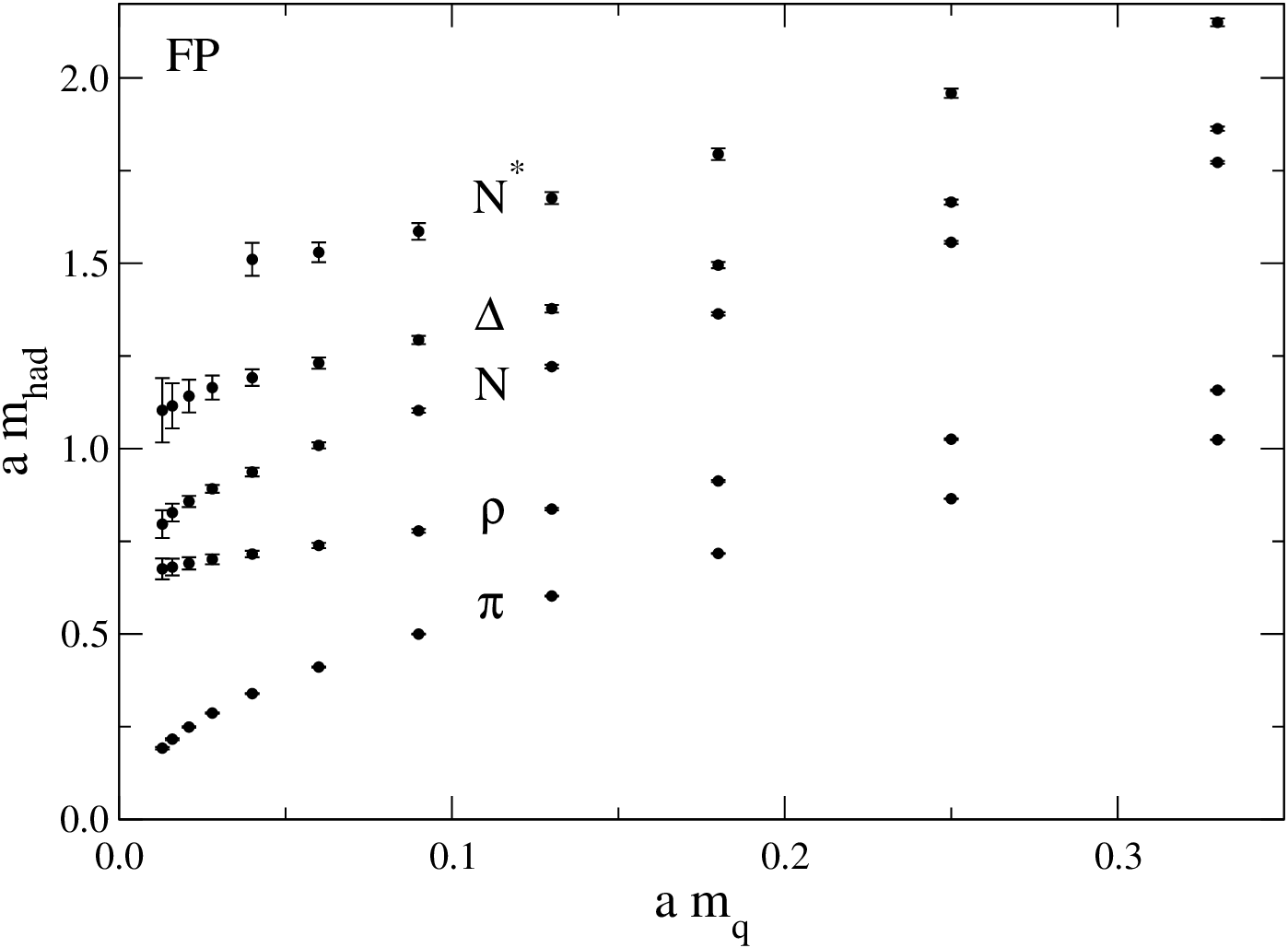}
\end{center}
\vspace{-7mm}
\caption{{}An overview of the hadron spectrum vs.~the bare quark mass 
as measured with the FP action at $a=0.153\,\mathrm{fm}$ and 
$L=2.5\,\mathrm{fm}$.
For the negative parity partner of the nucleon $N^*$ no
reasonable mass could be obtained for the four smallest quark masses.
% m\_16x32\_b3.00\_fp.eps
} 
\label{fig:overview_FP}
\begin{center}
\vspace{3mm}
\includegraphics[width=9.9cm,clip]{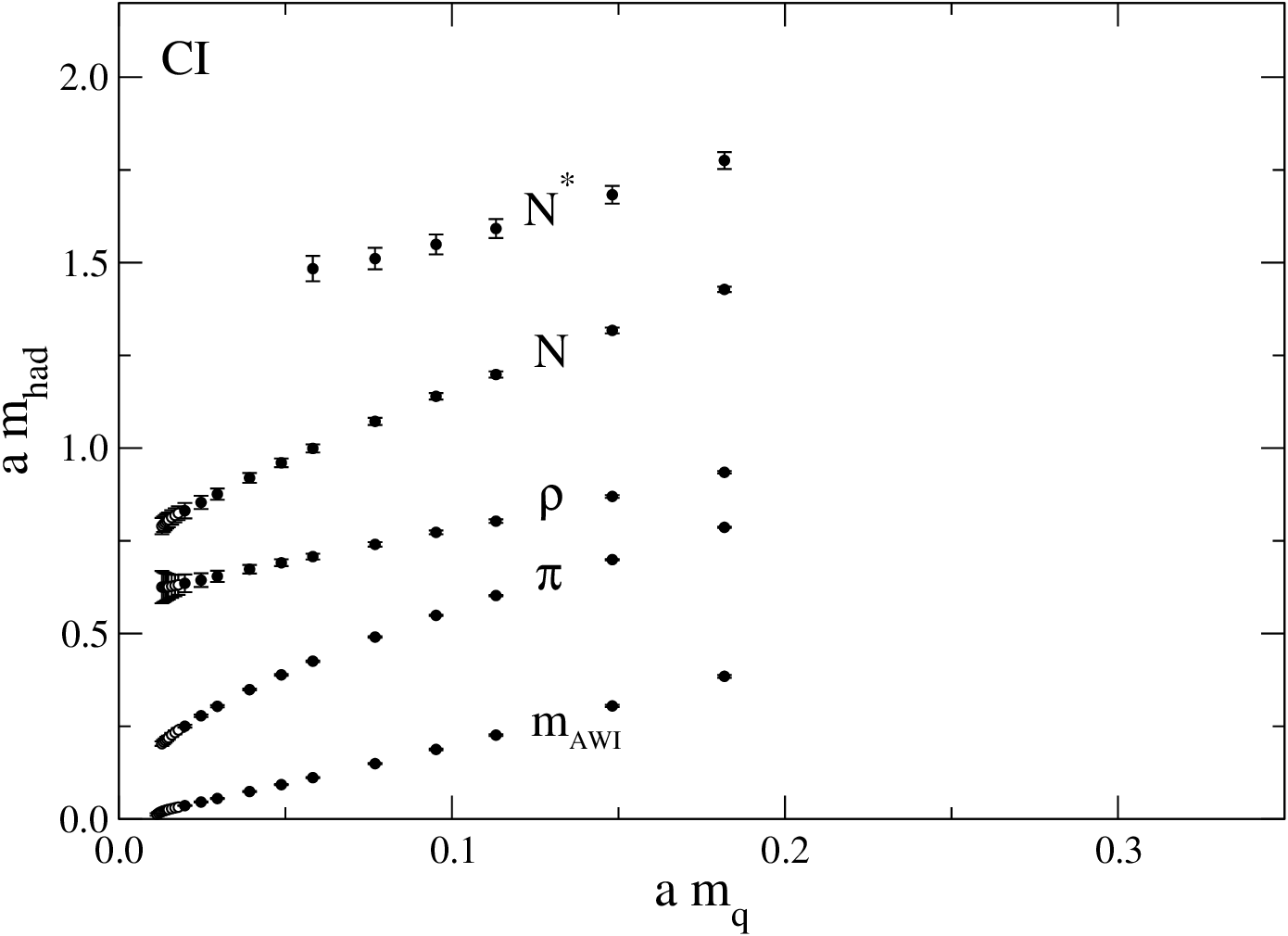}
\end{center}
\vspace{-7mm}
\caption{{}The same as Fig.~\protect{\ref{fig:overview_FP}} now
for the CI action
($a=0.148\,\mathrm{fm}$ and $L=2.4\,\mathrm{fm}$). 
% m\_16x32\_b7.90\_ci.eps
 } 
\label{fig:overview_CI}
\end{figure}
For the vector meson the last correlator in Eq.~\eqref{mesoncorrelators}
was used for both the FP and the CI operator. 
For the FP operator the pseudoscalar mass was determined from 
$\langle PP \rangle - \langle SS \rangle$ for small quark masses and 
from $\langle PP \rangle$ for larger masses as discussed in Section 3. 
The switching points can be found in the tables in Appendix A. 
For the CI operator we used the $\langle A_4 A_4 \rangle$ correlator. 
  
Figures \ref{fig:overview_FP}, \ref{fig:overview_CI} give an
overview of our raw data on the largest volume. 
We plot the different hadron masses as a function of
the bare quark mass for the two different Dirac operators. 
Although the errors increase as we approach small quark mass values, 
they remain under control in the $\pi,\rho,N$ and $\Delta$
channels. For the $N^*$ no reasonable mass and error estimate could be
obtained for the smallest quark masses. 
The lowest data set in Fig.~\ref{fig:overview_CI}
is the axial Ward identity mass $m_\mathrm{AWI}$ which we will discuss 
in the next section.

\subsection{AWI masses}

\begin{figure}[p]
\begin{center}
\vspace{-7mm}
\includegraphics[width=7.5cm,clip]{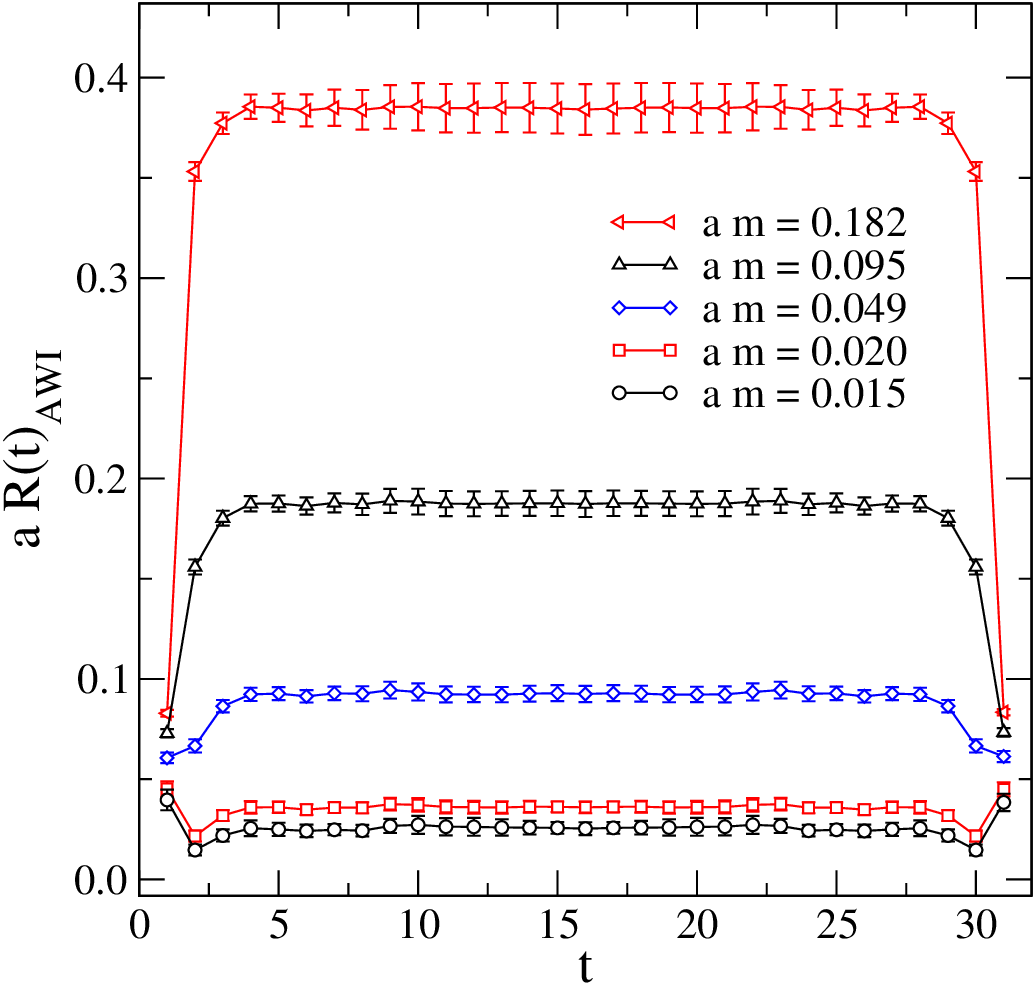}
\end{center}
\vspace{-7mm}
\caption{{}Time dependence of the ratio $R(t)_\mathrm{AWI}$ defined in 
Eq.~\eqref{awiratio} which we use to determine the axial Ward identity mass. 
CI, $16^3 \times 32$, $a = 0.148$~fm.
% ratio\_16x32b790\_hyp\_sp\_12\_2.eps
} 
\label{awiplateauplot}
\begin{center}
\vspace{5mm}
\includegraphics[width=9.9cm,clip]{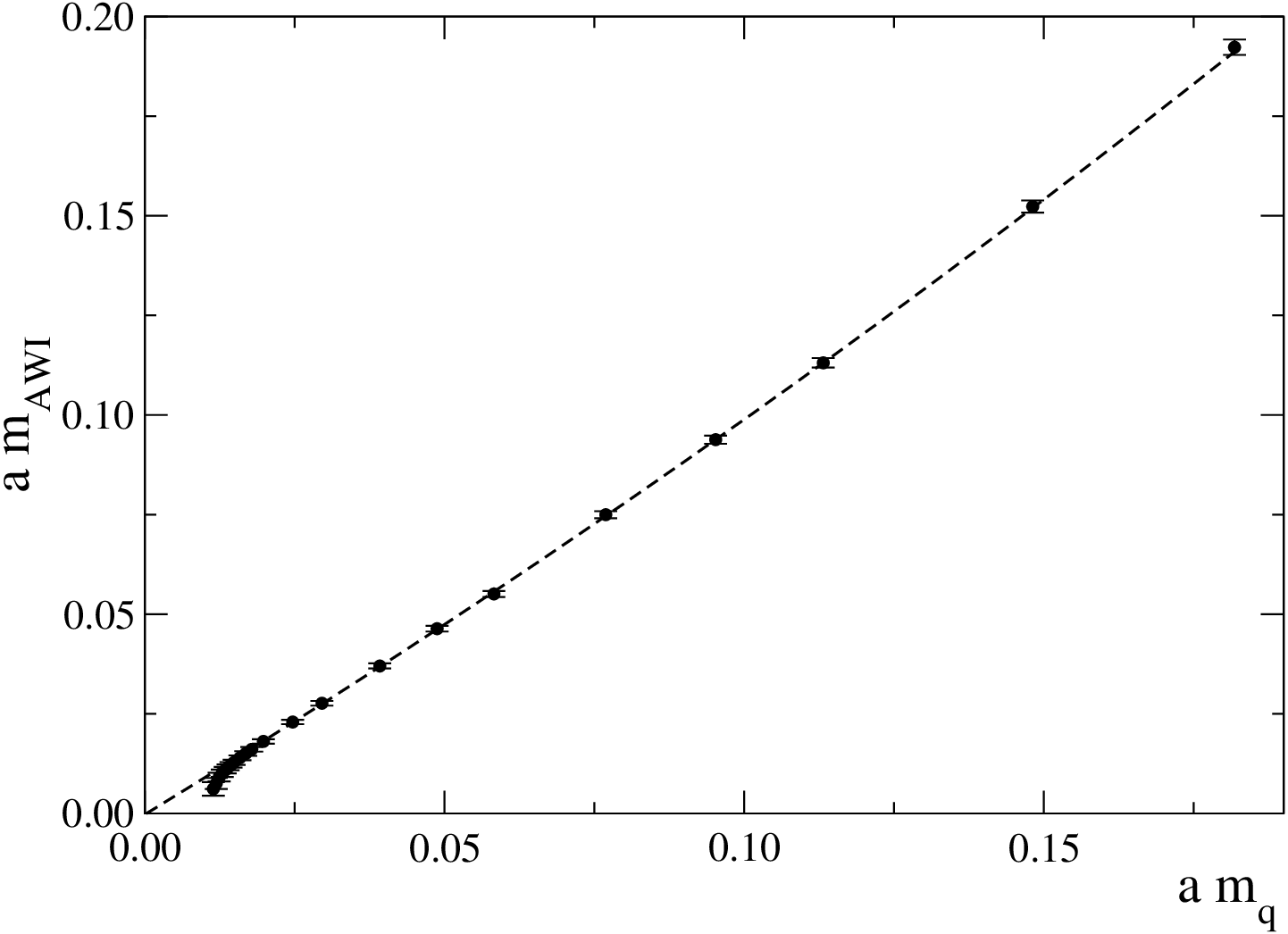}
\end{center}
\vspace{-7mm}
\caption{{}Bare axial Ward identity mass as a function of the bare quark mass
(both in lattice units) from the CI operator on the $16^3 \times 32$
lattice at $a = 0.148$ fm. 
The dashed line represents a quadratic fit in $m = m_q + m_\mathrm{res}$
to the data with a few small quark masses omitted.
% mAWI\_16x32\_b7.90\_ci0.eps
 } 
\label{awimassplot}
\end{figure}

An important observable are quark masses from the axial Ward identity.
To extract the axial Ward identity mass $m_\mathrm{AWI}$ we study the 
ratio 
\begin{equation}
R(t)_\mathrm{AWI} \; \; = \; \; 
\frac{\langle \partial_4 A_4(t) \, P(0) \rangle}{\langle P(t) \, P(0) \rangle}
\; .
\label{awiratio}
\end{equation}
The axial Ward identity bare mass is then given by $a \, m_\mathrm{AWI} =  
\frac{1}{2}R(\infty)_\mathrm{AWI}$.
In Eq.~\eqref{awiratio} the sink operators are projected to zero 
momentum and no smearing was applied to the sink. The factor from the smearing 
of the source operator ($P(0)$) cancels 
in the ratio. Note that for obtaining renormalized masses the r.h.s.\ of
Eq.~\eqref{awiratio} has to be multiplied with renormalization factors; 
these will be discussed elsewhere \cite{renormalizationci}. 
Here we will use the bare values which for our purposes (fits in 
the pseudoscalar channel, estimate of the residual quark mass) are
sufficient. 
Note that we use the naive current $A_4$ given in Eq.~\eqref{mesonoperators}.
The time derivative is implemented by a nearest neighbor difference.

We find that the ratio $R(t)$ exhibits excellent plateaus all the way 
down to our smallest quark masses. This is demonstrated in 
Fig.~\ref{awiplateauplot} where we show $R(t)$ for several quark masses. 
We fold the ratio around $T/2$ and subsequently fit the plateaus from 
$t = 6 \, ... \, 12$ to determine $a m_\mathrm{AWI}$. 
The results for $a \, m_\mathrm{AWI}$ from the CI operator 
as a function of the quark mass are shown in Fig.~\ref{awimassplot}. 
The data show essentially a linear behavior with some positive curvature at
large masses. At very small quark masses 
we find a downward trend which is a 1-3 standard deviation effect.

In the quenched approximation the hadron propagators might contain pieces, 
for example contributions from topological finite size artifacts, which cannot 
be associated with normal contributions from energy eigenstates.
In our large box these are small but can still be seen.
As Tables~\ref{tab:16x32_b3.0_PSV}, \ref{tab:16x32_b3.0_B}, \ref{tab:ci1}, 
\ref{tab:ci2} show, the pseudoscalar mass obtained from 
the $\langle PP \rangle$ correlator is larger than that from 
$\langle PP \rangle - \langle SS \rangle$ at the smallest quark 
masses by 1-2~$\sigma$.
The downward trend in Fig.~\ref{awimassplot} might be
related to such artifacts and/or statistical fluctuations.
The dashed line in this figure represents a quadratic fit
in $m = m_q + m_\mathrm{res}$
to the data with the smallest quark masses omitted.
Since the fit obviously does not describe the whole set of data
we do not quote a value for $m_\mathrm{res}$.

\subsection{The pseudoscalar channel}

\begin{figure}[p]
\vspace{-6mm}
\begin{center}
\includegraphics[width=9.5cm,clip]{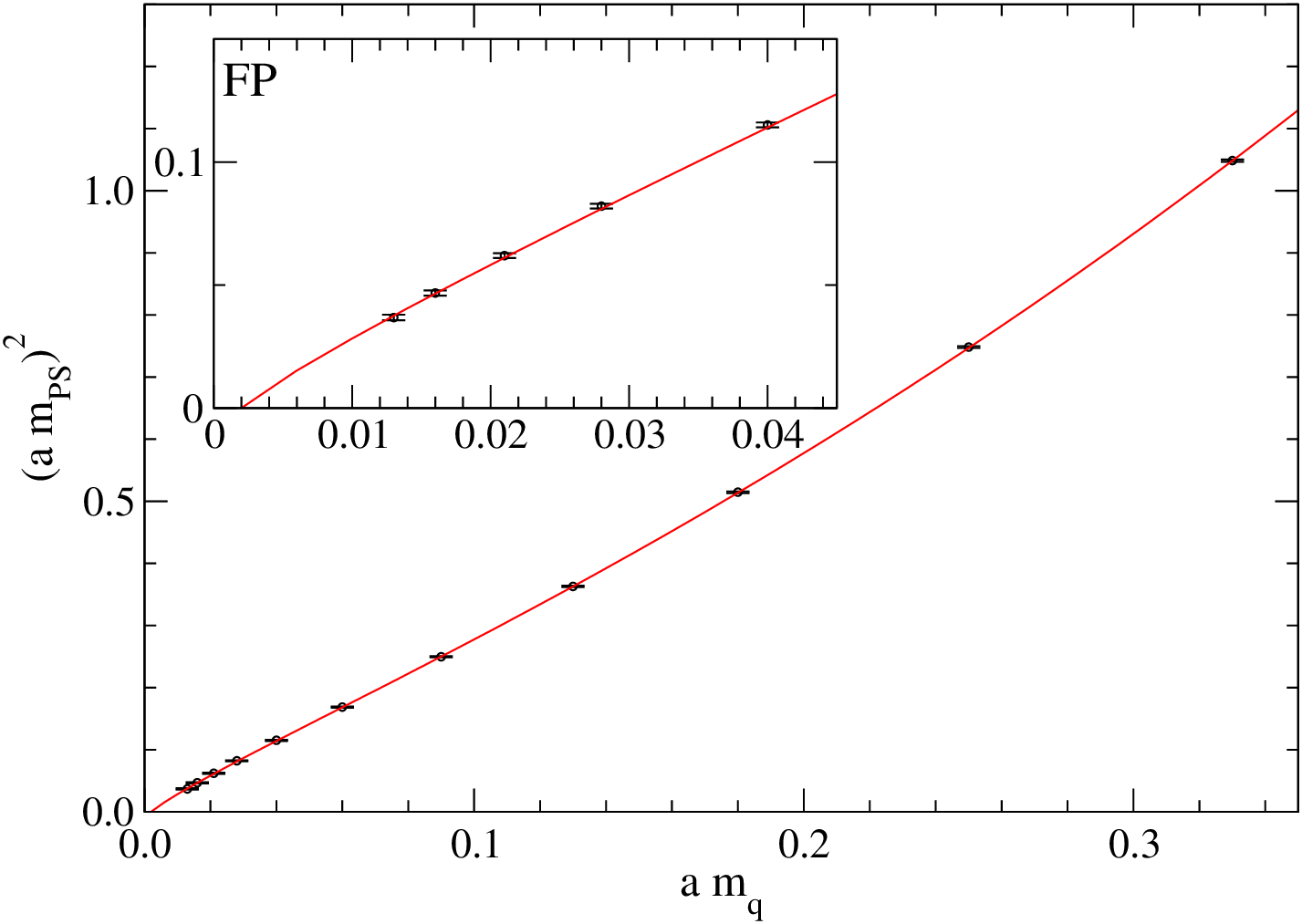}
\end{center}
\vspace{-4mm}
\caption{{}The square of the the pseudoscalar mass for degenerate 
quark masses vs. the bare quark mass (FP, $16^3\times 32, a = 0.15$ fm). 
The fit is of the form suggested by Q$\chi$PT, \eqref{psfitdeg}. 
% mpisq\_vs\_mq\_16x32\_b3.0.eps
}
\label{pionplot_FP}
\end{figure}

\begin{figure}[p]
\vspace{-6mm}
\begin{center}
\includegraphics[width=9.5cm,clip]{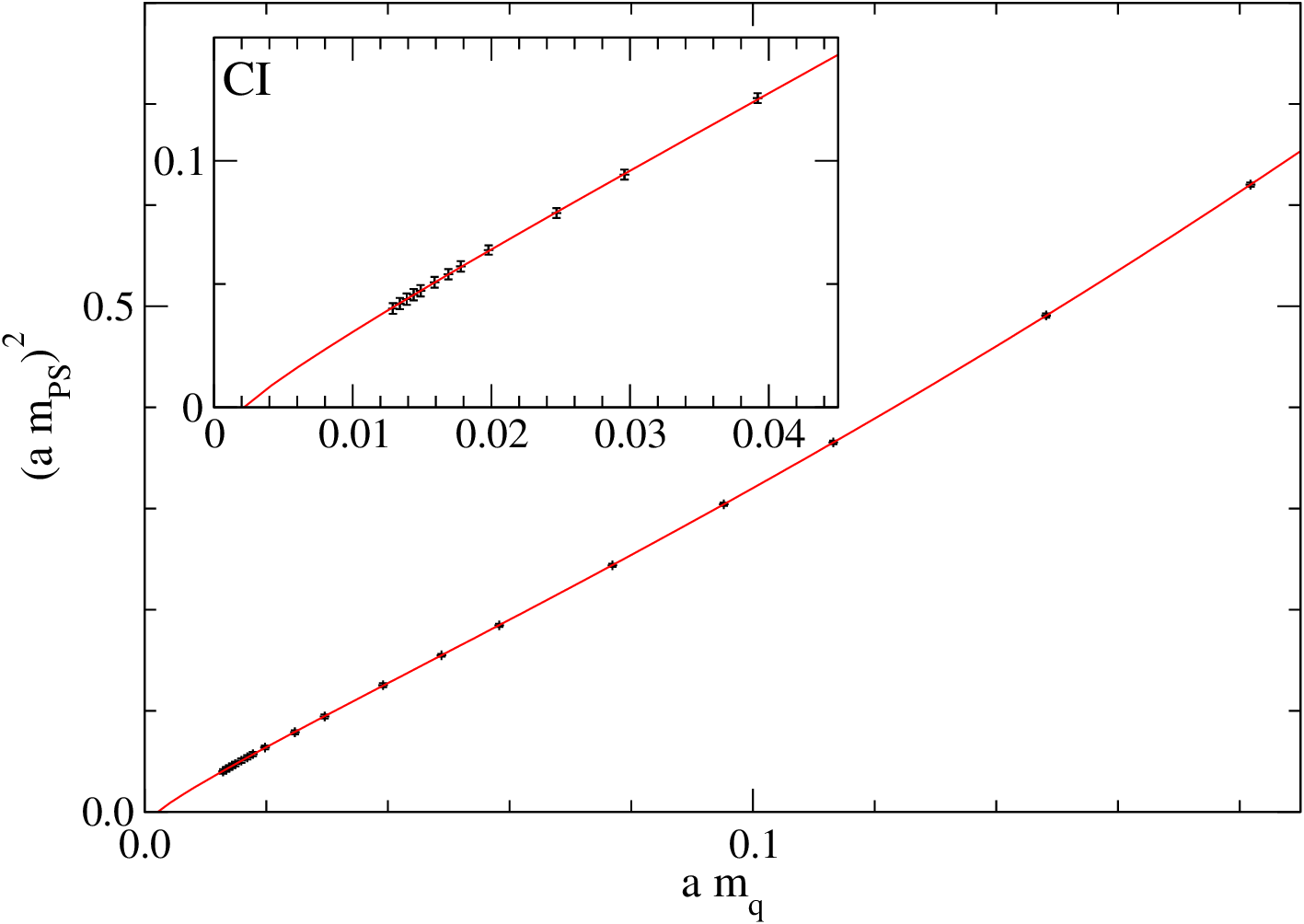}
\end{center}
\vspace{-4mm}
\caption{{}Same as the previous figure, now for the CI action.
($16^3\times 32, a = 0.15$~fm). Note that the scales differ from the ones used 
in Fig.\ \ref{pionplot_FP}.
% mpisq\_vs\_mq\_ci.eps
}
\label{pionplot_CI}
\end{figure}

Let us now address in detail our results for the pseudoscalar mass
in our largest volume. 
Quenched chiral perturbation theory (Q$\chi$PT) predicts
\cite{Bernard:1992mk,Colangelo,Yoshie:2001ts}:
\begin{equation}
m_\mathrm{PS}^2  \; =  \; A \,m  \; + \; B \,m \log m \; + \; C \, m^2 \;,
\label{psfitdeg} 
\end{equation}
where $m =m_q +m_\mathrm{res}'$, $m_q$ is the bare mass of the simulation
and $m_\mathrm{res}'$ is an effective residual additive quark mass
renormalization. As we shall discuss in the second part of this section
the first two terms in \eqref{psfitdeg} can be considered as the result of
expanding the sum of the leading logarithms in the chiral log parameter
$\delta$. For fixed  $\delta$ and at very small quark masses this expansion 
is not valid anymore and the fit parameters obtained from 
Eq.~\eqref{psfitdeg} should be interpreted accordingly. 
In particular, the fit parameter $m_\mathrm{res}'$ cannot be considered 
as a reliable prediction for the residual mass (defined as 
the value of $-m_q$ where $m_\mathrm{PS}$ vanishes).
As shown in Figs.~\ref{pionplot_FP} and \ref{pionplot_CI}
the fit describes the data very well, yielding 
$m_\mathrm{res}'=-0.0020(5)$(FP) and 
$m_\mathrm{res}'=-0.002(1)$(CI).

Note that the same fit parameters
determine also the pseudoscalar masses for non-degenerate 
quarks \cite{Bernard:1992mk,Yoshie:2001ts}:
\begin{multline} 
m_\mathrm{PS}^2  =  A \frac{m_1+m_2}{2} \; + \; 
B \frac{m_1+m_2}{2} 
\left( \frac{m_2 \log m_2 - m_1 \log m_1}{m_2-m_1} -1 \right)  
\\
 +  C\left( \frac{m_1+m_2}{2} \right)^2 \,.
\label{psfitnondeg}
\end{multline}

\begin{figure}[tbp]
\begin{center}
\includegraphics[width=9.9cm,clip]{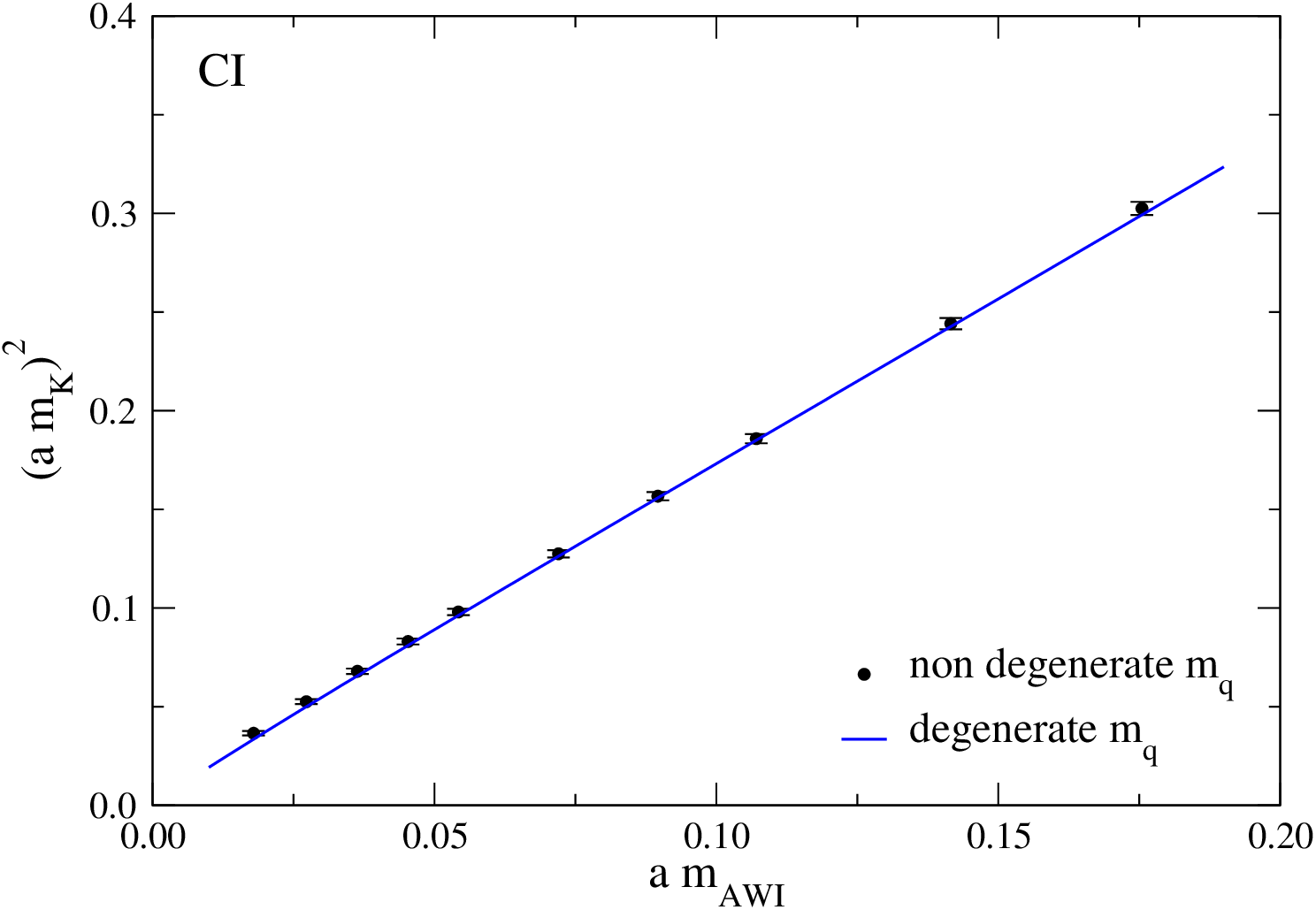}
\end{center}
\vspace{-7mm}
\caption{{}Comparison of pseudoscalar masses from correlators
with degenerate and non-degenerate quark masses 
(CI, $16^3\times 32, a = 0.15$ fm). The symbols
represent the kaon mass as a function of the heavy quark mass. 
The full curve comes from Eq.~\eqref{psfitnondeg} but the parameters 
$m_\mathrm{res}', A, B, C$  were determined in a fit of 
Eq.~\eqref{psfitdeg} to data  with degenerate quark masses.
% kaon.eps
 } 
\label{kaonplot}\end{figure}
We compared the measured values of $m_\mathrm{PS}(m_1,m_2)$ 
for the CI operator to those obtained from Eq.~\eqref{psfitnondeg}
(with the coefficients determined from the degenerate mass case,
Eq.~\eqref{psfitdeg}). The agreement was within 1-2\% and within 
the statistical errors.
Another test is shown in Fig.~\ref{kaonplot}.
The data represented by the symbols were obtained from correlators with 
non-degenerate quark masses as follows: One quark mass (the 'heavy') 
was held fixed, the other quark mass was extrapolated to the chiral limit 
and the procedure was repeated for different 'heavy' quark masses. 
Thus the symbols essentially represent the kaon mass as a function 
of the $s$-quark mass.
The full curve was generated from Eq.~\eqref{psfitnondeg} with $m_1 = 0$ 
but the parameters $m_\mathrm{res}'$, $A$, $B$, $C$ were determined from 
fitting our data with degenerate quark masses using Eq.~\eqref{psfitdeg}. 
It is obvious that the curve falls quite well on the data points and 
the figure demonstrates that our degenerate and non-degenerate mass 
data are well compatible with each other.

\begin{figure}[tbp]
\begin{center}
\includegraphics[width=9.5cm,clip]{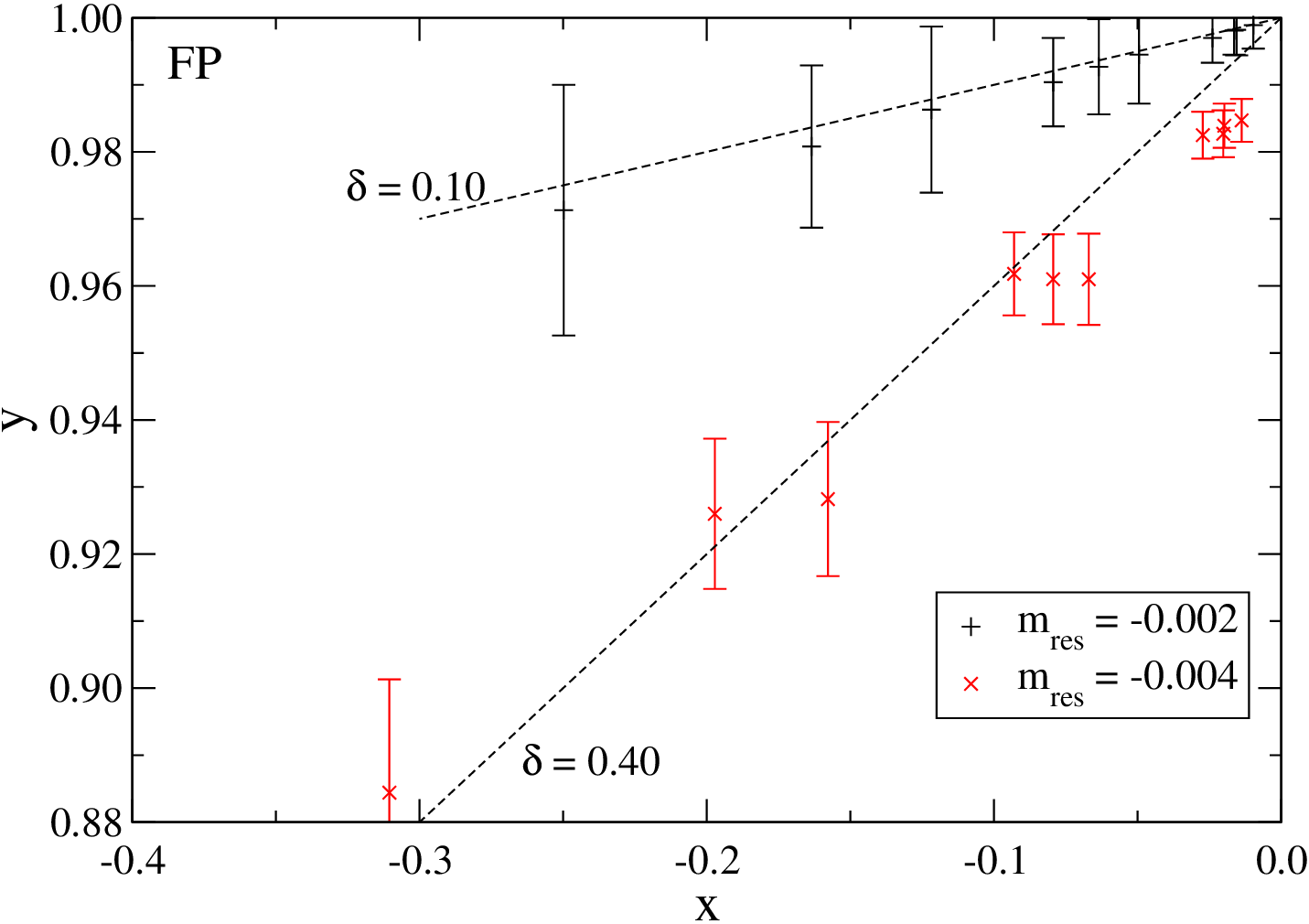}
\end{center}
\vspace{-4mm}
\caption{{}The plot corresponding to eqs.~\eqref{y_chiral},
\eqref{x_chiral} for two extreme values of the residual quark mass 
$m_\mathrm{res}$. The slope is given by the quenched parameter $\delta$
(FP, $16^3\times 32, a = 0.15$ fm).
% xy\_chir.eps
}
\label{xyplot}
\end{figure}

Let us finally comment on the quenched chiral log parameter $\delta$.
Several calculations of this parameter can be found in the literature
\cite{cppacsquench,domain1,domain2,MILC13,boston1,boston2,hauswirth,
kimohta,cppacsdelta,deltapapers} including also a previous result from our 
collaboration. 
These numbers range from $\delta = 0.05$ to $\delta = 0.23$ indicating that
this is a rather difficult problem. 

The $m \log m$ term in Eq.~\eqref{psfitdeg} is a quenched peculiarity which is
proportional to the chiral log parameter. Nevertheless, this equation is not
very useful to 
fix the value of $\delta$. Firstly, after expressing $B$ in terms of $\delta$
and inserting a scale parameter  $\Lambda$ in the log, this equation contains
too many parameters to be fitted to the data points which form a simple curve
in Figs.~\ref{pionplot_FP}, \ref{pionplot_CI}. 
Secondly, the first two terms in 
Eq.~\eqref{psfitdeg} are obtained from an expansion of \cite{sharpe_delta}
\begin{equation}
m^2_\mathrm{PS} \propto m^{1/(1+\delta)}
\label{llog}
\end{equation}
in the parameter $\delta$. In the interesting small quark mass
region the first two terms of the expansion do not give a good approximation
to the function in Eq.~\eqref{llog}. Thus, at least for the determination
of such sensitive
quantities as $m_\mathrm{res}$ and $\delta$, we will not use 
Eq.~\eqref{psfitdeg}.

Trying to use Eq.~\eqref{llog} to determine $\delta$ we have to decide 
which points to include in the fit. Since Eq.~\eqref{llog} refers 
to chiral perturbation theory with degenerate quarks, we have chosen 
to stay below the kaon mass, i.e.~$m \; < \; m_\mathrm{s}/2$, where 
$m_\mathrm{s}$ is the strange quark mass. 
In our spectroscopy with the FP operator on the coarsest lattice with large
volume the lightest 5 masses satisfy this condition. 
The fit contains 3 parameters: amplitude, $m_\mathrm{res}$ and $\delta$. 
The results with the 4 and 5 lightest quark masses included read 
$m_\mathrm{res}=-0.0043(10)$, $\delta = 0.25(7)$ and  
$m_\mathrm{res}=-0.0036(8)$,
$\delta = 0.19(4)$, respectively. Note that the error is statistical only. 
The deviation of the fit from the (correlated) data points is a small 
fraction of their statistical error, $\chi^2$ is small and gives 
no really useful information on the quality of the ansatz 
(related to the systematic error of $\delta$).

For heavy quarks we expect $m^2_\mathrm{PS} \propto m^2$ with a coefficient
approaching 4 for very heavy quarks. The ansatz
\begin{equation}
m^2_\mathrm{PS} \; = \; C_1\,m^{ 1/(1+\delta) } \; + \; C_2 \, m^2 
\label{llogcorr}  
\end{equation}
describes all the data points in Figs.~\ref{pionplot_FP}, \ref{pionplot_CI}
well and leads to
$m_\mathrm{res}^\mathrm{FP} =  -0.0027(6)$ and 
$\delta = 0.19(1)$ in consistency with the numbers above. Again, the errors
are statistical only. From these fits we quote
$m_\mathrm{res}^\mathrm{FP}=-\,0.003(1)$.
Note that this value is consistent with $m_\mathrm{res}'$ obtained 
from Eq.~\eqref{psfitdeg}.

Another possibility to determine $\delta$ is to start from 
the non-degenerate case Eq.~\eqref{psfitnondeg} and to reduce the number 
of free parameters by forming ratios where the unknown scale $\Lambda$ 
and the last (quadratic) term cancel. Beyond these features 
the combination \cite{cppacsquench}
\begin{equation}
y \; = \; \frac{2 m_1}{m_1+m_2} \; 
\frac{m_{\mathrm{PS},12}^2}{m_{\mathrm{PS},11}^2} \; 
\frac{2 m_2}{m_1+m_2} \;
\frac{m_{\mathrm{PS},12}^2}{m_{\mathrm{PS},22}^2}
\label{y_chiral}
\end{equation}
is equal to 1, up to 
small corrections proportional to $\delta$ over our data range which
justifies the expansion  $y = 1+\delta x + O(m^2,\delta^2)$, where 
\begin{equation}
x \; = \; 2 \; + \; \frac{m_1+m_2}{m_1-m_2}\ln\left(\frac{m_2}{m_1}\right) \,.
\label{x_chiral}
\end{equation}

With the FP we did not measure $m_\mathrm{PS}$
with non-degenerate quark masses needed in Eq.~\eqref{y_chiral}. 
However, Eq.~\eqref{psfitnondeg} allows us to obtain these numbers by 
interpolation from measured data for the degenerate case, 
using the fit parameters 
$m_\mathrm{res}'$, $A$, $B$, $C$ of Eq.~\eqref{psfitdeg}. 
On the other hand, $m_1$, $m_2$ in Eqs.~\eqref{y_chiral}, \eqref{x_chiral}
are the chiral mass parameters (vanishing together with
$m_\mathrm{PS}$), with the corresponding residual quark mass
$m_\mathrm{res}$ determined from Eqs.~\eqref{llog}, \eqref{llogcorr}.
Taking $m_\mathrm{res}=-0.003(1)$ from above and including data only where 
the pseudoscalar mass is not heavier than the physical kaon mass 
we obtain values ranging from $\delta=0.10$ 
(for $m_\mathrm{res}=-0.002$) to  $\delta=0.40$ (for $m_\mathrm{res}=-0.004$)
as shown in Fig.~\ref{xyplot}. Obviously, this method reacts strongly to 
the uncertainty in the residual quark mass. The same program of different
strategies to determine $\delta$ was also performed for the $CI$ operator
giving similar results.  

As a conclusion, we can confirm the introductory remark: The determination of
the chiral
log parameter is a difficult problem. It goes hand-in-hand with a reliable
determination of the residual quark mass $m_\mathrm{res}$. 
In order to reduce the systematic errors one needs precise data close to 
the chiral limit. In light of this conclusion the error estimate in our 
earlier determination $\delta=0.17(2)$ for CI, $\delta=0.18(3)$ for FP
\cite{boston1,boston2,hauswirth} is far too optimistic.

\subsection{Chiral extrapolation of vector meson and nucleon masses}

We experimented with different ways of extrapolating our vector meson and 
nucleon data to the physical region. 
Firstly, we performed fits suggested by
quenched chiral perturbation theory Q$\chi$PT. For vector mesons and 
baryons we fitted the data to the form \cite{1997hk,1996jy}:
\begin{equation} 
\label{eq:chirfit}
 m(m_\mathrm{PS}) \; \; = \; \; m_0 \; + 
\; A_{1/2} \, m_\mathrm{PS} \; + \; A_1  \, m_\mathrm{PS}^2 \;.
\end{equation}
For comparison we considered the following simple fit for 
the vector meson (V) and baryons (B)
\begin{eqnarray} 
\label{eq:powerfit}
 m_\mathrm{V}(m^2_\mathrm{PS}) & \; = \; & m_0 \; + \;  A_1 \, m^2_\mathrm{PS} 
\;+\;  A_2 \, m^4_\mathrm{PS} \; , 
\nonumber
\\
m_\mathrm{B}^2(m^2_\mathrm{PS}) & \; = \; & 
m_0^2  \; +  \; A_1  \, m^2_\mathrm{PS}  \; +  \;A_2  \, m^4_\mathrm{PS} \; .
\end{eqnarray}
(The motivation for this parametrization is the observation 
that the dependence of $m_\mathrm{V}$ and $m_\mathrm{B}^2$ 
from $m^2_\mathrm{PS}$ is nearly linear \cite{pleiter}.)
In all cases both types of fits are good and provide a smooth 
inter- and extrapolation in the vector meson, nucleon, delta and
negative parity nucleon channels. 

\begin{figure}[tbp]
\vspace{-6mm}
\begin{center}
\includegraphics[width=9.5cm,clip]{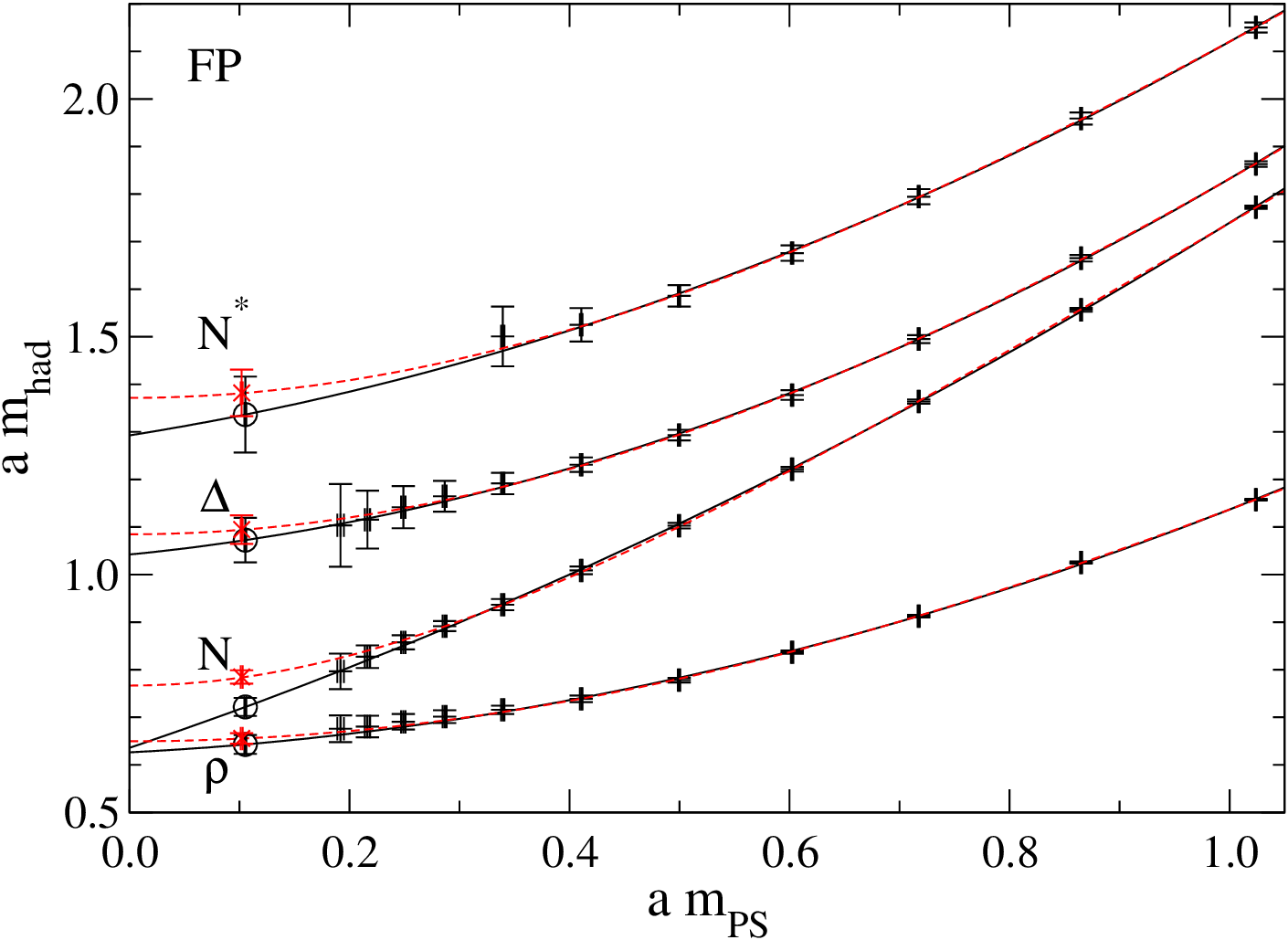}
\end{center}
\vspace{-6mm}
\caption{{}Chiral extrapolation of the FP data ($16^3 \times 32, a = 0.15$).
We extrapolate to the point 
where $m_\pi/m_\rho$ takes its physical value (open circle). Two different
fits are compared here: the one suggested by Q$\chi$PT theory
\eqref{eq:chirfit} (solid lines) and an ad-hoc form with powers of
$m_\mathrm{PS}^2$ \eqref{eq:powerfit} (dashed lines).
All quantities are measured in lattice units. 
% mh\_vs\_mPS\_FP.eps
}
\label{fig:h_vs_PS_FP}
\vspace{3mm}
\begin{center}
\includegraphics[width=9.5cm,clip]{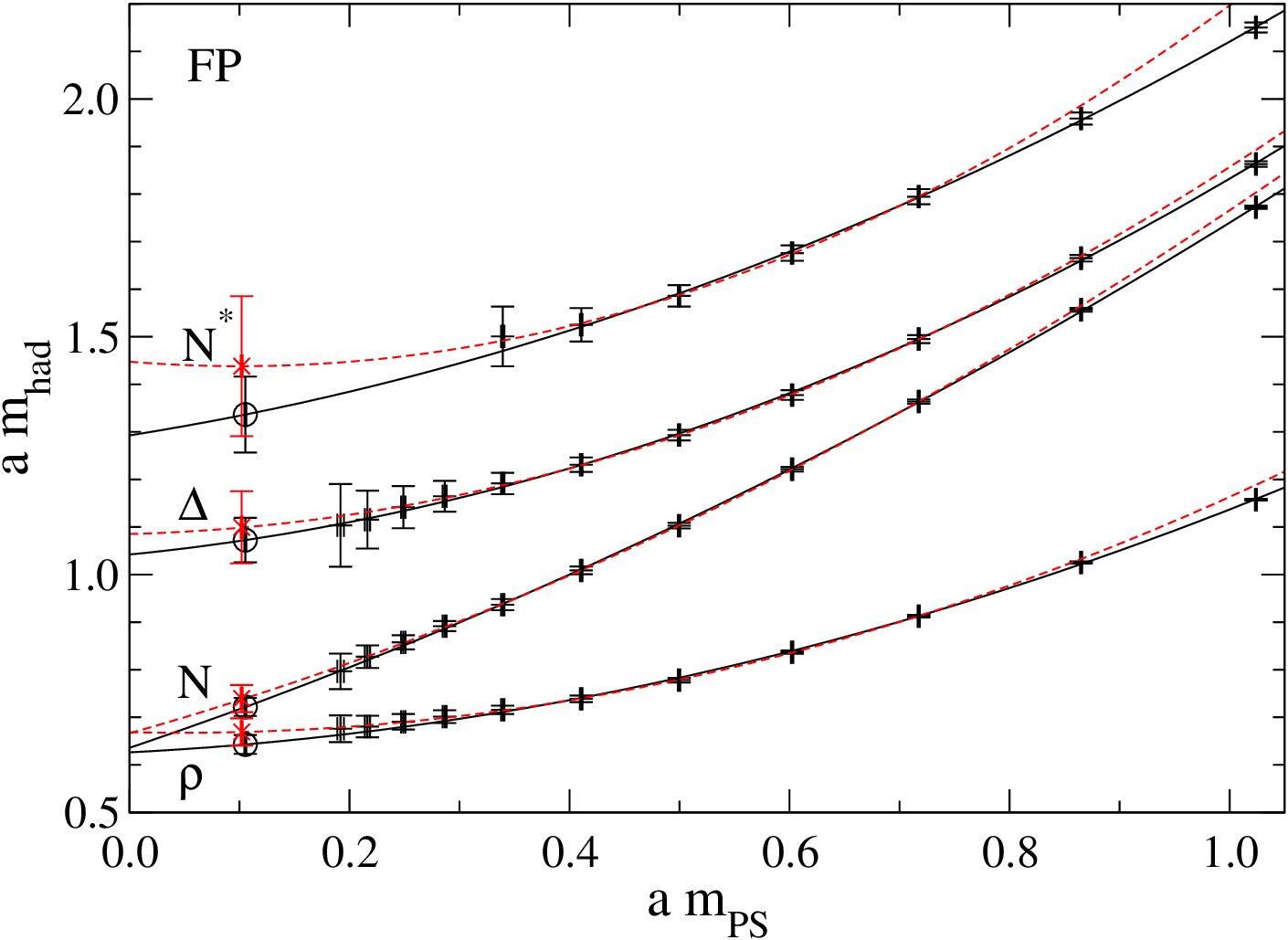}
\end{center}
\vspace{-6mm}
\caption{{} Chiral extrapolation for FP ($16^3 \times 32, a = 0.15$), 
similar to Fig.~\ref{fig:h_vs_PS_FP}. Both fits are of the
form \eqref{eq:chirfit}, but in one of them (dashed line) the
two heaviest quark masses are not included. 
% mh\_vs\_mPS\_FPa.eps
}
\label{fig:h_vs_PS_FPa}
\end{figure}

Although the form of Eq.~\eqref{eq:chirfit} is suggested 
by Q$\chi$PT we use it here beyond the range of its validity, 
i.e.\ as an effective parametrization.\footnote{Indeed we find 
$A_{1/2} > 0$ while Q$\chi$PT predicts $A_{1/2} < 0$.}
In Fig.~\ref{fig:h_vs_PS_FP} we give a graphical
view of the fits, of their extrapolation and of 
the physical mass prediction
in lattice units for the FP action. The full curves were 
fitted with the Q$\chi$PT formula from Eq.~\eqref{eq:chirfit}
while the dashed curves correspond to the ad-hoc form of 
Eq.~\eqref{eq:powerfit}. The extrapolated nucleon mass differs by more
than 2 $\sigma$ for these two extrapolations, while the shifts in the other
channels are small.

Although the fits using the ansatz from Q$\chi$PT seem to be
excellent, the way we used it above does not take the
theoretical background of this ansatz seriously: It was applied for the
whole range of our quark masses, although some of them are obviously outside 
the validity of Q$\chi$PT. Thus in order to estimate this systematic error
we experimented also with reducing the fit 
region for the bare quark masses and analyzed the effect on the extrapolated 
hadron masses. Fig.~\ref{fig:h_vs_PS_FPa} illustrates the effect of
excluding the two largest quark masses from the fit on the
extrapolated mass values for the FP operator. We find that the effect 
is less than 1 $\sigma$ in all the channels.

\begin{figure}[t]
\vspace{-1mm}
\begin{center}
\includegraphics[width=9.2cm,clip]{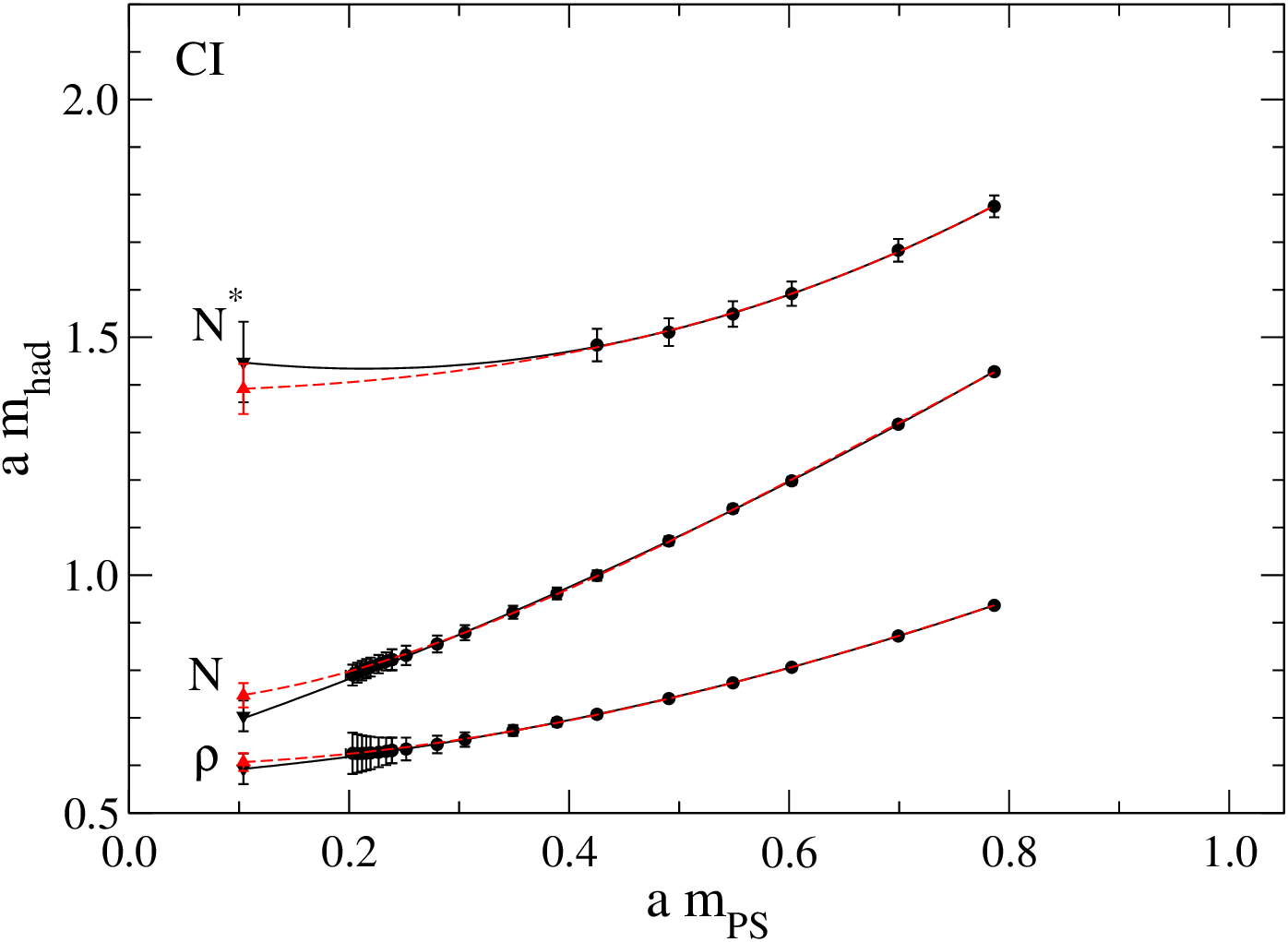}
\end{center}
\vspace{-7mm}
\caption{{}Chiral extrapolation for the CI operator 
($16^3 \times 32, a = 0.15$). 
We show the Q$\chi$PT fit \eqref{eq:chirfit} using solid lines 
and the fit to \eqref{eq:powerfit} with dotted lines.
All quantities are measured in lattice units. 
% mh\_vs\_mPS\_CI.eps
}
\label{fig:h_vs_PS_CI}
\end{figure}

For the CI operator we also performed a series of fits experimenting
with both the Q$\chi$PT formula (\ref{eq:chirfit}) and 
the ansatz (\ref{eq:powerfit}). In our fits we also varied
the number of large quark mass data taken into account and the
degree of the polynomials in (\ref{eq:chirfit}), 
(\ref{eq:powerfit}). Similarly to the FP operator we found that
the variation of the extrapolated masses is a 1 $\sigma$ effect for 
the vector meson and the negative parity nucleon and slightly larger 
(1.5 $\sigma$) for the nucleon. In Fig.~\ref{fig:h_vs_PS_CI}
we show the two extremal fits which are given by Q$\chi$PT
(\ref{eq:chirfit}) with all masses (full curve) and 
the ansatz (\ref{eq:powerfit}) with all masses (dashed curve).

\subsection{Setting the scale - results in physical units}

Three parameters enter light hadron spectroscopy: the
lattice spacing $a$, the average mass\footnote{Here $m_\mathrm{ud}$ 
and $m_\mathrm{s}$ denote the values of 
$m = m_q + m_\mathrm{res}'$ entering Eqs.~\eqref{psfitdeg},
\eqref{psfitnondeg} as determined by the input physical masses.}
of the light quarks $m_\mathrm{ud} = (m_u + m_d)/2$ and the strange 
quark mass $m_\mathrm{s}$.
In order to determine their values, one has to identify observables
measured on the lattice with experimental values in the
continuum. However, as the quenched spectrum does not coincide with the
physical one, the determination of the parameters is
ambiguous. As is well known, depending on the choice of experimental data
to fit the parameters 
the quenching errors in the prediction can be shifted around. Nevertheless, we
want to argue below that the quenching errors show an intuitively
understandable pattern in hadron spectroscopy and the spectrum has more
consistency than it is usually believed to have. 
Similar observations were made earlier in \cite{sommer_rho}.  
Here we compare three different procedures to set the physical input:

\vspace{3mm}
\begin{itemize}

\item[I:]
The masses $m_\pi,m_\rho$ and $m_K$ or $m_\Phi$ are used as experimental
input data.
The scale $a(\rho)$ is set using the $\rho$ mass.
To determine the light quark mass the pseudoscalar and vector masses
are extrapolated to that value $m_\mathrm{ud}$,  where
the physical value of $m_\pi/m_\rho$ is reached.
To determine the strange quark mass, we either use the $K$ or the $\Phi$ 
mass as input. This is a widely used procedure.

\item[II:]
The Sommer parameter $r_0$ (=0.5 fm) and the masses $m_\pi$ and  $m_K$, or
$m_\pi$ and $m_\Phi$ are used as experimental input data.
The scale $a(r_0)$ is set from the 
Sommer parameter. To determine the light quark mass the 
pseudoscalar mass was extrapolated to the value $m_\mathrm{ud}$,  where
the physical value of $r_0 m_\pi$ is reached. To determine the
strange quark mass, we again either use the $K$ or the $\Phi$ mass as input.
\\

\item[III:]
The masses $m_\pi,m_K$ and  $m_\Phi$ are used as experimental
input data.
The scale $a(\Phi)$ is set by requiring $m_\mathrm{V}(m_\mathrm{s})=m_\Phi$.
The values $m_\mathrm{s}$ and $m_\mathrm{ud}$ are obtained from the ratios
$m_\mathrm{PS}(m_\mathrm{s},m_\mathrm{ud})/m_\mathrm{V}(m_\mathrm{s})=
m_K/m_\Phi$ and 
$m_\mathrm{PS}(m_\mathrm{ud})/m_\mathrm{V}(m_\mathrm{s})=m_\pi/m_\Phi$.
\end {itemize}

\begin{table}[tbp]
\begin{center}
\begin{tabular}{c|c|c}
 & $ a_\mathrm{FP} $   & $a_\mathrm{CI}$ \\
\hline
$\rho$ & $0.165(5)\,\mathrm{fm}$  & $0.154(8)~\mathrm{fm}$  \\
$r_0$  & $0.153(2)\,\mathrm{fm}$  & $0.148(2)~\mathrm{fm}$  \\
$\Phi$ & $0.1536(14)\,\mathrm{fm}$ & $0.146(2)~\mathrm{fm}$
\end{tabular}
\end{center}
\caption{{}The  lattice constant extracted from the $\rho$, the $r_0$
and the $\Phi$ scale (methods I, II and III). 
The errors quoted are the statistical ones. \label{atable}}
\end{table}

Before discussing the results, let us have a look at the sources of
systematic errors. In method I the scale is set by an
{\it extrapolated} value of the vector meson mass.
The systematic error of these
extrapolations is discussed in the previous Section 4.4. 
Also the statistical error in the vector channel is increasing rapidly with
decreasing quark mass.
Concerning method II one has to remark that also the determination of
$r_0$ is plagued by uncertainties. 
Firstly, the physical value of the Sommer parameter is not known precisely and,
secondly, its extraction becomes more difficult
on coarse lattices. Method III seems to be better protected from such
systematic problems. 

\begin{figure}[tbp]
\vspace{-5mm}
\begin{center}
\epsfig{file=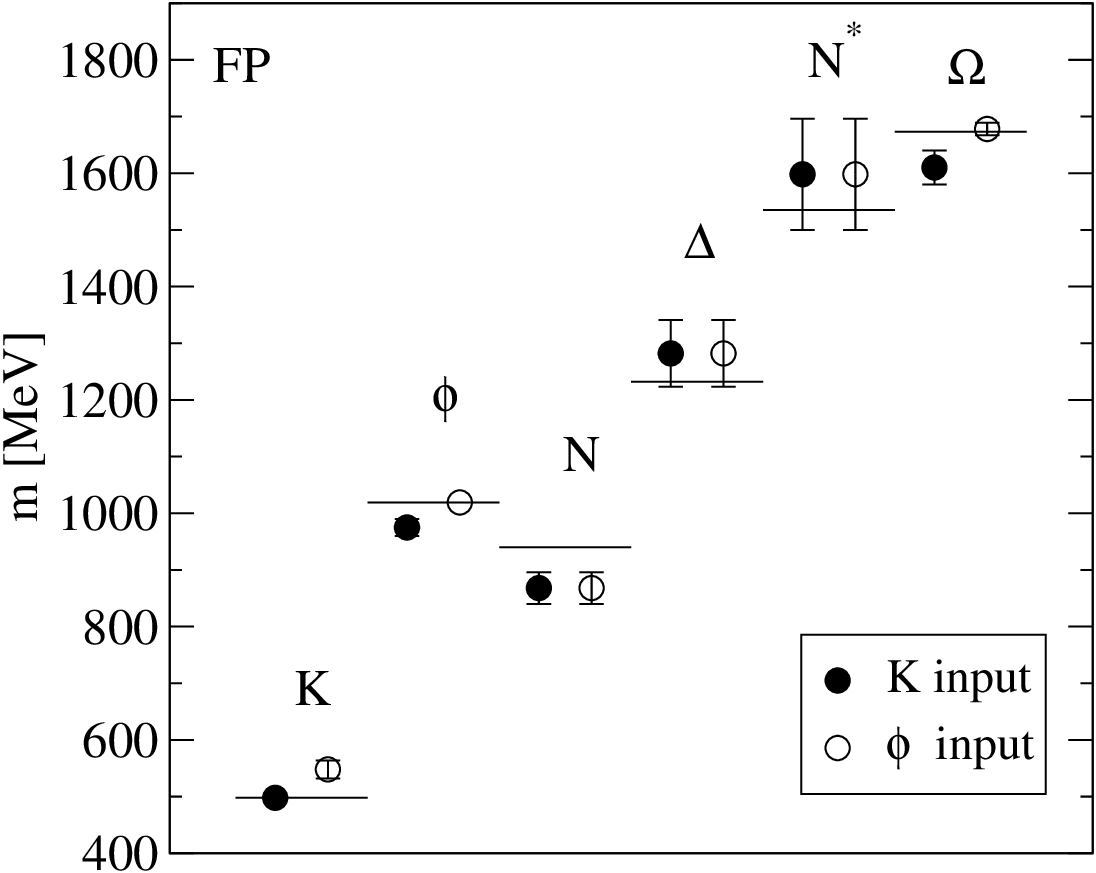,height=5cm,clip}
\epsfig{file=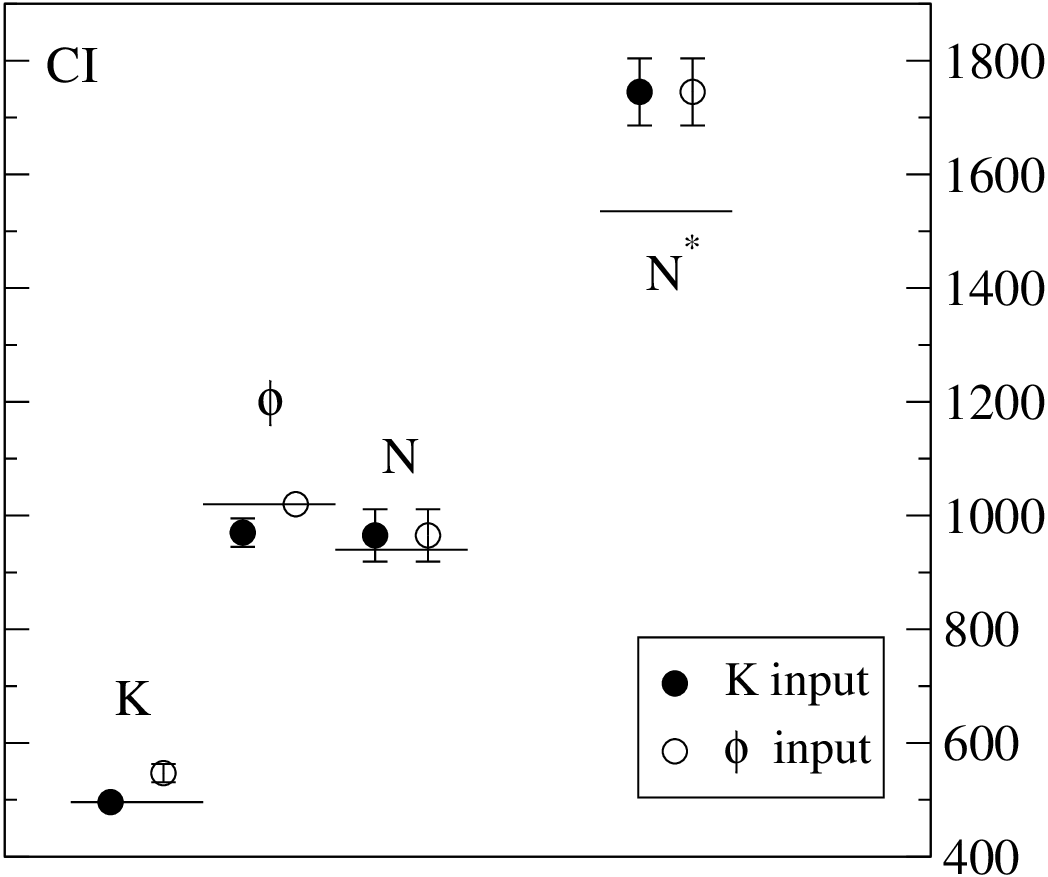,height=5cm,clip}
\end{center}
\vspace{-4mm}
\caption{{}The prediction for the hadron masses using the conventional
way to fix the parameters (method I): The lattice spacing $a(\rho)$ and the light 
quark mass are fixed by $m_\pi$ and $m_\rho$, while the strange quark mass 
is taken from $m_K$ or $m_\phi$. The l.h.s.\ plot shows the FP results,
the CI data can be found on the r.h.s. 
The horizontal lines indicate the experimental numbers.
% hadron\_spectrum\_rho\_FP.eps
% hadron\_spectrum\_rho\_CI.eps
\label{fig:spectrum_rho}
}
\vspace{5mm}
\begin{center}
\epsfig{file=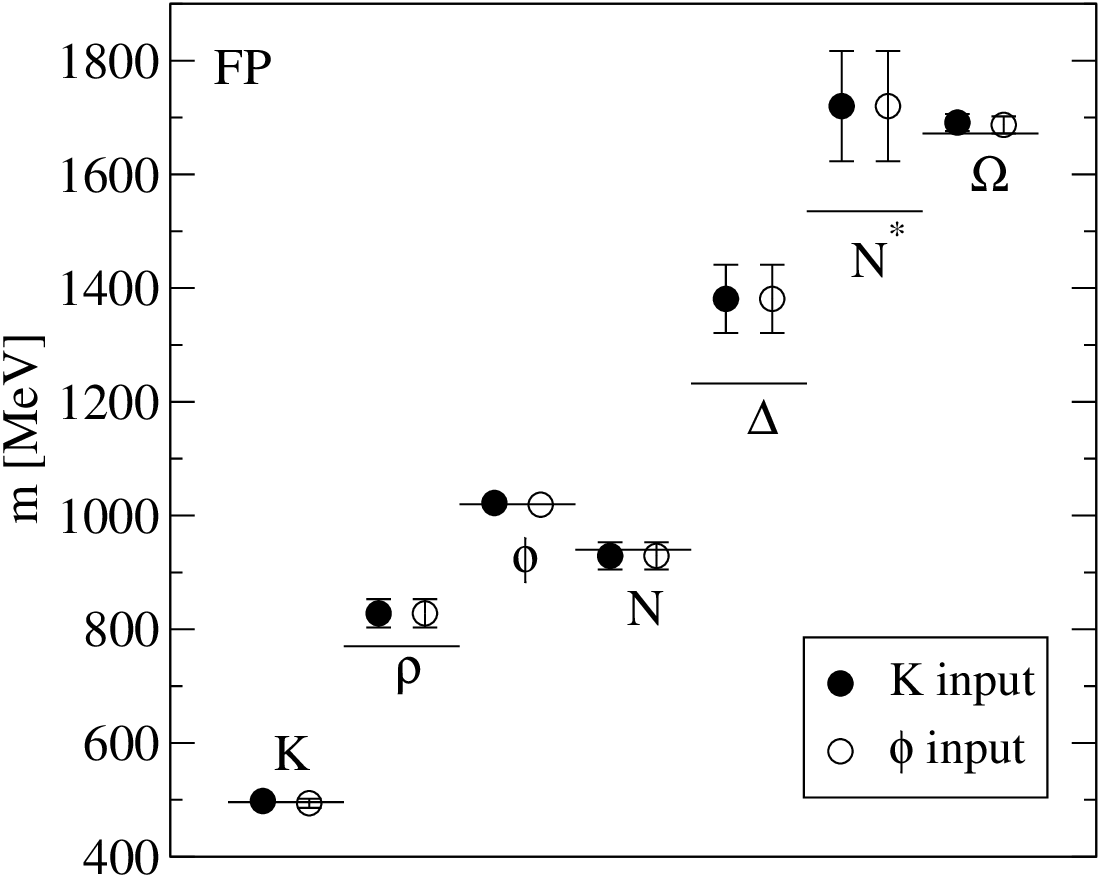,height=5cm,clip}
\epsfig{file=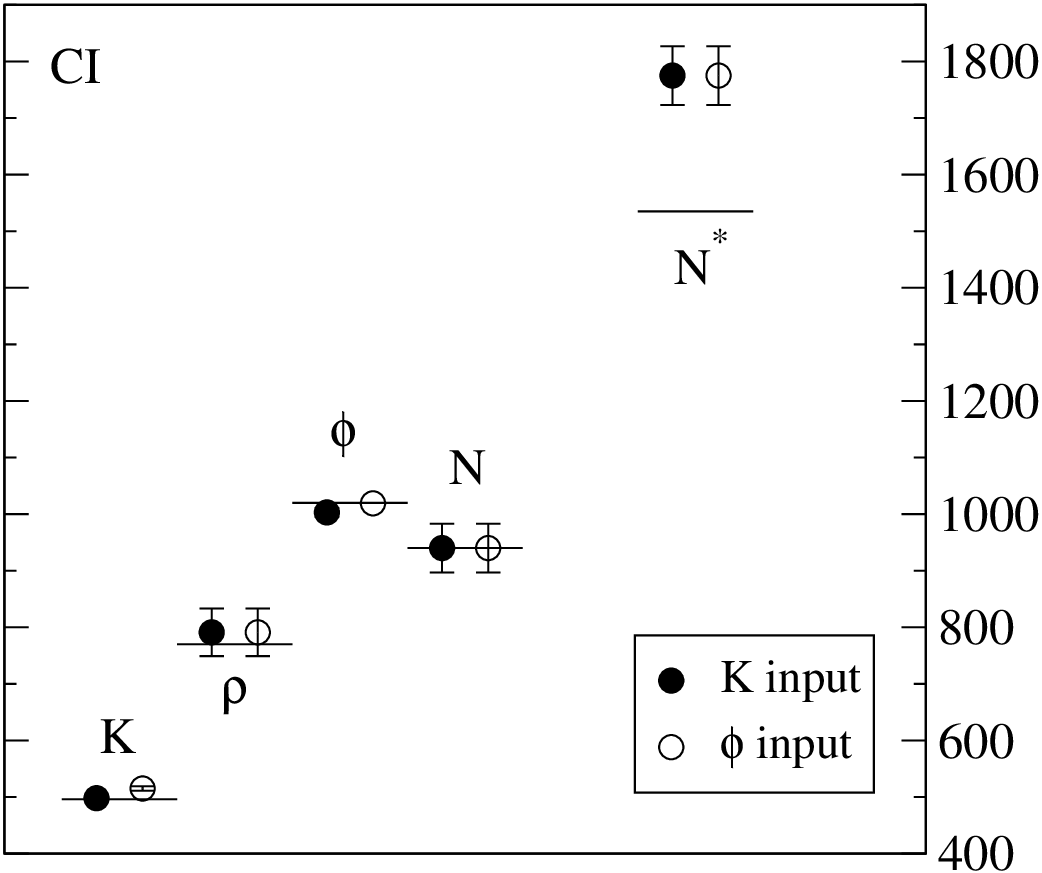,height=5cm,clip}
\end{center}
\vspace{-4mm}
\caption{{}The hadron masses with the lattice spacing $a(r_0)$ taken from 
the Sommer parameter. The light and strange quark masses are fixed using 
$m_\pi$ and $m_K$ or $m_\rho$ (method II).
% hadron\_spectrum\_a\_FP.eps
% hadron\_spectrum\_a\_CI.eps
}
\label{fig:spectrum_a}
\end{figure}
\begin{figure}[t]
\vspace{5mm}
\begin{center}
\epsfig{file=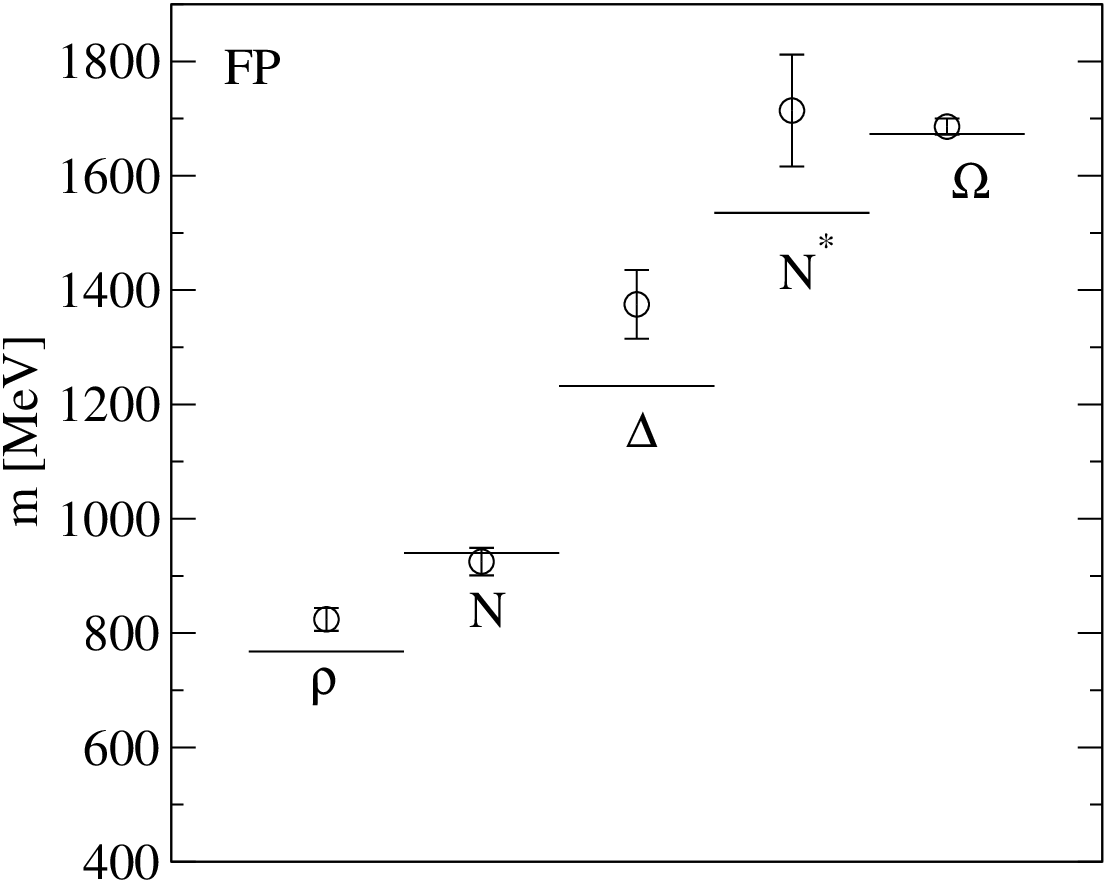,height=5cm,clip}
\epsfig{file=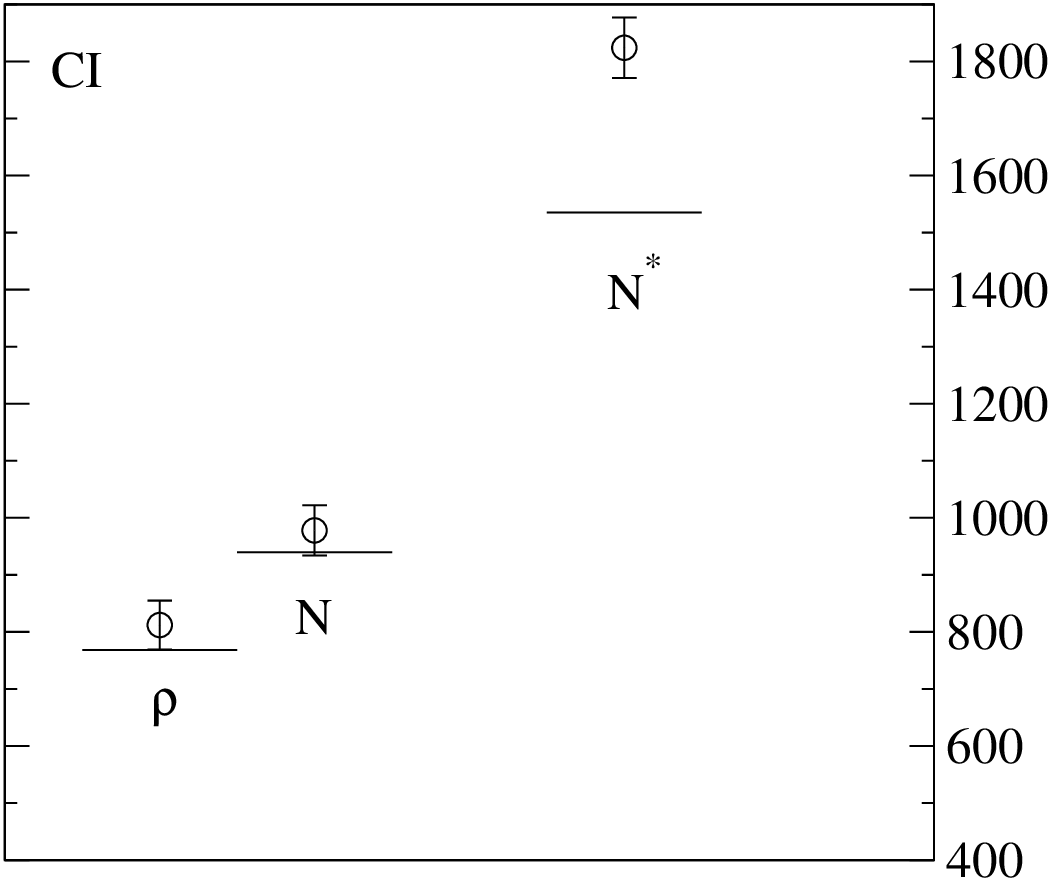,height=5cm,clip}
\end{center}
\vspace{-4mm}
\caption{{}Here the lattice spacing $a(\Phi)$, the light and strange 
quark masses are fixed by $m_\pi$, $m_K$ and $m_\Phi$ (method III). 
The Sommer parameter is a prediction in this case: $r_0=0.49(1)$ fm 
for both FP and CI.
% hadron\_spectrum\_Phi\_FP.eps
% hadron\_spectrum\_Phi\_CI.eps
}
\label{fig:spectrum_Phi}
\end{figure}

The results for the spectrum obtained with these three procedures are 
presented in Figs.~\ref{fig:spectrum_rho}, \ref{fig:spectrum_a} 
and Fig.~\ref{fig:spectrum_Phi}. The l.h.s.\ plots give 
the FP data, while the r.h.s.\ plots show the results from CI. 
The errors shown are only statistical errors and were estimated by jackknife
resampling: All the necessary steps (finding the fit parameters to
the data, solving equations, etc.) were repeated on every jackknife
sample. Our results for the scales are collected in Table~\ref{atable}. 
Again the errors given are only the statistical ones. 

Let us now discuss the merits and problems of the different methods.
Using method I (Fig.~\ref{fig:spectrum_rho} and Table~\ref{pi_rho_K_Phi})
we find the well-known qualitative features of this standard approach. 
(The fit parameters are: 
FP: $a=0.165(5)\,{\rm fm}$, $am_{\rm ud}=0.0033(2)$, 
$am_{\rm s}=0.113(7)$[K-input],
$am_{\rm s}=0.139(17)$[$\Phi$-input];
CI: $a=0.154(8)~\mathrm{fm}$,
$am_{\rm s}=0.093(6)$[K-input],
$am_{\rm s}=0.115(12)$[$\Phi$-input].)
The predictions for the masses of hadrons containing strange quarks depend 
rather strongly on using the $K$ or the $\Phi$ meson mass as an input. 
The scale, set by the $\rho$ in this method, lies higher than that obtained 
in the gauge sector by $r_0$.
(Although, as Table~\ref{atable} shows, for CI all the three scales are
compatible within the statistical errors.)

Setting the scale by $r_0$ 
(method II, Fig.~\ref {fig:spectrum_a} and Table~\ref{a1530} ) 
the discrepancy between the spectra 
with $m_K$ vs.\ $m_\Phi$ input disappears. Even more, these masses and the
$\Omega$ are shifted to their right place. Since they have small
statistical errors this is a non-trivial coincidence. 
(The parameters are: 
FP:  $a=a(r_0)=0.153\,{\rm fm}$,
$am_{\rm ud}=0.0027(1)$, $am_{\rm s}=0.0954(13)$[K-input],
$am_{\rm s}=0.0939(35)$[$\Phi$-input];
CI: $a=a(r_0)=0.148(2)\,{\rm fm}$, 
$am_{\rm s}=0.083(1)$[K-input], 
$am_{\rm s}=0.090(3)$[$\Phi$-input].)

From the consistency of the $m_K$ vs.\ $m_\Phi$ input above it follows that 
method III (Fig.~\ref {fig:spectrum_Phi} and Table~\ref{tab:res_new_fp}) 
predicts essentially the same spectrum as method II. 
In this case the $r_0$ is a {\it prediction}. We obtain 
$r_0=0.49(1)\,\mathrm{fm}$ for both FP and CI. 
(Fit parameters:
FP: $a=0.1536(14)\,{\rm fm}$, $am_{\rm ud}=0.0028(1)$, 
$am_{\rm s}=0.0962(22)$;
CI: $a=0.146(2)~\mathrm{fm}$,
$am_{\rm s}=0.078(27)$.)

We note at this point that unlike the quenched $\delta$ parameter,
the hadron spectrum is quite insensitive to the actual choice of
$m_\mathrm{res}'$. Indeed $m_\mathrm{res}'$ enters our analysis 
only through the use of Eq.~\eqref{psfitnondeg} in the determination of
$m_K$. But even taking the extreme value of $m_\mathrm{res}'=0$,
the hadron masses change only by a fraction of their statistical
errors while $a$ changes by $1\,\sigma$ (1\%) and
$am_{\rm ud}$, $am_{\rm s}$ are changed by 15\% and 8\%, respectively.

The hadron spectrum in Figs.~\ref{fig:spectrum_a} and \ref{fig:spectrum_Phi}
shows a feature which is easy to understand intuitively: Quenched spectroscopy
works fine for the narrow $K$ (or $\Phi$), $N$ and $\Omega$, while it has
problems with the broad resonances $\rho, \Delta$ and $N^*$ with widths 
$\approx$150, 120 and 150 MeV, respectively. 
It would be interesting to test this intuitive picture on other resonances, 
in particular on baryons with non-degenerate quarks which were not treated 
in our analysis.

\begin{table}[p]
\begin{center}
\begin{tabular}{|c|c|cc|cc|}
\hline
Hadron &Exp.&\multicolumn{2}{c|}{K-input}&\multicolumn{2}{c|}{$\Phi$-input}\\
       &    &FP&CI&FP&CI\\
\hline
K        & 0.498 & ---        & ---       & 0.548(16) & 0.547(16) \\
$\Phi$   & 1.019 & 0.975(15)  & 0.970(25) & ---       & ---       \\
N        & 0.940 & 0.868(28)  & 0.965(46) & 0.868(28) & 0.965(46) \\
$\Delta$ & 1.232 & 1.282(59)  & ---       & 1.282(59) & ---       \\
$N^*$    & 1.535 & 1.598(98)  & 1.745(59) & 1.598(98) & 1.745(59) \\
$\Omega$ & 1.673 & 1.610(30)  & ---       & 1.678(11) & ---       \\
\hline
\end{tabular}
\caption{Hadron masses in GeV with the scale and quark masses set by
  $\pi$-$\rho$-K and $\pi$-$\rho$-$\Phi$ input (method I).
%The parameters are
%$a=0.165(5)\,{\rm fm}$, $am_{\rm ud}=0.0033(2)$, 
%$am_{\rm s}=0.113(7)$[K-input],
%$am_{\rm s}=0.139(17)$[$\Phi$-input].
\label{pi_rho_K_Phi}
}
\end{center}
%\label{tab:res_conv}
\vspace{6mm}
\begin{center}
\begin{tabular}{|c|c|cc|cc|}
\hline
Hadron &Exp.&\multicolumn{2}{c|}{K-input}&\multicolumn{2}{c|}{$\Phi$-input}\\
       &    &FP&CI&FP&CI\\
\hline
K        & 0.498 & ---        & ---       & 0.494(8)  & 0.515(4)  \\
$\rho$   & 0.768 & 0.828(25)  & 0.791(42) & 0.828(25) & 0.791(42) \\
$\Phi$   & 1.019 & 1.022(6)   & 1.003(8)  &  ---      & ---       \\
N        & 0.940 & 0.928(24)  & 0.940(43) & 0.928(24) & 0.940(43) \\
$\Delta$ & 1.232 & 1.381(60)  & ---       & 1.381(60) & ---       \\
$N^*$    & 1.535 & 1.720(97)  & 1.775(52) & 1.720(97) & 1.775(52) \\
$\Omega$ & 1.673 & 1.691(15)  & ---       & 1.687(15) & ---       \\
\hline
\end{tabular}
\caption{
Hadron masses in GeV when the scale is set using the Sommer parameter
$r_0=0.5~{\rm fm}$ (method II).
%Fixed lattice spacing: $a=0.153\,{\rm fm}$.
%$am_{\rm ud}=0.0027(1)$, $am_{\rm s}=0.0954(13)$[K-input],
%$am_{\rm s}=0.0939(35)$[$\Phi$-input].
\label{a1530}
}
\end{center}
%\label{tab:res_fixeda}
\vspace{6mm}
\begin{center}
\begin{tabular}{|c|c|cc|}
\hline
Hadron  &   Exp.  &  FP & CI \\
\hline
$\rho$   &   0.768 &   0.824(20) &0.812(43)\\
N        &   0.940 &   0.925(24) &0.978(44)\\
$\Delta$ &   1.232 &   1.375(60) &---\\
$N^*$    &   1.535 &   1.714(98) &1.824(53)\\
$\Omega$ &   1.673 &   1.686(14) &---\\
\hline
\end{tabular}
\caption{Hadron masses in GeV with the scale $a$ 
and the quark masses fixed by the $\pi, K, \Phi$
masses (method III).
%$a=0.1536(14)\,{\rm fm}$, $am_{\rm ud}=0.0028(1)$, $am_{\rm s}=0.0962(22)$.
\label{tab:res_new_fp}
}
\end{center}

\end{table}

\begin{figure}[tbp]
\vspace{-5mm}
\begin{center}
\epsfig{file=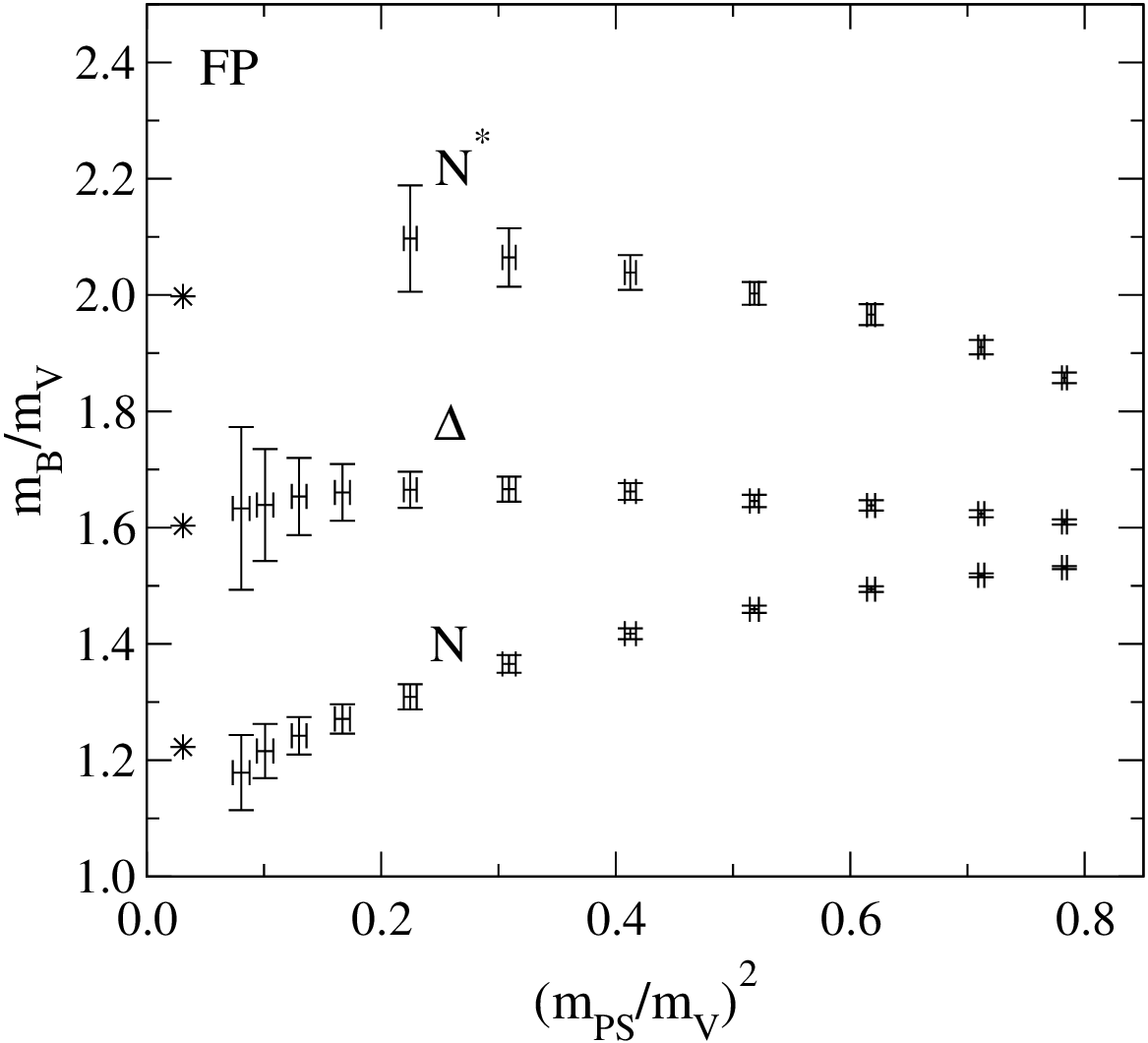,width=8.5cm,clip}
\end{center}
\vspace{-4mm}
\caption{{}APE plot for $N$, $\Delta$ and $N^*$ in the FP case
($16^3 \times 32, a = 0.15$ fm). The
asterisks indicate the experimental numbers.
% APE\_all\_FP.eps
}
\label{fig:APE_all_FP}
\vspace{5mm}
\begin{center}
\epsfig{file=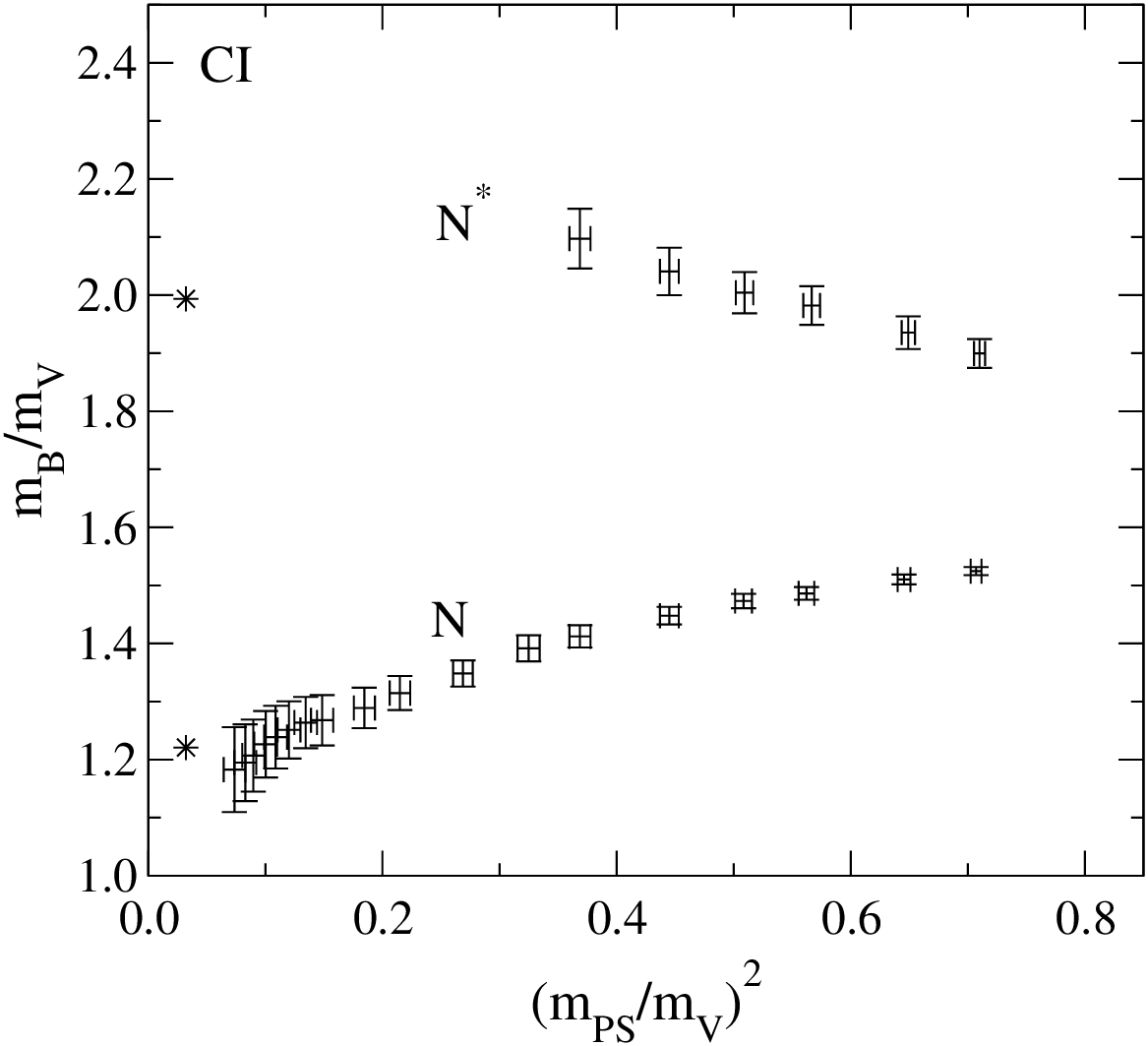,width=8.5cm,clip}
\end{center}
\vspace{-4mm}
\caption{{}APE plot for $N$ and $N^*$ from the CI operator
($16^3 \times 32, a = 0.15$~fm). 
% APE\_all\_CI.eps
}
\label{fig:APE_all_CI}
\end{figure}

To complete our presentation of hadron spectroscopy in a large volume we 
present standard APE plots in Figs.\ \ref{fig:APE_all_FP} and 
\ref{fig:APE_all_CI}, where the mass ratios 
$m_N/m_\mathrm{V}$, $m_\Delta/m_\mathrm{V}$, 
$m_{N^*}/m_\mathrm{V}$ are plotted against $(m_\mathrm{PS}/m_\mathrm{V})^2$ 
for the FP action and the same for the CI action (without the $\Delta$). 

\section{Volume dependence and scaling properties}
\label{sect:a_v}
The results of our large volume spectroscopy discussed in the previous
section were obtained on a rather coarse $a = 0.15\,\mathrm{fm}$
lattice. It is, therefore, essential to check the size of cut-off
effects. In this section we shall also compare our
results with those of recent large scale simulations.
Similarly, we want to
see, whether our $L = 2.5\,\mathrm{fm}$ lattice used for spectroscopy is
sufficiently large to neglect physical\footnote{as opposed to the topological
finite size artifacts discussed in Section 3 which are caused by 
the quenched approximation.} finite size effects. In full QCD the
pion cloud around the hadrons would dominate the finite size effects
in such a large box. In the quenched approximation this is expected to be
strongly suppressed. Indeed, the finite size effects were found to be
significantly smaller in quenched QCD than in full QCD \cite{fin_size}.
The authors of \cite{fs_milc} found no finite 
size effects for $L > 2 \,\mathrm{fm}$ on the 2\%
level. In \cite{fin_size} a $6\pm 3$\% decrease was observed in 
the nucleon mass in $L=(1.6-2.4) \,\mathrm{fm}$ (see also \cite{ukqcd1} 
and the summary in \cite{gottlieb}).   
For the vector meson the corrections are
smaller. Although the situation is not completely coherent, the earlier
results suggest that the finite size effects in our $L = 2.5\,\mathrm{fm}$ box
should be small. However, since our pions
are light we explicitly checked for finite size effects.

We tried to plan our simulations carefully, but we made errors
nevertheless. Mentioning them here might help others to avoid them. 
A large part of the scaling test was done in a fixed small 
$L=1.2 \,\mathrm{fm}$ volume and the number of configurations was chosen 
to be approximately the same as in the large volume simulations. 
The fluctuations in the small volume are, however, stronger which makes 
our scaling tests less stringent. In addition, we were running with 
the same quark mass set on the small volume (scaling test) as on 
the large volume (spectroscopy). This made little sense since the scaling 
test is most interesting for the heavy, compact objects, while most of 
the computer time was spent on the light objects.

\subsection{Scaling tests}

The parameters of our simulation allow us to compare spectroscopic data
in a fixed $L=1.2\,\mathrm{fm}$ volume at three different
resolutions (see Table~\ref{table:1}): 
$a(r_0)=0.153\,\mathrm{fm}$, $0.102\,\mathrm{fm}$, $0.077\,\mathrm{fm}$ 
and $a(r_0)=0.148\,\mathrm{fm}$, $0.102\,\mathrm{fm}$, $0.078\,\mathrm{fm}$ 
for the FP and CI actions, respectively. 

As discussed in Section 3, the topological finite size
artifacts are large in such a small volume at small quark masses. This
is the case not only for the pion, but also for the nucleon as 
has been illustrated in Fig.~\ref{fig:4Nop}. 
Here it is mandatory to use correlators for which these artifacts are 
suppressed.
 
Our statistics in this small volume simulation is poorer than 
in the large volume spectroscopy. For reasons discussed in Section 2.3 
we decided to use uncorrelated fits with the weight 
$\mathrm{Cov}_\mathrm{measd}$ in (\ref{chi2}) for the FP operator
while we continued to use correlated fits for the CI case.
In the Appendix we collect the pseudoscalar, vector and nucleon mass 
predictions for different quark masses. In a few cases at small 
quark masses, our statistics and the difficulty with the zero mode 
artifacts did not allow us to get a mass prediction with a reliable error 
estimate. No numbers are quoted in these cases.

\begin{figure}[p]
\vspace*{-6mm}
\begin{center}
\includegraphics[width=9cm,clip]{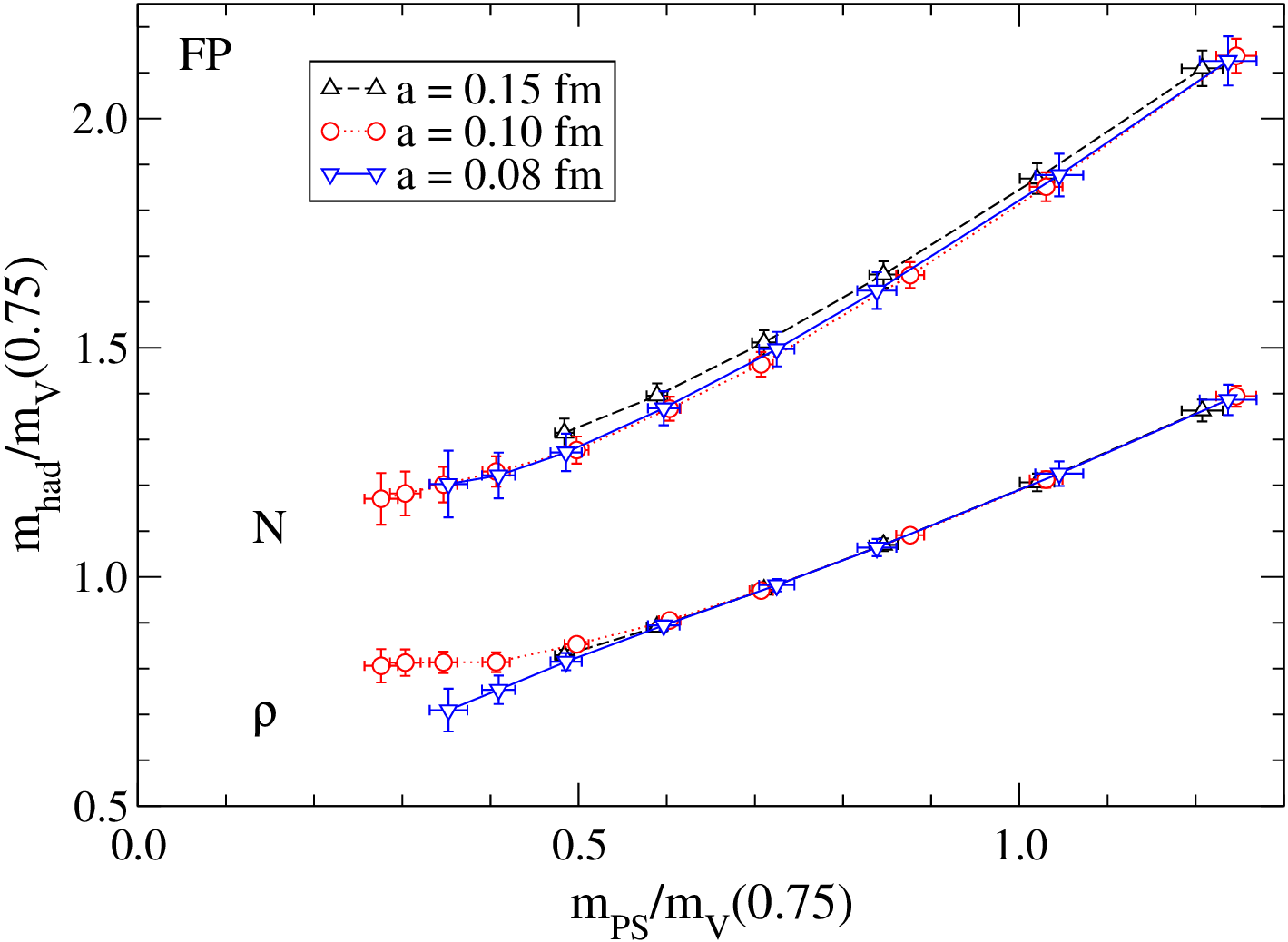}
\end{center}
\vspace{-6mm}
\caption{{}
Scaling test in a small volume of $L=1.2\,\mathrm{fm}$
for vector meson (lower curve) and the nucleon at three values of the
lattice spacing. 
The hadron masses are measured in $m_\mathrm{V}(x_0), x_0=0.75$ units,
i.e.~in terms of the mass in the vector meson channel at
$m_\mathrm{PS}/m_\mathrm{V}=0.75$. 
We used correlators with reduced quenched zero mode 
artifacts. The strange behavior in the vector channel at small quark
masses on the finest lattice might be related to the remaining 
${\cal O}((am_q)^{-1})$ zero mode effects.
% scaling\_test.eps
}
\label{fig:scaling}
\vspace{6mm}
\begin{center}
\includegraphics[width=9cm,clip]{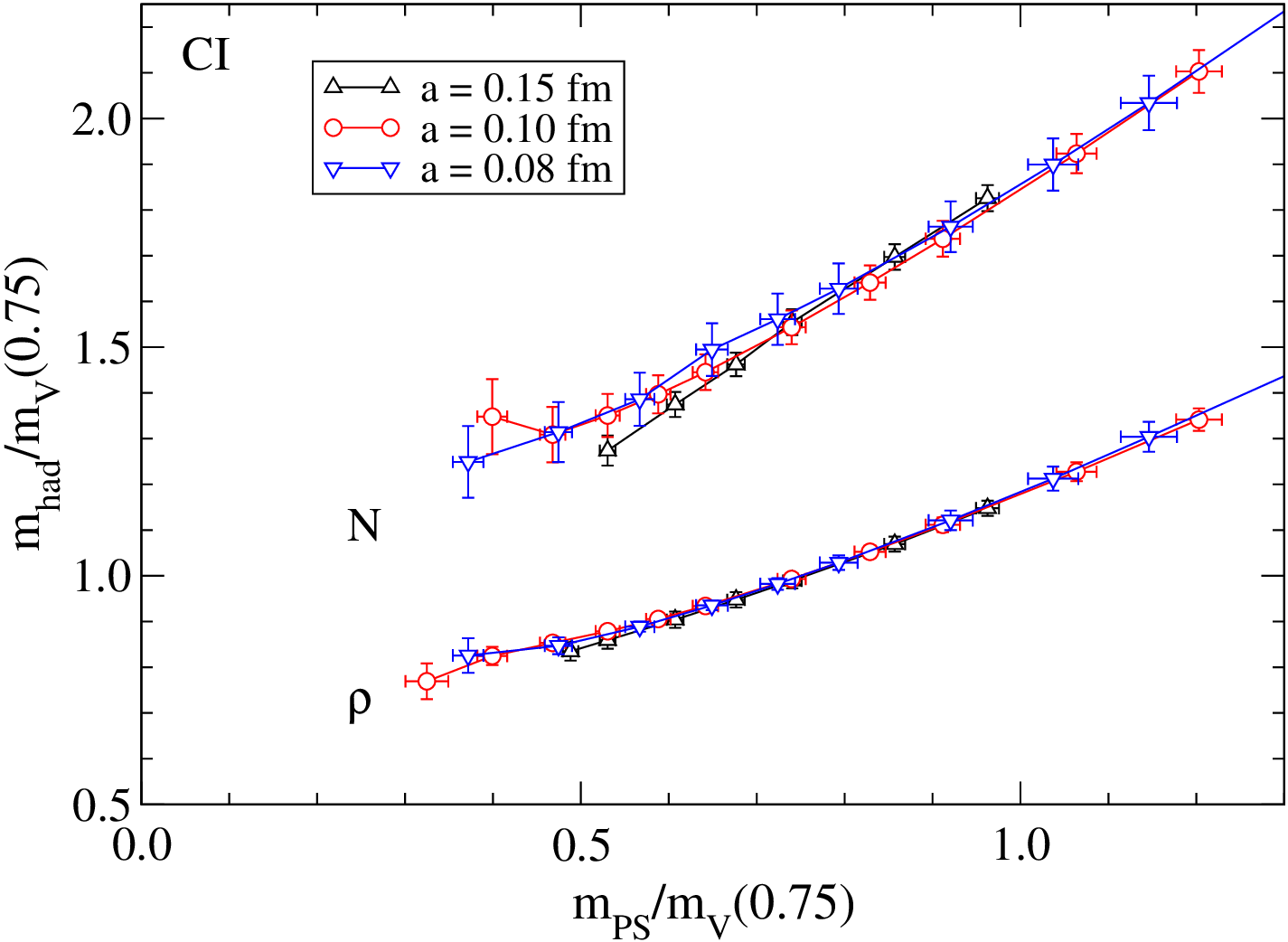}
\end{center}
\vspace{-6mm}
\caption{{}
Same as Fig.~\ref{fig:scaling} for the CI operator.
The standard nucleon correlator was used here.
% scaling\_test\_CI.eps
}
\label{fig:scaling_CI}
\end{figure}

Fig.~\ref{fig:scaling} shows the dependence of the vector meson
and the nucleon mass on the lattice resolution for different quark
masses. The hadron masses are measured in $m_\mathrm{V}(x_0), x_0=0.75$ units,
i.e.~in terms of the mass of the vector meson at
$m_\mathrm{PS}/m_\mathrm{V}=0.75$. For large quark masses the hadrons are
expected to be more compact and so more sensitive to the lattice
resolution. Within the errors, the nucleon channel shows no cut-off 
effects neither for large nor for small quark masses.
We used appropriate correlators to reduce/cancel the quenched zero mode 
artifacts. The strange behavior in the vector channel at small quark
masses for $a= 0.08\,\mathrm{fm}$ might be related to the remaining 
${\cal O}((am_q)^{-1})$ zero mode effects. 
It should be noted that for the vector meson curve the
(0.75,1.0) point is fixed independently of $a$ which makes this case
less informative for testing scaling violations. Furthermore, considering
mass ratios might cancel parts of the cut-off effects in general. 
Fig.~\ref{fig:scaling_CI} is the equivalent plot for the CI action and 
similarly to the FP case the data for the different values of $a$ 
are compatible within error bars. 

\begin{figure}[tbp]
\vspace{-4mm}
\begin{center}
\includegraphics[width=9.5cm,clip]{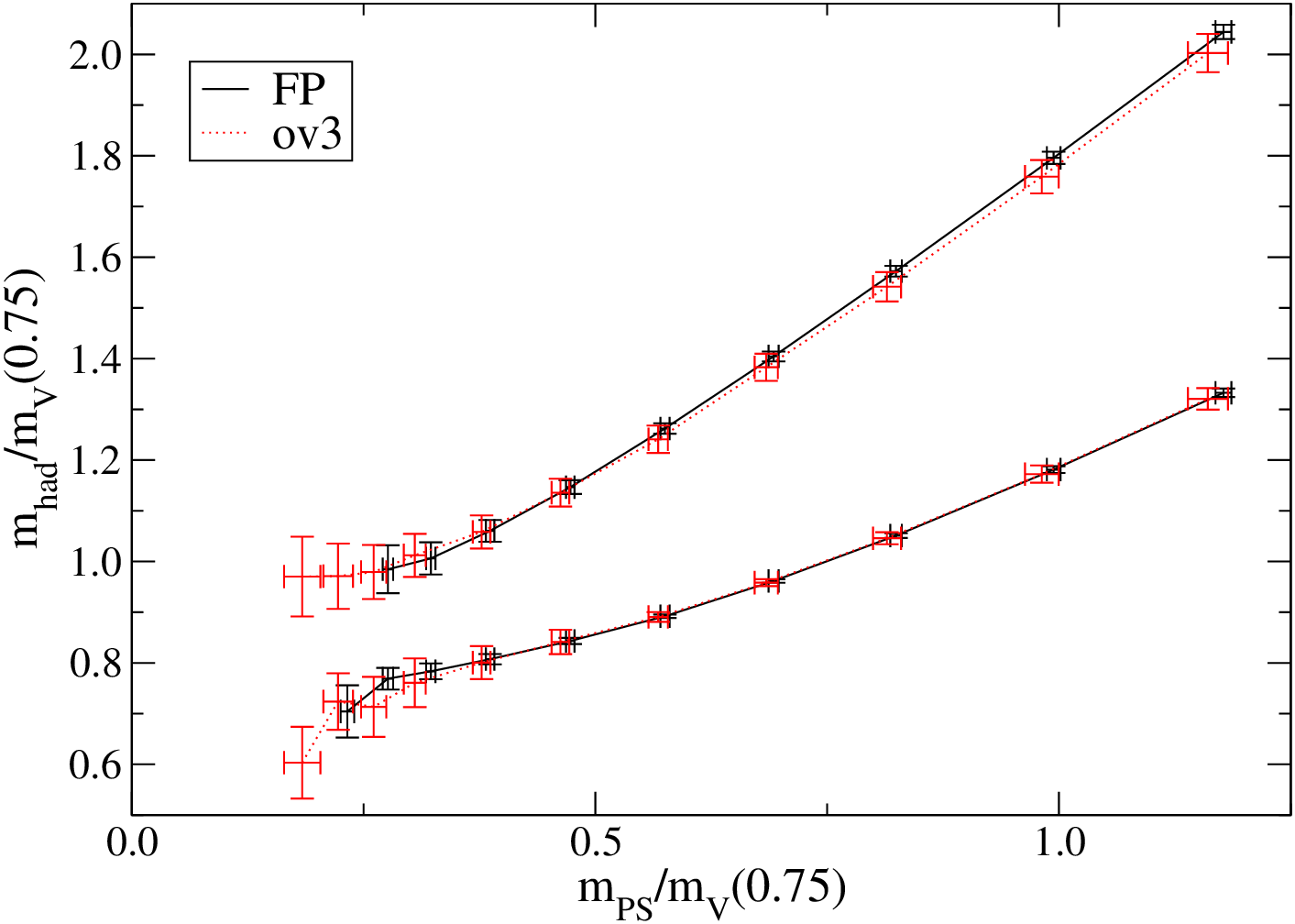}
\end{center}
\vspace{-4mm}
\caption{{} Results obtained with two different Dirac operators
(FP and FP+3 overlap steps) are compared here at fixed
$a(r_0)=0.153\,\mathrm{fm}$ in a box of $L=1.8\,\mathrm{fm}$. 
The nucleon (upper curve) and vector meson masses are measured in 
$m_\mathrm{V}(x=0.75)$ units like in Fig.~\ref{fig:scaling}.
% mxhat\_fp\_o3.eps
}
\label{fig:mxhat_fp_o3}
\vspace{5mm}
\begin{center}
\includegraphics[width=9.5cm,clip]{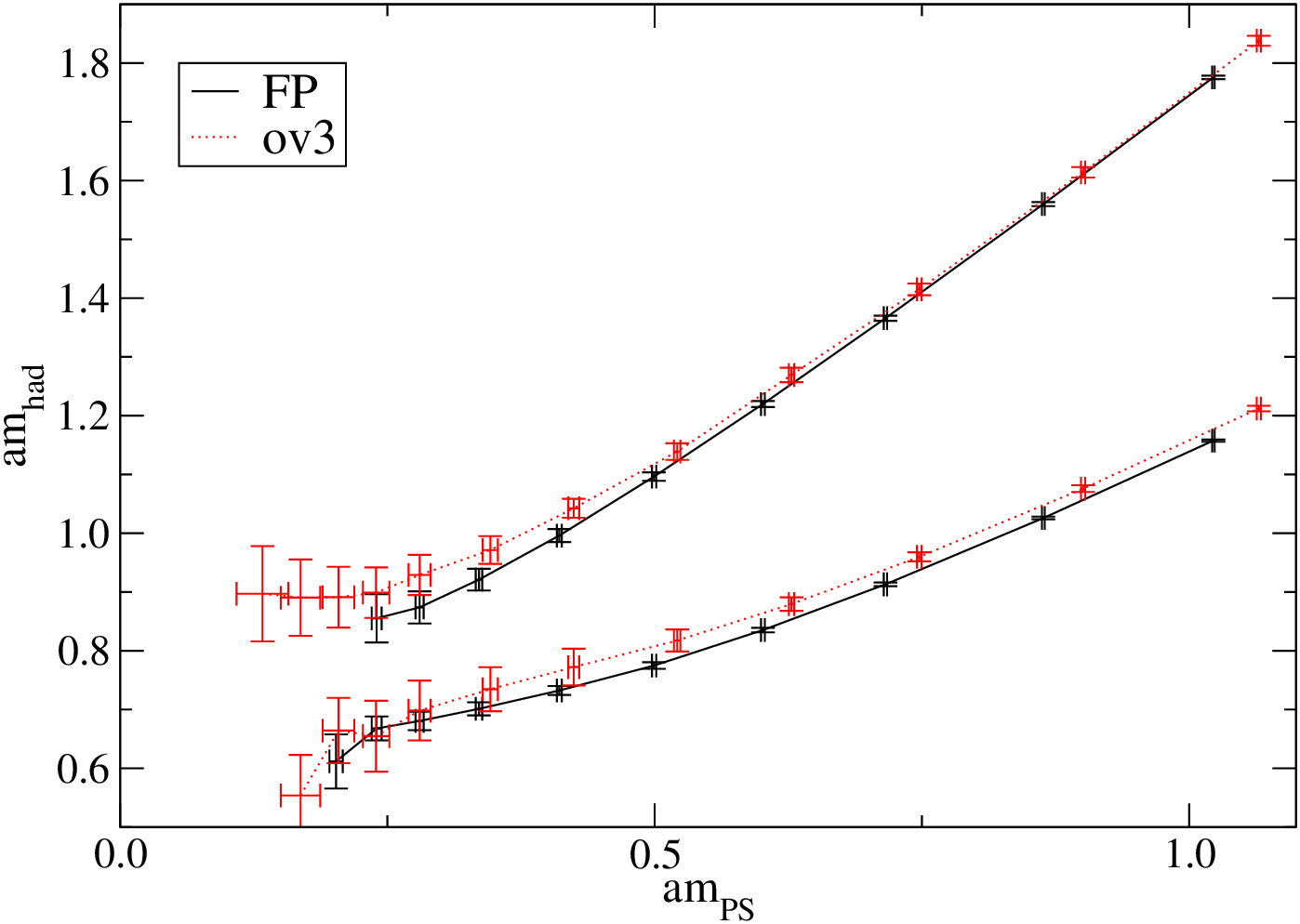}
\end{center}
\vspace{-4mm}
\caption{{} The same as Fig.~\ref{fig:mxhat_fp_o3} but in terms of the
dimensionless hadron masses. This is a correct comparison since they
were measured on the same gauge configurations in this quenched study.
% amx\_fp\_ov.eps
}
\label{fig:amx_fp_ov}
\end{figure}

\begin{figure}[tbp]
%\vspace{-6mm}
\begin{center}
\includegraphics[width=9cm,clip]{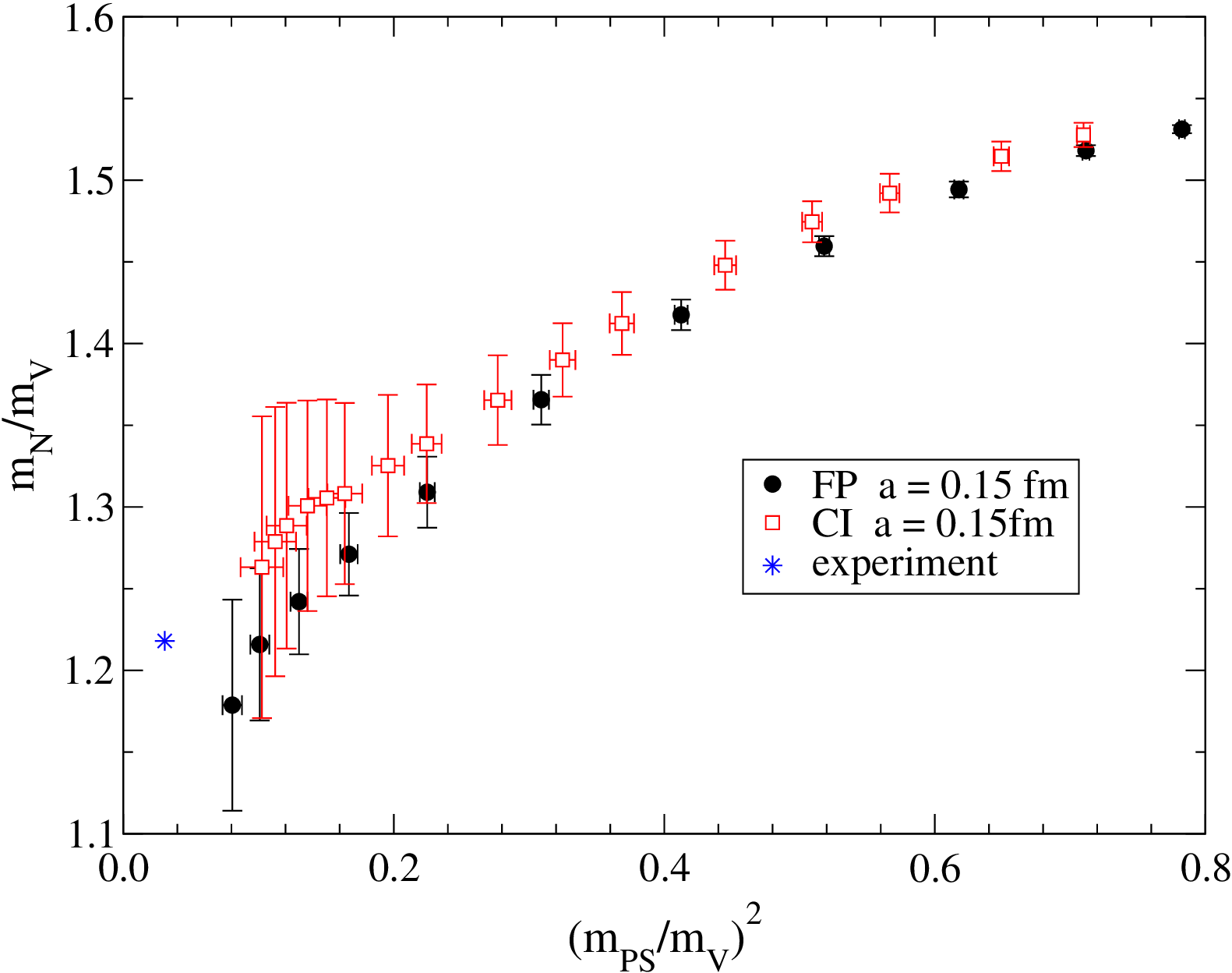}
\end{center}
\vspace{-6mm}
\caption{{}The APE plots for the nucleon with the FP and CI actions. 
% APE\_largeV\_FP\_CI.eps
}
\label{fig:APE_N_largeV1}
\end{figure}

\begin{figure}[tbp]
\begin{center}
\vspace{-10mm}
\includegraphics[width=12cm,clip]{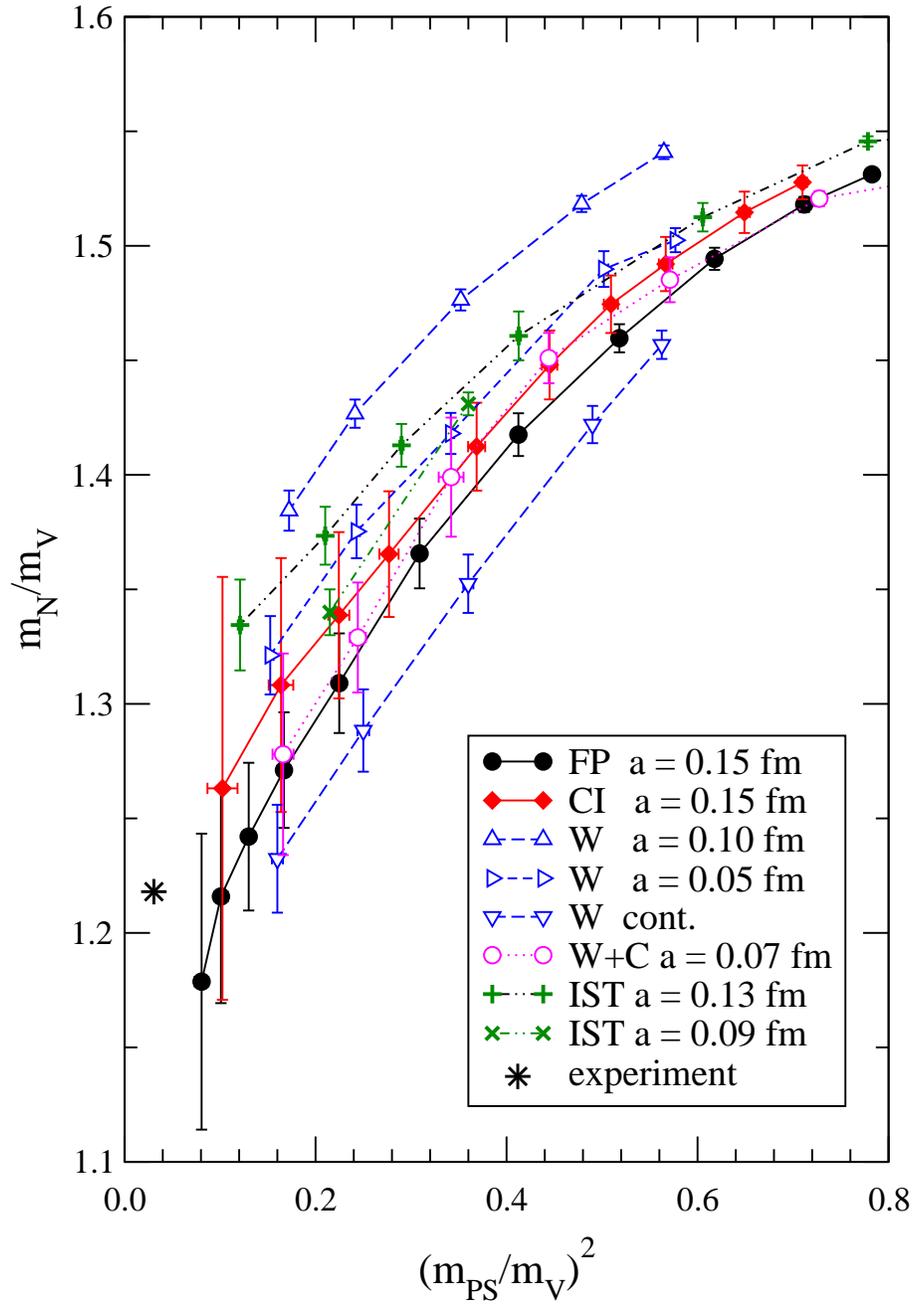}
\end{center}
\vspace{-6mm}
\caption{{}The APE plots for the nucleon with the FP and CI actions are 
compared to those obtained by CP-PACS at $a=0.10\,\mathrm{fm}$, 
$a=0.05\,\mathrm{fm}$ and their continuum extrapolation,
by MILC at $a=0.13\,\mathrm{fm}$, $a=0.09\,\mathrm{fm}$ and with clover 
improved Wilson fermions at $a = 0.07$~fm.
The asterisk indicates the experimental number.
% APE\_largeV\_FP\_CP\_PACS\_IST\_1.eps
}
\label{fig:APE_N_largeV2}
\end{figure}

\begin{figure}[tbp]
\begin{center}
\vspace{-8mm}
\includegraphics[width=9.3cm,clip]{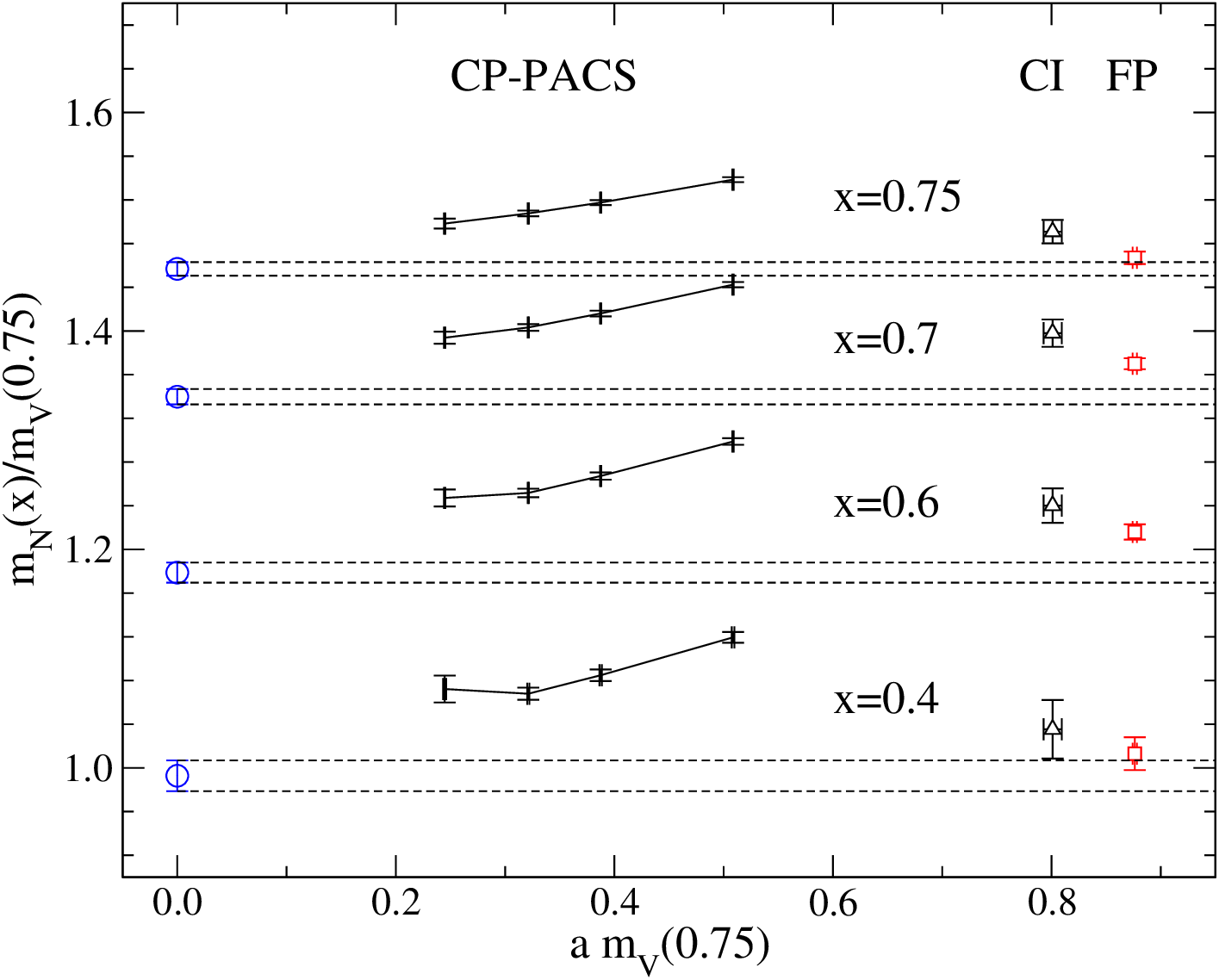}
\end{center}
\vspace{-8mm}
\caption{{}
The nucleon mass $m_\mathrm{N}(x)$ interpolated to 
$m_\mathrm{PS}/m_\mathrm{V}=x$ vs. the lattice spacing  at four different 
values of $x$, in units of $m_\mathrm{V}(0.75)$.
For the Wilson action the four data points and their continuum
extrapolations (open circles) are taken from \cite{cppacsquench}(CP-PACS). 
The values with the FP action are on the right at 
$a m_\mathrm{V}(0.75) \approx 0.87$ ($a=0.153\,\mathrm{fm}$),
the CI data ($a=0.148\,\mathrm{fm}$) slightly left of the FP
numbers.
% CP\_PACS\_mNxh1.eps
}
\label{fig:CPPACS_cont}
%\end{figure}

%\begin{figure}[tbp]
\begin{center}
\includegraphics[width=9.3cm, clip]{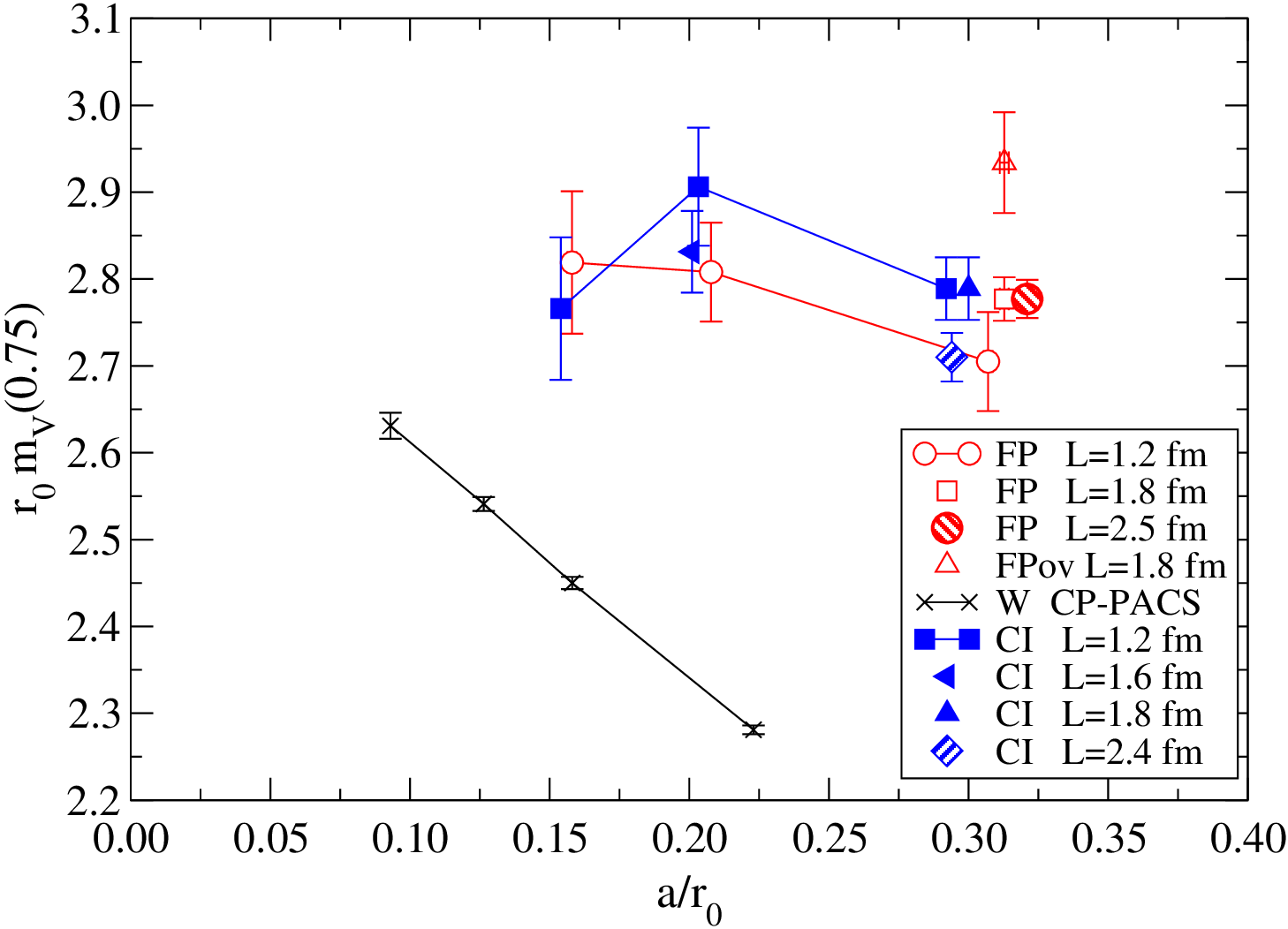}
\end{center}
\vspace{-8mm}
\caption{{} The scaling behavior of $r_0m_\mathrm{V}(x=0.75)$. Since no
finite volume effects were observed in $m_\mathrm{V}$ in the region 
$L = 1.2-2.5\,\mathrm{fm}$ (see Fig.~\ref{fig:h_size_eff_fp})
we included data from different volumes in this figure. Results from
four different actions are compared here: FP, FP+3 overlap steps (1
data point only), CI and Wilson fermions from CP-PACS.
% r0\_mrhohat.eps
}
\label{fig:r0_scaling}
\end{figure}

Figs.~\ref{fig:mxhat_fp_o3} and \ref{fig:amx_fp_ov} compare the
results from two different Dirac operators: the parametrized FP and
the one obtained from this after 3 overlap projection steps. The latter has
smaller deviations from a GW solution, but might be driven away from
the fixed-point of the renormalization group transformation used to construct
the FP action.  
Fig.~\ref{fig:mxhat_fp_o3} compares the vector and nucleon masses in 
$m_\mathrm{V}(x_0), x_0=0.75$ units at
$a(r_0)=0.153\,\mathrm{fm}$ in a box of size $L=1.8\,\mathrm{fm}$. 
No difference can be seen within the errors. Since the gauge configurations 
are the same for the two Dirac operators (i.e.~$a(r_0)$ is fixed and 
the same) we can compare the dimensionless hadron masses $am_h$ directly. 
In this way we avoid taking mass ratios which might cancel part of the cut-off
effects. The corresponding Fig.~\ref{fig:amx_fp_ov} indicates some deviations
beyond the statistical error in the vector channel for
larger quark masses. This might be a (small) cut-off effect coming from one 
(or both) of the Dirac operators, but 
Fig.~\ref{fig:amx_fp_ov} does not reveal which Dirac operator
is responsible for it.

Figs.~\ref{fig:APE_N_largeV1} and \ref{fig:APE_N_largeV2} are APE
plots where the large volume FP and CI results (Fig.~\ref{fig:APE_N_largeV1})
and the FP, CI, CP-PACS
\cite{cppacsquench}, improved staggered MILC
\cite{MILC13,MILC09} and clover improved Wilson results \cite{pleiter} 
(Fig.~\ref{fig:APE_N_largeV2}) are
compared. The continuum extrapolation of the CP-PACS data is also
given in Fig.~\ref{fig:APE_N_largeV2}. This non-trivial continuum
extrapolation is illustrated in Fig.~\ref{fig:CPPACS_cont}, where the
FP and CI numbers are repeated again. 
We can draw the following conclusions from
these figures. The FP and CI APE-plots are consistent with each other.
The improved staggered results at $a=0.13\,\mathrm{fm}$ and
$a=0.09\,\mathrm{fm}$ are lying above the 
FP curve beyond the statistical errors. If we assume that
the CP-PACS continuum extrapolation is correct then the FP results at
$a=0.15\,\mathrm{fm}$ show cut-off effects, but they are closer to the
continuum than the Wilson results at $a=0.05\,\mathrm{fm}$ and the improved
staggered results at $a=0.13\,\mathrm{fm}$ and $a=0.09\,\mathrm{fm}$.

Finally, we connect the hadronic observables with the Sommer parameter
$r_0$ obtained in the gauge sector. Fig.~\ref{fig:r0_scaling} shows 
$r_0 m_\mathrm{V}(x=0.75)$ as a function of $a/r_0$. This figure compares
data for three different resolutions at fixed $L=1.2\,\mathrm{fm}$. 
As we shall see in the next section we observe no finite size effects in 
the vector meson channel in the region $L = 1.2-2.5\,\mathrm{fm}$. 
For this reason we included in this figure data from different volumes also.

The FP and CI data
obtained in a fixed $L=1.2\,\mathrm{fm}$ box show no cut-off effects beyond 
the errors and are consistent with each other\footnote{In both cases 
the coarsest $a=0.15 \,\mathrm{fm}$ point is somewhat shifted downwards 
although within the statistical errors. In the FP case this point is also 
below the large volume result at the same resolution (again within errors). 
Since the masses are expected to increase rather then decrease
with decreasing volume, the  downward shift of the small-volume 
mass for FP at $a/r_0\approx 0.3$ is not a real effect.}. 
The same conclusion can be drawn if we include points obtained  at larger 
volumes. On the other hand, the prediction 
of the Dirac operator after 3 overlap steps at $a(r_0)=0.153\,\mathrm{fm}$ 
in a box $L=1.8\,\mathrm{fm}$ deviates from that of the FP operator on 
the same lattice (this was already shown by Fig.~\ref{fig:amx_fp_ov}). 
For comparison, the CP-PACS Wilson action results \cite{cppacsquench}
for $a = 0.05-0.1\,\mathrm{fm}$ are also given. 

\subsection{Finite size effects}

\begin{figure}[tbp]
\vspace{-5mm}
\begin{center}
\includegraphics[width=9.5cm,clip]{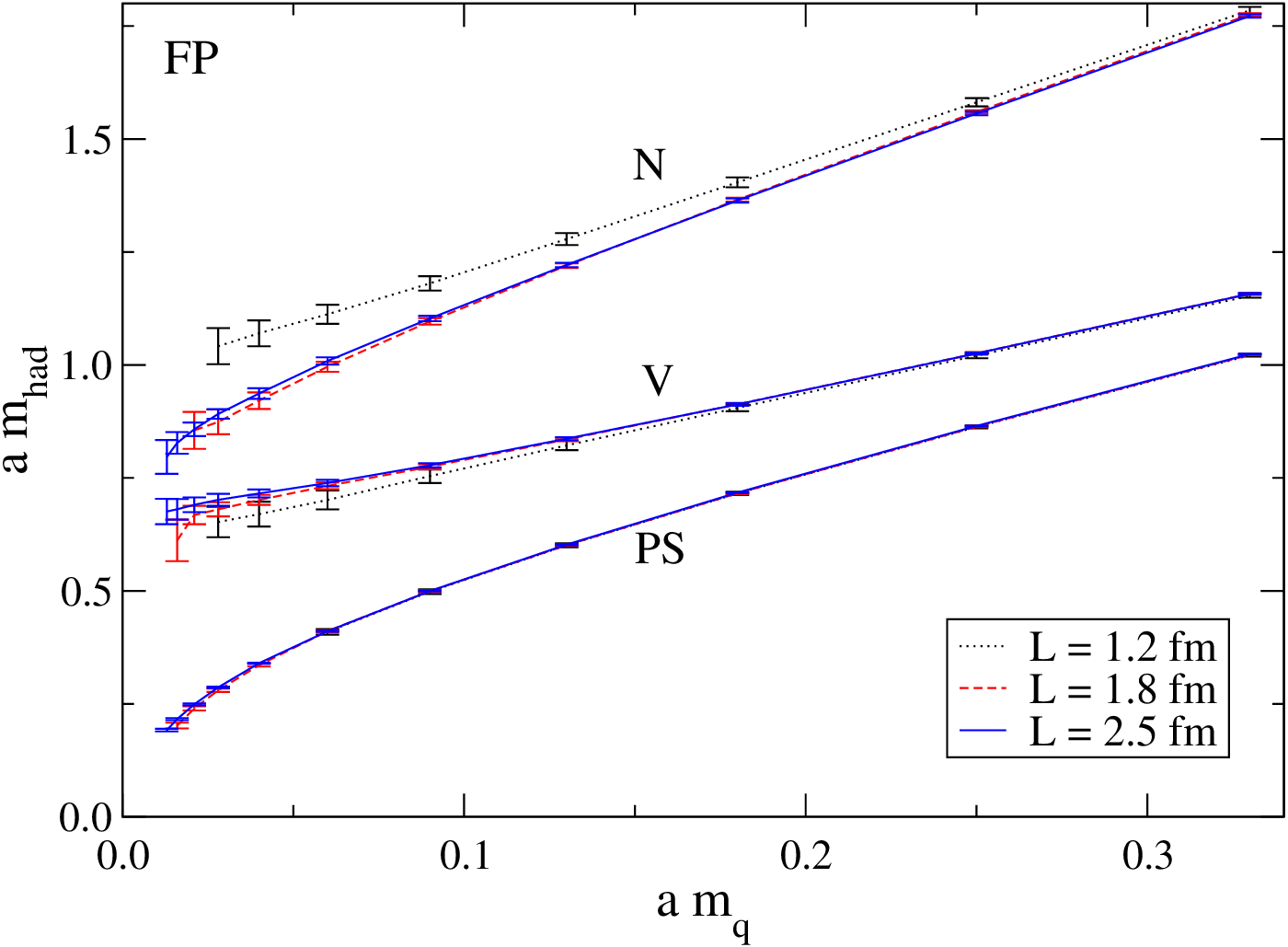}
\end{center}
\vspace{-4mm}
\caption{{}
Hadron masses in the PS, V and N channels vs.~the bare quark mass,
measured with the FP action at $a=0.15\,\mathrm{fm}$ and at
three different lattice volumes.
The nucleon mass shows finite size effects at $L=1.2\,\mathrm{fm}$,
while the PS and V meson data indicate no finite size effects even
in such a small lattice volume in our quenched simulation. 
% mhad\_vs\_mq\_b3.0\_FP.eps
}
\label{fig:h_size_eff_fp}
\vspace{5mm}
\begin{center}
\includegraphics[width=9.5cm,clip]{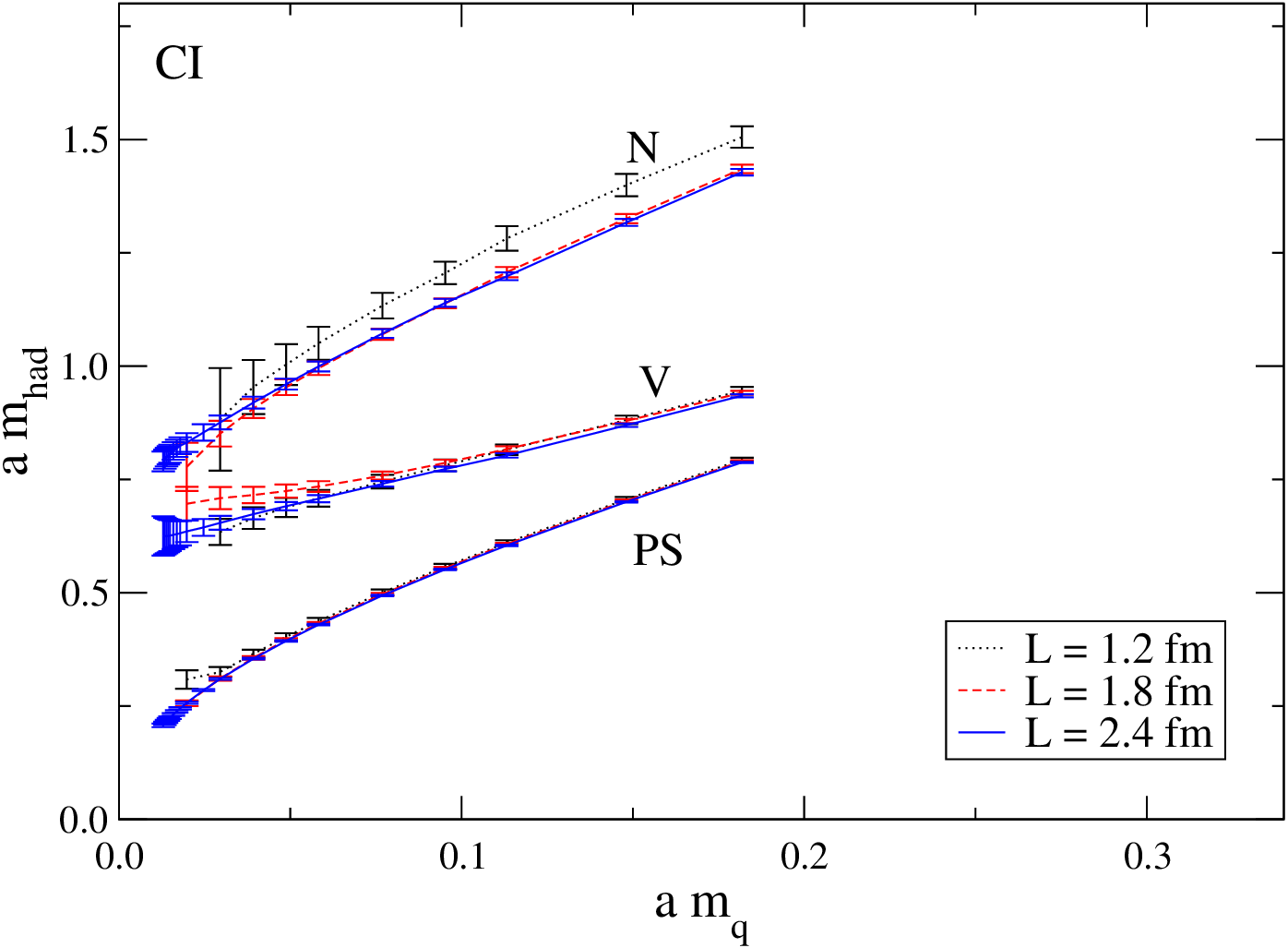}
\end{center}
\vspace{-4mm}
\caption{{}
Same as Fig.~\ref{fig:h_size_eff_fp}, now for the CI operator.\hfill
% mhad\_vs\_mq\_b7.90\_CI\_1.eps
}
\label{fig:h_size_eff_ci}
\end{figure}

We studied the spectrum in three different volumes $L=2.5\,\mathrm{fm}$,
$L=1.8\,\mathrm{fm}$ and $L=1.2\,\mathrm{fm}$ at fixed lattice unit 
$a~= 0.15~\mathrm{fm}$. Figs.~\ref{fig:h_size_eff_fp}
and \ref{fig:h_size_eff_ci} show the volume dependence of the N, V and
PS masses as a function of the quark mass. No finite size effects
are seen beyond the errors, except for the nucleon in the smallest 
$L=1.2\,\mathrm{fm}$ box. In this small box the nucleon mass is pushed 
upwards as the quark mass is decreased.

\begin{figure}[tbp]
\vspace{-6mm}
\begin{center}
\includegraphics[width=9cm,clip]{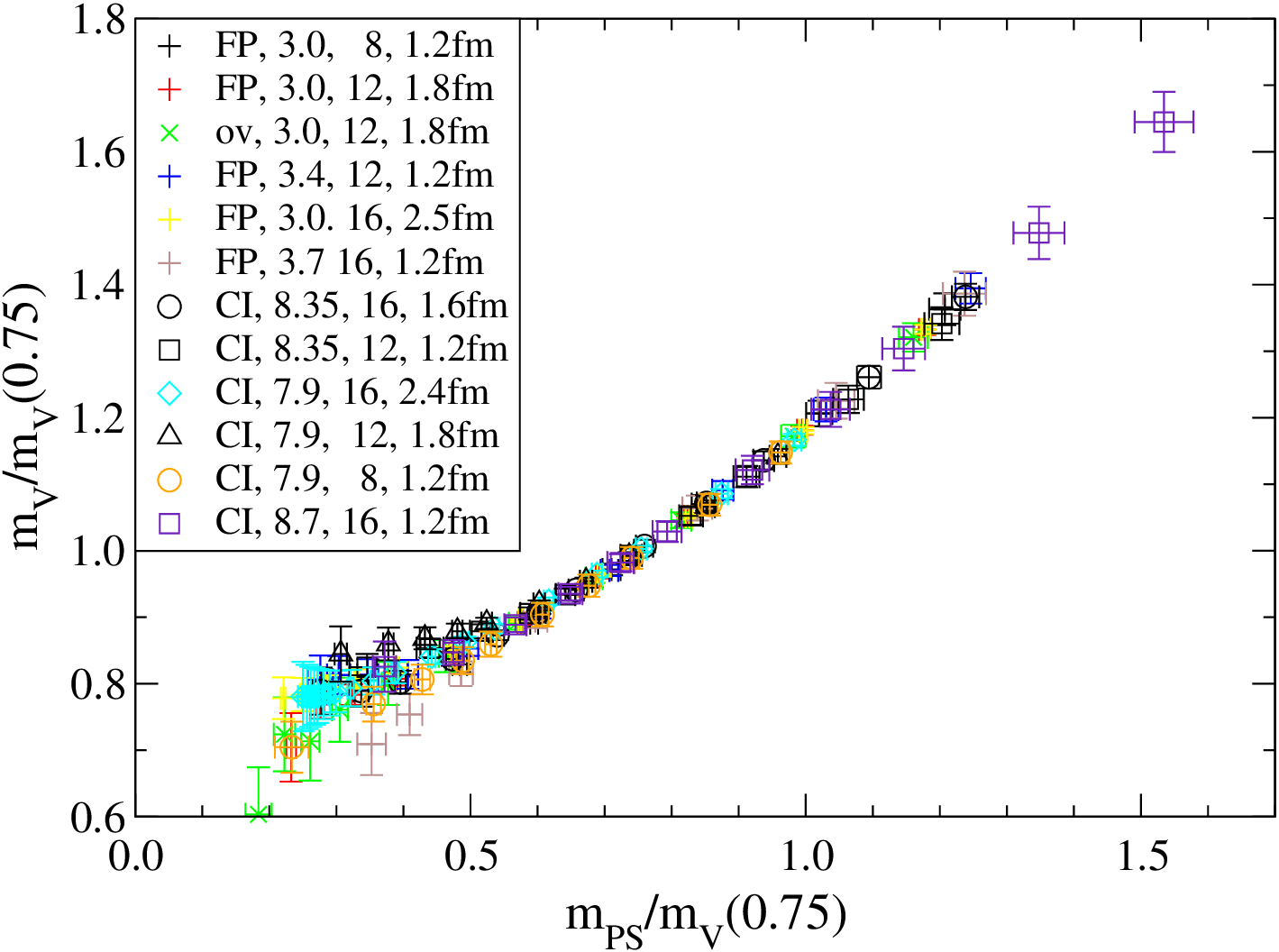}
\end{center}
\vspace{-4mm}
\caption{{}
The masses $m_\mathrm{V}$ vs. $m_\mathrm{PS}$, both in 
$m_\mathrm{V}(0.75)$ units. We show all our results for 
the FP and CI operators.
% mrhohat\_FP\_CI.eps
\label{fig:our_rho}
}
\vspace{5mm}
\begin{center}
\includegraphics[width=9cm,clip]{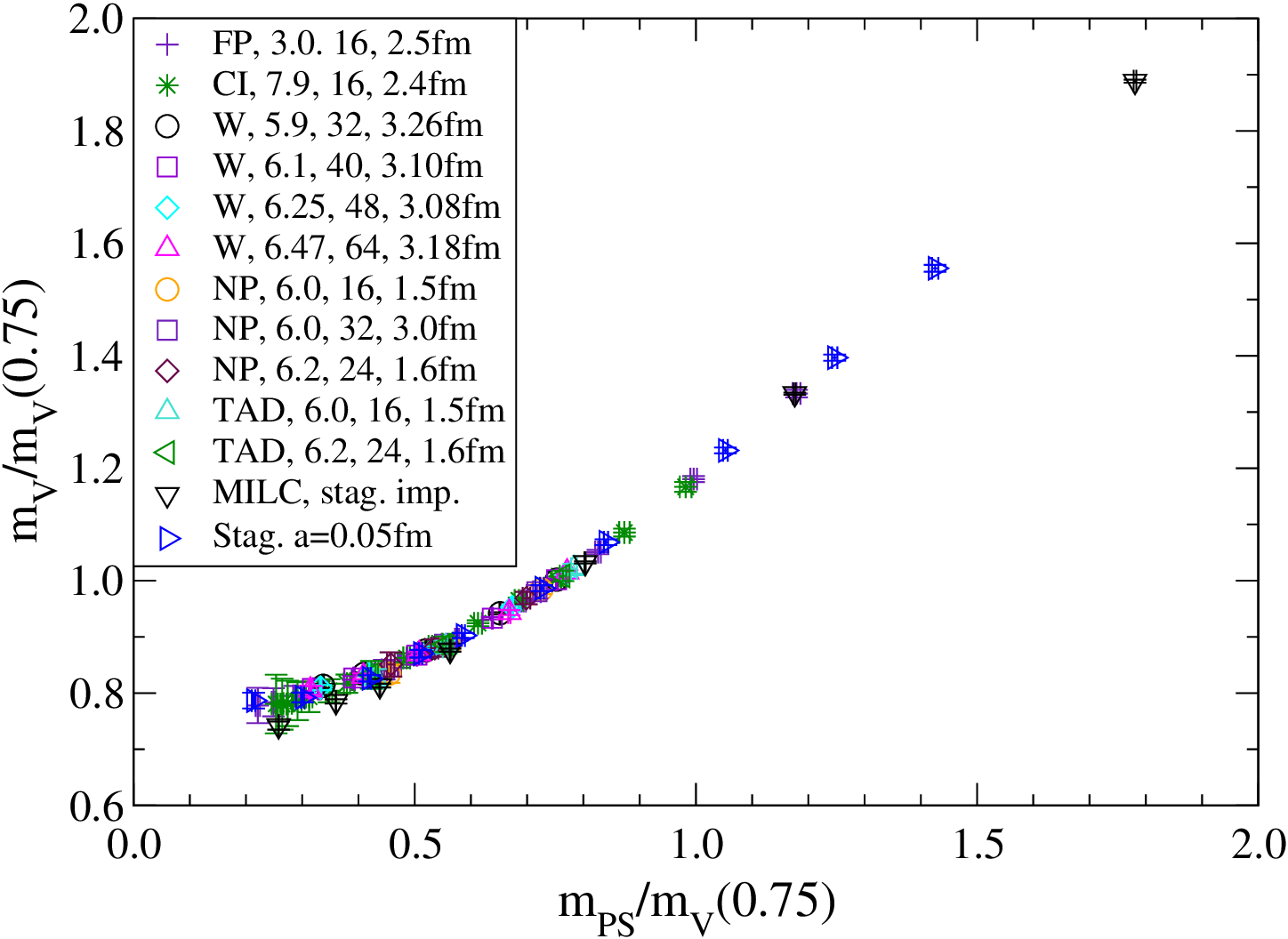}
\end{center}
\vspace{-4mm}
\caption{{}
The masses $m_\mathrm{V}$ vs. $m_\mathrm{PS}$, both in 
$m_\mathrm{V}(0.75)$ units. The Wilson data (W) are taken from 
\cite{cppacsquench}, the nonperturbatively improved (NP) and tadpole
improved (TAD) data from \cite{ukqcd1}, the staggered improved data (MILC)
from \cite{MILC13}, and the staggered data from \cite{kimohta}.
In the legend we indicate the $\beta$ values of the corresponding
gauge action, the spatial lattice size in lattice units and in fm.
% mrhohat.eps
\label{fig:all_rho}
}
\end{figure}

\begin{figure}[tbp]
\vspace{-5mm}
\begin{center}
\includegraphics[width=9.5cm,clip]{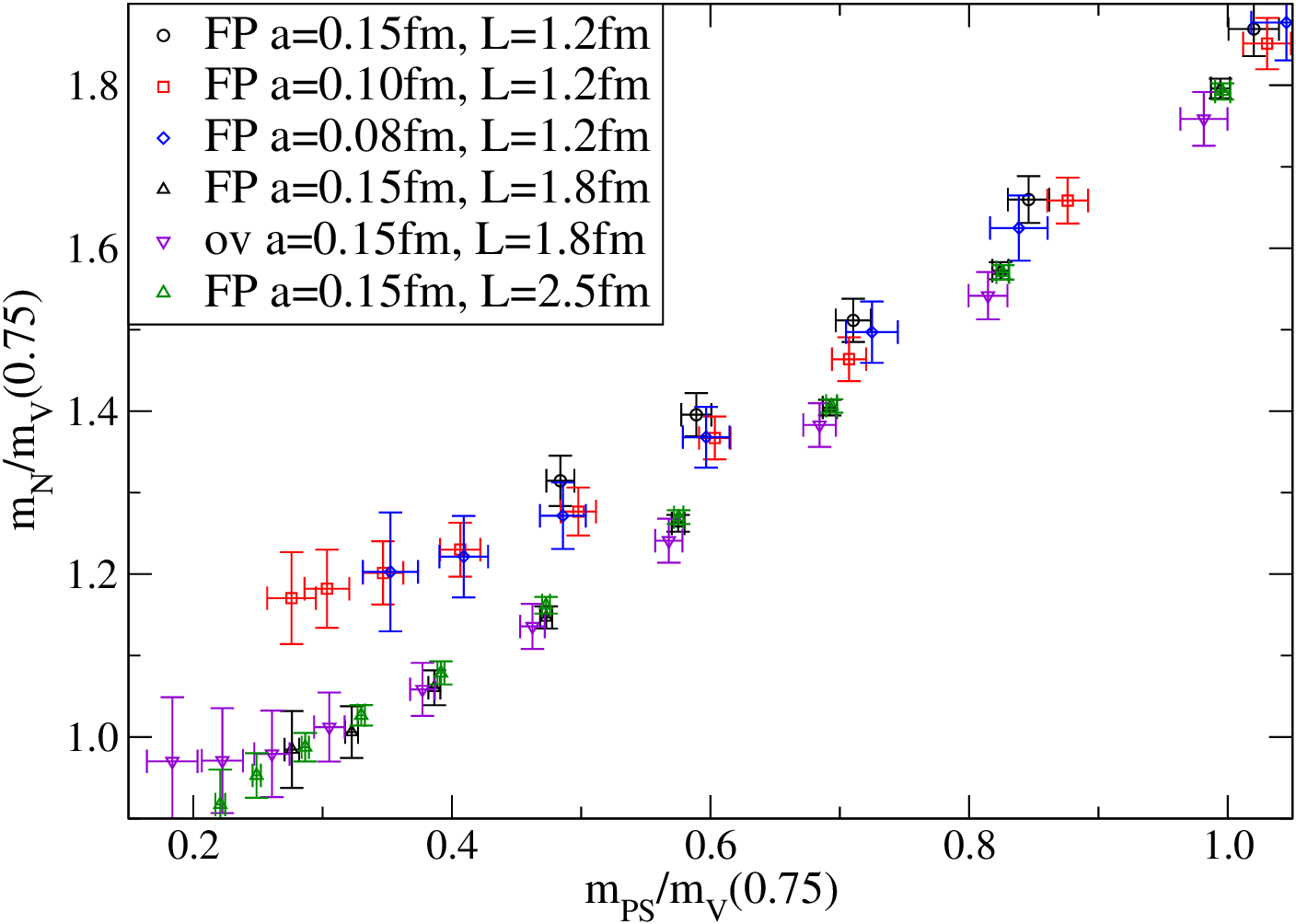}
\end{center}
\vspace{-4mm}
\caption{{}
The nucleon mass vs.~the pseudoscalar mass, both measured in 
$m_\mathrm{V}(0.75)$ units for the FP action  at different lattice 
spacings and physical volumes. The small-volume data at 
$L=1.2\,\mathrm{fm}$ show scaling. The three volumes at $a=0.15\,\mathrm{fm}$
shows finite size effects at $L=1.2\,\mathrm{fm}$, but not for
$L\ge 1.8\,\mathrm{fm}$.
% mnhat\_FP.eps
}
\label{fig:mn_fp}
\vspace{5mm}
\begin{center}
\includegraphics[width=9.5cm,clip]{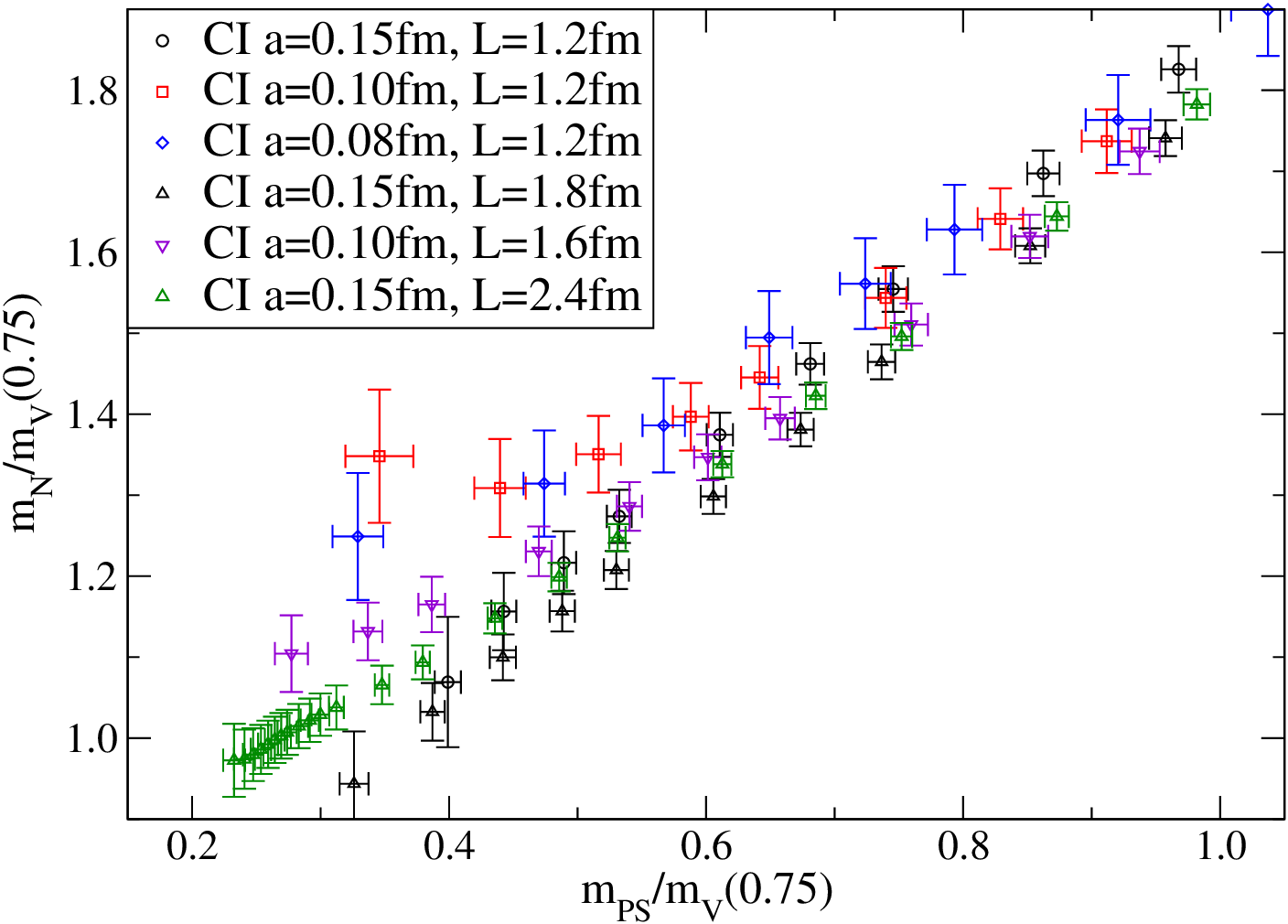}
\end{center}
\vspace{-4mm}
\caption{{}
Same as Fig.~\ref{fig:mn_fp} now for the CI operator.
% mnhat\_CI.eps
}
\label{fig:mn_ci}
\end{figure}

Figs.~\ref{fig:our_rho} and \ref{fig:all_rho} show data in
different volumes and lattice resolutions for the vector meson and the
nucleon mass as a function of the pseudoscalar mass, where all the
masses are measured in $m_\mathrm{V}(0.75)$ units.
The somewhat unexpected
message from Fig.~\ref{fig:all_rho} is that the vector meson mass in
these units is largely independent of the cut-off, of the volume and of
the action used, i.e.~the cut-off effects are practically absorbed 
by $m_\mathrm{V}(0.75)$. 

Fig.~\ref{fig:mn_fp} summarizes what we already saw in
earlier figures of the nucleon mass from the FP action. The points referring to
a fixed $L=1.2\,\mathrm{fm}$ box at three different resolutions run together,
no cut-off effects are seen. On the other hand, in this smallest box the
nucleon shows significant finite size effects which are decreasing as the quark
mass is increasing. There is no finite size effect seen when comparing the
$L=1.8\,\mathrm{fm}$  and $L=2.5\,\mathrm{fm}$ results. At heavy quark masses
all the points join to form a single curve indicating the absence both of
finite size and cut-off effects.

In the case of the CI operator the interpretation of the corresponding 
Fig.~\ref{fig:mn_ci} is similar, although the separation into two sets of data,
one suffering from finite size effects, one free of them is not as clear-cut 
as for the FP operator. The main reason is that for the CI medium size volumes
the physical size is slightly different. We have $L = 1.6$ fm and $L = 1.8$ 
for our $16^3\times 32, \beta = 8.35$, respectively $12^3\times 24, 
\beta = 7.90$ CI lattices. For the FP operator the two corresponding lattices 
both have the somewhat larger size $L = 1.8$. Fig.~\ref{fig:mn_ci}
shows that the $L = 1.6$ fm data lie in between the $L = 1.2$ results
clearly suffering from finite size effects and the $L = 1.8$ fm 
and $L = 2.4$ fm data essentially free of them. The drop of the $L = 1.8$ fm
data below the $L = 2.4$ fm results at small quark masses is probably due to the
topological finite size artifacts (we did not use improved operators here). 
We remark that in \cite{cinstar} where a linear combination of three 
nucleon operators was used, this discrepancy is resolved. 
\vskip5mm
\noindent
{\bf Acknowledgements:} 
The calculations were done on the Hitachi SR8000 at the Leibniz
Rechenzentrum in Munich and at the Swiss Center for Scientific
Computing in Manno. We thank the LRZ and CSCS staff for training 
and support. 
This work was supported in parts by DFG, BMBF, SNF, the
European Community's Human Potential Programme under contract
HPRN-CT-2000-00145, the DOE under grant DOE-FG03-97ER40546
and the Fonds zur F\"orderung der Wissenschaftlichen Forschung in 
\"Osterreich, project FWF-P16310-N08.
C.\ Gattringer acknowledges support by the Austrian Academy of Sciences 
(APART 654). 
The authors thank Gunnar Bali, Gilberto Colangelo, Christine Davies, 
Tom DeGrand, Anna Hasenfratz, Rainer Sommer and Stewart Wright 
for useful discussions.

\clearpage

\newpage

\begin{appendix}

\section{Data obtained with the FP action}
\label{FP:data}
We list here the measured hadron masses for different input bare quark
masses. All the masses are in lattice units. The numbers in brackets
give the jackknife errors.
We give also the value of $\chi^2_\mathrm{df}$ of the fit and the fit 
range. In cases where we could not use a correlated fit with the measured
covariance matrix (see, Section 2.3), the value of
$\chi^2_\mathrm{df}$ does not characterize the quality of the fit. 
In these cases no $\chi^2_\mathrm{df}$ is quoted.

The pseudoscalar mass was determined from the 
$\langle PP \rangle - \langle SS \rangle$ and 
the $\langle PP \rangle$ correlators for small and large
quark masses, respectively (see Section 3). The
corresponding switching-point (the quark mass at and above which the 
$\langle PP \rangle $ correlator is used) is quoted in the table caption.
In a few cases (mainly in small volumes at small quark masses) we did not
quote the hadron mass since the error was very large and difficult to pin down.

We measured the baryon spectrum using several different correlators
(see Section 3). In most of the cases (in particular in large volumes)
the correlator in \eqref{Ncorr1} proved to be the best choice 
for the nucleon and and the analogous one for $\Delta$. 
The exceptions are indicated in the table caption.
\vspace{10mm}

\begin{table}[hb]
\small
\begin{center}
\begin{tabular}{l|l|c|c|l|c|c|l|c} \hline\hline
\multicolumn{9}{c}{$16^3\times 32$, $\beta=3.0$, parametrized FP} \\ \hline
$am_q$ & $am_{\rm PS}$(P) & $\chi^2_{df}$ & $t$ & $am_{\rm PS}$(P-S) &
$\chi^2_{df}$ & $t$ & $am_{\rm V}$ & $t$ \\
\hline
0.013 & 0.1992(31)& 1.5& [6,13]& 0.1919(30)& 0.9 & [6,16] & 0.676(28)& [5,10]\\
0.016 & 0.2227(20)& 1.1& [6,16]& 0.2163(24)& 0.9 & [6,14] & 0.681(23)& [5,10]\\
0.021 & 0.2535(15)& 0.9& [6,16]& 0.2489(20)& 0.8 & [6,14] & 0.691(16)& [5,9] \\
0.028 & 0.2891(14)& 0.7& [6,16]& 0.2865(17)& 0.7 & [6,13] & 0.701(13)& [5,10]\\
0.04  & 0.3401(13)& 0.5& [6,16]& 0.3393(15)& 0.5 & [6,16] & 0.716(9) & [5,9] \\
0.06  & 0.4107(12)& 0.5& [6,16]& 0.4118(14)& 0.4 & [6,16] & 0.739(7) & [6,9] \\
0.09  & 0.4997(11)& 0.5& [6,16]& 0.5023(13)& 0.6 & [7,16] & 0.778(5) & [6,10]\\
0.13  & 0.6024(11)& 0.6& [6,16]& 0.6049(13)& 1.0 & [8,16] & 0.837(3) & [6,10]\\
0.18  & 0.7173(10)& 0.8& [6,16]& 0.7196(12)& 1.6 & [9,16] & 0.913(3) & [7,12]\\
0.25  & 0.8650(9) & 0.8& [6,16]& 0.8665(12)& 1.7 & [11,16]& 1.025(2) & [7,12]\\
0.33  & 1.0240(9) & 0.6& [6,16]& 1.0257(11)& 1.1 & [11,16]& 1.158(2) & [7,12]\\
 \hline
\end{tabular}
\end{center}
\caption{Pseudo-scalar and vector meson masses on a $16^3\times 32$ lattice at
$\beta=3.0$ with $D^{\rm FP}$. The switching-point is $am_q=0.06$.
\label{tab:16x32_b3.0_PSV}
} 
\end{table}
\vspace{1cm}

\begin{table}[tbp]
\small
\begin{center}
\begin{tabular}{l|l|c|c|l|c|l|c} \hline\hline
 \multicolumn{8}{c}{$16^3\times 32$, $\beta=3.0$, parametrized FP} \\
\hline
$am_q$ & $am_{\rm Oct}$(N) & $\chi^2_{df}$ & $t$ & $am_{\rm Dec}(\Delta)$ &
$t$ &  $am_{\rm Oct}$ $(N^*)$ & $t$  \\
\hline
0.013 & 0.797(37) & 0.4 & [5,7]  & 1.104(87) &  [5,8] &           &        \\ 
0.016 & 0.828(24) & 0.2 & [5,7]  & 1.115(61) &  [5,8] &           &        \\ 
0.021 & 0.858(15) & 0.7 & [5,8]  & 1.142(44) &  [5,8] &           &        \\ 
0.028 & 0.892(11) & 1.1 & [5,10] & 1.165(32) &  [5,8] &           &        \\ 
0.04  & 0.937(12) & 1.1 & [7,12] & 1.192(22) &  [5,8] & 1.511(45) &  [4,6] \\ 
0.06  & 1.009(8)  & 1.3 & [7,16] & 1.231(15) &  [5,8] & 1.530(27) &  [4,6] \\ 
0.09  & 1.103(6)  & 0.4 & [7,16] & 1.293(11) &  [5,8] & 1.586(23) &  [4,6] \\ 
0.13  & 1.222(5)  & 0.2 & [7,14] & 1.377(10) &  [6,9] & 1.676(16) &  [4,6] \\ 
0.18  & 1.364(4)  & 0.4 & [7,14] & 1.495(8)  &  [6,9] & 1.795(16) &  [4,8] \\ 
0.25  & 1.557(4)  & 0.6 & [7,14] & 1.665(7)  &  [6,9] & 1.959(13) &  [4,8] \\ 
0.33  & 1.772(4)  & 0.7 & [7,14] & 1.863(6)  &  [6,9] & 2.150(11) &  [4,8] \\ 
 \hline
\end{tabular}
%\vspace{-2mm}
\end{center}
\caption{{}The masses of the nucleon $N$, the $\Delta$ and the negative 
parity partner of the nucleon $N^*$ on a $16^3\times 32$ lattice 
at $\beta=3.0$ with $D^{\rm FP}$.
\label{tab:16x32_b3.0_B}
} 
\end{table}

\vspace{1cm}

\begin{table}[tbp]
\small
\begin{center}
\begin{tabular}{l|l|c|l|c|l|c} 
\hline\hline
 \multicolumn{7}{c}{$12^3\times 24$, $\beta=3.0$, parametrized FP} \\ 
\hline
$am_q$ & $am_{\rm PS}$(P) & $t$ & $am_{\rm PS}$(P-S) 
 & $t$ & $am_{\rm V}$ & $t$ \\
\hline
0.016 & 0.2484(144)& [6,12] & 0.2019(63) & [5,12] & 0.612(46) & [6,10] \\ 
0.021 & 0.2621(40) & [5,12] & 0.2399(46) & [5,12] & 0.668(20) & [5,10] \\ 
0.028 & 0.2941(31) & [5,12] & 0.2801(38) & [5,12] & 0.681(16) & [5,10] \\ 
0.04  & 0.3428(25) & [5,12] & 0.3356(30) & [5,12] & 0.701(11) & [5,10] \\ 
0.06  & 0.4108(23) & [5,10] & 0.4108(23) & [5,12] & 0.732(8)  & [5,10] \\ 
0.09  & 0.4994(20) & [6,10] & 0.5029(19) & [6,12] & 0.775(6)  & [5,12] \\ 
0.13  & 0.6013(17) & [6,10] & 0.6075(15) & [6,12] & 0.835(4)  & [5,12] \\ 
0.18  & 0.7159(15) & [6,10] & 0.7235(14) & [7,12] & 0.913(3)  & [6,12] \\ 
0.25  & 0.8636(13) & [6,10] & 0.8717(12) & [7,12] & 1.026(2)  & [6,12] \\ 
0.33  & 1.0227(12) & [6,10] & 1.0307(11) & [7,12] & 1.158(2)  & [6,12] \\ 
 \hline
\end{tabular}
%\vspace{-2mm}
\end{center}
\caption{{}Pseudo-scalar and vector meson masses on a $12^3\times 24$ 
lattice at $\beta=3.0$ with $D^{\rm FP}$. The switching-point is $am_q=0.06$.
\label{tab:12x24_b3.0_PSV}
} 
\end{table}

\vspace{1cm}

\begin{table}[tbp]
\small
\begin{center}
\begin{tabular}{l|l|c|l|c|l|c}
\hline\hline
 \multicolumn{7}{c}{$12^3\times 24$, $\beta=3.0$, parametrized FP} \\ \hline
$am_q$ & $am_{\rm Oct}$(N) & $t$ & $am_{\rm Dec}(\Delta)$ 
 & $t$ & $am_{\rm Oct}(N^*)$ & $t$ \\
\hline
0.016 &            &        &           &        &           &       \\ 
0.021 & 0.855(41)  & [6, 9] &           &        &           &       \\ 
0.028 & 0.874(27)  & [6, 9] & 0.984(69) & [6,10]  &           &       \\ 
0.04  & 0.921(18)  & [6,10] & 1.016(67) & [7,10]  &           &       \\ 
0.06  & 0.996(11)  & [6,10] & 1.146(43) & [7,10] &           &       \\ 
0.09  & 1.096(7)   & [6,10] & 1.264(25) & [7,10] &           &       \\ 
0.13  & 1.220(5)   & [6,10] & 1.374(14) & [7,10] &           &        \\ 
0.18  & 1.366(4)   & [6,12] & 1.497(9)  & [7,10] & 1.738(33) & [6,10] \\ 
0.25  & 1.560(3)   & [6,12] & 1.667(7)  & [7,10] & 1.927(23) & [6,10] \\ 
0.33  & 1.776(3)   & [6,12] & 1.862(6)  & [7,12] & 2.131(18) & [6,10] \\ 
 \hline
\end{tabular}
%\vspace{-2mm}
\end{center}
\caption{{}The masses of the nucleon $N$, the $\Delta$ and 
the negative parity partner of the nucleon $N^*$ on a $12^3\times 24$ 
lattice at $\beta=3.0$ with $D^{\rm FP}$. For the nucleon at
$am_q=0.021$ and 0.028 the correlator Eq.~\eqref{Ncorr4} was used.
\label{tab:12x24_b3.0_B}
}
\end{table}

\vspace{1cm}

\begin{table}[tbp]
\small
\begin{center}
\begin{tabular}{l|l|c|l|c|l|c} 
\hline\hline
 \multicolumn{7}{c}{$12^3\times 24$, $\beta=3.0$, 
                       parametrized FP + 3 overlap steps} \\ 
\hline
$am_q$ & $am_{\rm PS}$(P) & $t$ & $am_{\rm PS}$(P-S) & $t$ & 
                                             $am_{\rm V}$ & $t$ \\
\hline
0.009 &            &        & 0.133(24)  & [6, 9] &           &        \\
0.012 &            &        & 0.169(18)  & [6, 9] & 0.554(69) & [6, 9] \\
0.016 & 0.2301(83) & [6,10] & 0.204(15)  & [6, 9] & 0.664(56) & [6, 8] \\ 
0.021 & 0.2605(70) & [6,10] & 0.239(12)  & [6, 9] & 0.655(60) & [6,10] \\ 
0.028 & 0.2981(59) & [6,10] & 0.280(10)  & [6, 9] & 0.698(51) & [6,10] \\ 
0.04  & 0.3528(48) & [6,10] & 0.3462(71) & [6,10] & 0.735(37) & [6,10] \\ 
0.06  & 0.4278(39) & [6,10] & 0.4243(51) & [6,10] & 0.772(31) & [6,12] \\ 
0.09  & 0.5210(31) & [6,10] & 0.5213(37) & [6,10] & 0.818(19) & [6,12] \\ 
0.13  & 0.6280(26) & [6,10] & 0.6319(31) & [6,10] & 0.879(11) & [6,12] \\ 
0.18  & 0.7476(22) & [6,10] & 0.7545(27) & [6,10] & 0.960(8)  & [6,12] \\ 
0.25  & 0.9008(20) & [6,10] & 0.9116(21) & [6,12] & 1.076(6)  & [6,12] \\ 
0.33  & 1.0652(19) & [6,10] & 1.0757(21) & [6,12] & 1.212(5)  & [6,12] \\ 
 \hline
\end{tabular}
%\vspace{-2mm}
\end{center}
\caption{{}Pseudo-scalar and vector meson masses on a $12^3\times 24$ 
lattice at $\beta=3.0$ with $D^{\rm FP}$+ 3 overlap steps. 
The switching-point is $am_q=0.09$.
\label{tab:12x24_b3.0_ov_PSV}
} 
\end{table}

\vspace{1cm}

\begin{table}[tbp]
\small
\begin{center}
\begin{tabular}{l|l|c|l|c} 
\hline\hline
 \multicolumn{5}{c}{$12^3\times 24$, $\beta=3.0$, 
parametrized FP}+3 overlap steps \\ 
\hline
$am_q$ & $am_{\rm Oct}$(N) & $t$ & $am_{\rm Dec}(\Delta)$  & $t$  \\
\hline
0.009 & 0.897(81)  & [4, 6] & 0.87(15)  & [4, 6]  \\
0.012 & 0.890(65)  & [4, 6] & 0.92(10)  & [4, 6]  \\
0.016 & 0.891(52)  & [4, 6] & 1.004(79) & [4, 6]  \\ 
0.021 & 0.899(43)  & [4, 7] & 1.071(76) & [4, 7]  \\ 
0.028 & 0.929(34)  & [4, 8] & 1.134(70) & [4, 7]  \\ 
0.04  & 0.971(24)  & [4, 8] & 1.191(60) & [4, 7]  \\ 
0.06  & 1.042(16)  & [4, 8] & 1.256(48) & [4, 8]  \\ 
0.09  & 1.139(14)  & [5, 8] & 1.336(44)& [5, 9]  \\ 
0.13  & 1.269(12)  & [5, 9] & 1.434(31) & [5,10]  \\ 
0.18  & 1.415(10)  & [5, 9] & 1.560(22) & [5,10]  \\ 
0.25  & 1.614(9)   & [5, 9] & 1.738(16) & [5,10]  \\ 
0.33  & 1.838(9)   & [5, 9] & 1.944(14) & [5,10]  \\ 
 \hline
\end{tabular}
%\vspace{-2mm}
\end{center}
\caption{{}The masses of the nucleon $N$ and the $\Delta$ 
 on a $12^3\times 24$ lattice at $\beta=3.0$ with 
$D^{\rm FP}$ + 3 overlap steps. 
For the nucleon  and $\Delta$ the operator in Eq.~\eqref{Ncorr3} 
was used.
\label{tab:12x24_b3.0_ov_B}
}
\end{table}

%\vspace{1cm}

\begin{table}[tbp]
\small
\begin{center}
\begin{tabular}{l|l|c|l||c|l|c} 
\hline\hline
\multicolumn{7}{c}{$12^3\times 24$, $\beta=3.4$, parametrized FP} \\ 
\hline
$am_q$ & $am_{\rm PS}$(P)  & $t$ & $am_{\rm PS}$(P-S) &
  $t$ & $am_{\rm V}$  & $t$ \\
\hline
0.029 & 0.193(10)& [7,10] & 0.161(11) & [7,12]  & 0.470(23) & [7,9] \\ 
0.032 & 0.201(9) & [7,12] & 0.177(10) & [7,12]  & 0.474(18) & [7,9] \\ 
0.037 & 0.218(7) & [7,12] & 0.202(8)  & [7,12]  & 0.468(19) & [7,10] \\ 
0.045 & 0.247(6) & [7,12] & 0.237(8)  & [8,12]  & 0.475(17) & [8,12] \\ 
0.058 & 0.292(4) & [7,12] & 0.291(6)  & [8,12]  & 0.498(11) & [8,12] \\ 
0.078 & 0.352(4) & [8,12] & 0.356(5)  & [9,12]  & 0.528(9)  & [9,12] \\ 
0.10  & 0.413(3) & [8,12] & 0.419(4)  & [9,12]  & 0.566(7)  & [9,12] \\ 
0.14  & 0.511(3) & [8,12] & 0.520(3)  & [9,12]  & 0.637(5)  & [9,12] \\ 
0.18  & 0.601(2) & [8,12] & 0.611(3)  & [9,12]  & 0.708(4)  & [9,12] \\ 
0.24  & 0.727(2) & [8,12] & 0.738(2)  & [9,12]  & 0.814(3)  & [9,12] \\ 
 \hline
\end{tabular}
%\vspace{-2mm}
\end{center}
\caption{{}Pseudo-scalar and vector meson masses on a $12^3\times 24$ lattice
 at $\beta=3.4$ with $D^{\rm FP}$. The switching-point is $am_q=0.078$.
\label{tab:12x24_b3.4_PSV}
} 
\end{table}

%\vspace{1cm}

\begin{table}[htb]
\small
\begin{center}
\begin{tabular}{l|l|c|l|c|l|c}
\hline\hline
 \multicolumn{7}{c}{$12^3\times 24$, $\beta=3.4$, parametrized FP} \\ \hline
$am_q$ & $am_{\rm Oct}$(N)  & $t$ & $am_{\rm Dec}(\Delta)$ 
 & $t$ & $am_{\rm Oct}$ $(N^*)$  & $t$ \\
\hline
0.029 & 0.683(32) & [6, 9] & 0.720(92) & [7,12] &           &       \\  
0.032 & 0.690(27) & [6, 9] & 0.748(64) & [7,10] &           &       \\ 
0.037 & 0.701(21) & [6, 9] & 0.769(52) & [7,10] &           &       \\ 
0.045 & 0.718(17) & [6,10] & 0.794(50) & [7,12] &           &       \\ 
0.058 & 0.745(15) & [6,12] & 0.835(35) & [7,12] & 1.088(56) & [5,8] \\ 
0.078 & 0.798(10) & [6,12] & 0.890(23) & [7,12] & 1.103(24) & [5,8] \\ 
0.10  & 0.854(10) & [8,12] & 0.949(16) & [7,12] & 1.135(20) & [6,8] \\ 
0.14  & 0.968(7)  & [8,12] & 1.048(12) & [8,12] & 1.238(13) & [6,8] \\ 
0.18  & 1.080(6)  & [8,12] & 1.151(9)  & [8,12] & 1.344(10) & [6,8] \\ 
0.24  & 1.247(5)  & [8,12] & 1.305(7)  & [8,12] & 1.503(8)  & [6,8] \\ 
 \hline
\end{tabular}
%\vspace{-2mm}
\end{center}
\caption{{}The masses of the nucleon $N$, the $\Delta$ and the negative 
parity partner of the nucleon $N^*$ on a $12^3\times 24$ lattice 
at $\beta=3.4$ with $D^{\rm FP}$. For the nucleon the correlator
Eq.~\eqref{Ncorr3} has been used.
\label{tab:12x24_b3.4_B}
} 
\end{table}
\vspace{1cm}

\begin{table}[tbp]
\small
\begin{center}
\begin{tabular}{l|l|c|l|c|l|c} 
\hline\hline
 \multicolumn{7}{c}{$16^3\times 32$, $\beta=3.7$, parametrized FP} \\ 
\hline
$am_q$ & $am_{\rm PS}$(P) & $t$ & $am_{\rm PS}$(P-S) 
 & $t$ & $am_{\rm V}$ & $t$ \\
\hline
0.0235 & 0.126(13) & [10,13] & 0.123(18)& [10,13] &            &         \\ 
0.026  & 0.139(11) & [10,13] & 0.141(13)& [10,13] &            &         \\ 
0.03   & 0.157(9)  & [10,13] & 0.163(10)& [10,13] & 0.316(23)  & [10,16] \\ 
0.036  & 0.182(7)  & [10,13] & 0.190(8) & [10,13] & 0.336(18)  & [10,16] \\ 
0.045  & 0.216(6)  & [10,13] & 0.227(6) & [10,16] & 0.363(14)  & [10,16] \\ 
0.06   & 0.266(4)  & [10,13] & 0.275(5) & [10,16] & 0.398(10)  & [10,16] \\ 
0.08   & 0.323(3)  & [10,16] & 0.330(5) & [12,16] & 0.437(7)   & [10,16] \\ 
0.1    & 0.373(3)  & [10,16] & 0.380(4) & [12,16] & 0.474(5)   & [10,16] \\ 
0.14   & 0.466(2)  & [10,16] & 0.471(3) & [12,16] & 0.546(4)   & [10,16] \\ 
0.18   & 0.551(2)  & [10,16] & 0.555(3) & [12,16] & 0.618(3)   & [10,16] \\ 
 \hline
\end{tabular}
%\vspace{-2mm}
\end{center}
\caption{{}Pseudo-scalar and vector meson masses on a $16^3\times 32$ 
lattice at $\beta=3.7$ with $D^{\rm FP}$. The switching-point is $am_q=0.026$.
\label{tab:16x32_b3.7_PSV}
} 
\end{table}
\vspace{1cm}

\begin{table}[tbp]
\small
\begin{center}
\begin{tabular}{l|l|c}
\hline\hline
 \multicolumn{3}{c}{$16^3\times 32$, $\beta=3.7$, parametrized FP} \\ \hline
$am_q$ & $am_{\rm Oct}$(N) & $t$ \\
\hline
0.0235 & 0.515(86)  & [8,11] \\ 
0.026  & 0.530(51)  & [8,11] \\ 
0.03   & 0.536(29)  & [8,11] \\ 
0.036  & 0.544(19)  & [8,11] \\ 
0.045  & 0.566(14)  & [8,11] \\ 
0.06   & 0.609(12)  & [8,11] \\ 
0.08   & 0.667(10)  & [8,11] \\ 
0.1    & 0.724(9)   & [8,11] \\ 
0.14   & 0.836(8)   & [8,11] \\ 
0.18   & 0.947(7)   & [8,11] \\
 \hline
\end{tabular}
%\vspace{-2mm}
\end{center}
\caption{{}The masses of the nucleon $N$ on a $16^3\times 32$ 
lattice at $\beta=3.7$ with $D^{\rm FP} $. 
The correlator Eq.~\eqref{Ncorr4} has been chosen here.
\label{tab:16x32_b3.7_B}
}
\end{table}
%\vspace{1cm}

\begin{table}
\small
\begin{center}
\begin{tabular}{l|l|c|l|c} 
\hline\hline
 \multicolumn{5}{c}{$8^3\times 24$, $\beta=3.0$, parametrized FP} \\ \hline
$am_q$ & $am_{\rm PS}$(P)  & $t$ & 
 $am_{\rm V}$  & $t$ \\
\hline
0.028 &            &          & 0.652(34)  & [5, 9] \\ 
0.04  &            &          & 0.670(28)  & [5,10] \\ 
0.06  & 0.409(6)   & [5,10]   & 0.701(21)  & [5,10] \\ 
0.09  & 0.498(5)   & [5,10]   & 0.754(15)  & [5,10] \\ 
0.13  & 0.601(4)   & [5,10]   & 0.822(11)  & [5,10] \\ 
0.18  & 0.716(4)   & [5,10]   & 0.905(8)   & [5,10] \\ 
0.25  & 0.863(3)   & [5,10]   & 1.021(6)   & [5,10] \\ 
0.33  & 1.022(3)   & [5,10]   & 1.153(4)   & [5,10] \\ 
\hline
\end{tabular}
%\vspace{-2mm}
\end{center}
\caption{{}Pseudo-scalar and vector meson masses on a $8^3\times 24$ lattice at
$\beta=3.0$ with $D^{\rm FP}$. We found it difficult to obtain
controlled mass predictions in the $P-S$ channel on that lattice with
our statistics. We do not give numbers for this channel. }
\end{table}
%\vspace{1cm}

\begin{table}[tbp]
\small
\begin{center}
\begin{tabular}{l|l|c} 
\hline\hline
 \multicolumn{3}{c}{$8^3\times 24$, $\beta=3.0$, parametrized FP} \\ \hline
$am_q$ & $am_{\rm Oct}$(N)  & $t$ \\
\hline
0.028 & 1.042(40)  & [4, 7]  \\
0.04  & 1.070(29)  & [4, 7]  \\ 
0.06  & 1.112(21)  & [4, 8]  \\ 
0.09  & 1.181(16)  & [4, 8]  \\ 
0.13  & 1.279(13)  & [4, 9]  \\ 
0.18  & 1.404(11)  & [4, 9]  \\ 
0.25  & 1.581(9)   & [4, 9]  \\ 
0.33  & 1.785(8)   & [4, 9]  \\
 \hline
\end{tabular}
%\vspace{-2mm}
\end{center}
\caption{The masses of the nucleon $N$ on a $8^3\times 24$ lattice 
at $\beta=3.0$ with $D^{\rm FP}$. 
For the nucleon the correlator in Eq.~\eqref{Ncorr4} was used.  
\label{tab:8x24_b3.0_B}
} 
\end{table}

\begin{table}[tbp]
\small
\begin{center}
\begin{tabular}{l|l|l} \hline\hline
 \multicolumn{3}{c}{$16^3\times 32$, $\beta=3.0$, parametrized FP} \\ 
\hline
$x$ & $m_\mathrm{N}(x)/m_\mathrm{V}(x)$ & $m_\Delta(x)/m_\mathrm{V}(x)$ \\
\hline
0.40 & 1.258(22)  & 1.644(44) \\
0.45 & 1.294(16)  & 1.651(34) \\
0.50 & 1.329(13)  & 1.658(28) \\
0.55 & 1.365(11)  & 1.666(24) \\
0.60 & 1.399(9)   & 1.669(20) \\
0.65 & 1.427(7)   & 1.665(15) \\
0.70 & 1.453(6)   & 1.658(11) \\
0.75 & 1.476(5)   & 1.646(9)  \\
0.80 & 1.498(5)   & 1.633(9)  \\
 \hline
\end{tabular}
%\vspace{-2mm}
\end{center}
\caption{{}Mass ratios interpolated to $m_\mathrm{PS}/m_\mathrm{V}=x$ on 
a $16^3\times 32$ lattice at $\beta=3.0$ with $D^\mathrm{FP}$.} 
\label{tab:hratios_x}
\end{table}

\begin{table}[tbp]
\small
\begin{center}
\begin{tabular}{l|l|l|l|l} \hline\hline
 \multicolumn{4}{c}{$16^3\times 32$, $\beta=3.0$, parametrized FP} \\ 
\hline
$am_q$ & $m_\mathrm{PS}/m_\mathrm{V}$ & $m_\mathrm{N}/m_\mathrm{V}$ & 
  $m_\Delta/m_\mathrm{V}$
 & $m_{\mathrm{N}^*}/m_\mathrm{V}$ \\
\hline
0.013 & 0.284(13) & 1.179(65) & 1.633(140)& \\ 
0.016 & 0.318(11) & 1.216(47) & 1.639(96) & \\ 
0.021 & 0.361(9)  & 1.242(32) & 1.654(66) & \\ 
0.028 & 0.408(8)  & 1.271(25) & 1.661(49) & \\ 
0.04  & 0.474(6)  & 1.309(22) & 1.665(31) & 2.111(66) \\ 
0.06  & 0.556(5)  & 1.366(15) & 1.666(22) & 2.071(40) \\ 
0.09  & 0.642(4)  & 1.418(9)  & 1.662(14) & 2.039(30) \\ 
0.13  & 0.720(3)  & 1.460(6)  & 1.646(11) & 2.003(20) \\ 
0.18  & 0.786(2)  & 1.494(5)  & 1.638(9)  & 1.966(18) \\ 
0.25  & 0.844(2)  & 1.518(3)  & 1.624(6)  & 1.910(12) \\ 
0.33  & 0.885(1)  & 1.531(3)  & 1.610(5)  & 1.858(9)  \\ 
 \hline
\end{tabular}
%\vspace{-2mm}
\end{center}
\caption{{}Mass ratios on a $16^3\times 32$ lattice at $\beta=3.0$ 
with $D^\mathrm{FP}$.}
\label{tab:hratios_mq}
\end{table}

\begin{table}[tbp]
\small
\begin{center}
\begin{tabular}{l|l|l|l} \hline\hline
 \multicolumn{4}{c}{$16^3\times 32$, $\beta=3.0$, parametrized FP} \\ 
\hline
$x$ & $m_\mathrm{V}(x)/m_\mathrm{V}(x_0)$ & 
 $m_\mathrm{N}(x)/m_\mathrm{V}(x_0)$ & $m_\Delta(x)/m_\mathrm{V}(x_0)$ \\
\hline
0.40 & 0.794(14) & 1.013(15)  & 1.324(32) \\
0.45 & 0.810(12) & 1.058(12)  & 1.350(28) \\
0.50 & 0.829(10) & 1.106(11)  & 1.379(24) \\
0.55 & 0.850(8)  & 1.158(9)   & 1.413(20) \\
0.60 & 0.875(6)  & 1.216(7)   & 1.452(17) \\
0.65 & 0.907(4)  & 1.286(6)   & 1.501(13) \\
0.70 & 0.947(2)  & 1.370(5)   & 1.563(11) \\
0.75 & 1.00      & 1.476(5)   & 1.646(9)  \\
0.80 & 1.076(3)  & 1.614(6)   & 1.760(8)  \\
 \hline
\end{tabular}
%\vspace{-2mm}
\end{center}
\caption{{}Hadron masses interpolated to $m_\mathrm{PS}/m_\mathrm{V}=x$ 
in units of $m_\mathrm{V}(x_0)$ with $x_0=0.75$ on 
a $16^3\times 32$ lattice at $\beta=3.0$ with $D^\mathrm{FP}$.} 
\label{tab:hper_mhat_fp}
\end{table}

\clearpage

\pagebreak

\section{Data obtained with the CI action
\label{CI:data}
}

In this appendix the extracted masses for the CI operator are
presented. The techniques used and the quantities computed are the 
same as in the case of 
the FP operator. However, there are a few differences: For the PS mass
we list the results for the $\langle A_4 A_4 \rangle$ correlator as
well which we used in the figures and in the fits to the data. Only in
Tables~\ref{tab:hratios_ci_mq} and \ref{tab:hratios_mq_CI} we use the
combination of $\langle PP \rangle - \langle SS\rangle$ 
and the $\langle PP \rangle$
results for the pseudo-scalar mass. 
There the switching point is at $a m_q= 0.04$.
We use Eq.~\eqref{ncorrlarge} to extract the nucleon and $N^*$ mass.
\vspace{10mm}
\begin{table}[hb]
\begin{center}
\hspace*{-11mm}
\begin{tabular}{l|l|c|c|l|c|c|l|c|c} \hline\hline
 \multicolumn{9}{c}{$16^3\times 32$, $\beta=7.9$, CI} \\ \hline
$am_q$ & $am_\mathrm{PS}$(P) & $\chi^2_\mathrm{df}$ & $t$ & 
$am_\mathrm{PS}$(A) &
$\chi^2_\mathrm{df}$ & $t$ & $am_\mathrm{PS}$(P-S) & 
  $\chi^2_\mathrm{df}$ & $t$ \\
\hline
0.0129&0.2074(39)& 0.6&[7,14]&0.2031(61)& 1.1&[7,12]&0.2002(56)& 2.9&[6,12] \\
0.0134&0.2120(36)& 0.6&[7,14]&0.2076(60)& 1.2&[7,12]&0.2051(55)& 2.9&[6,12] \\
0.0139&0.2163(33)& 0.6&[7,14]&0.2117(59)& 1.4&[7,12]&0.2095(54)& 3.0&[6,12] \\
0.0144&0.2204(31)& 0.6&[7,14]&0.2157(58)& 1.5&[7,12]&0.2137(53)& 3.0&[6,12] \\
0.0149&0.2242(30)& 0.6&[7,14]&0.2195(56)& 1.6&[7,12]&0.2176(52)& 3.0&[6,12] \\
0.0159&0.2315(28)& 0.6&[7,14]&0.2268(54)& 1.8&[7,12]&0.2251(49)& 3.0&[6,12] \\
0.0169&0.2384(26)& 0.7&[7,14]&0.2336(52)& 1.9&[7,12]&0.2323(46)& 3.0&[6,12] \\
0.0178&0.2450(25)& 0.7&[7,14]&0.2402(50)& 2.0&[7,12]&0.2392(43)& 3.0&[6,12] \\
0.0198&0.2572(24)& 0.6&[7,16]&0.2502(37)& 1.8&[7,16]&0.2526(38)& 2.9&[6,12] \\ 
0.0247&0.2850(22)& 0.8&[7,16]&0.2786(32)& 1.9&[7,16]&0.2808(36)& 2.5&[8,16] \\
0.0296&0.3099(21)& 1.0&[7,16]&0.3040(29)& 1.9&[7,16]&0.3073(33)& 2.3&[8,16] \\ 
0.0392&0.3544(19)& 1.1&[7,16]&0.3490(25)& 1.8&[7,16]&0.3541(28)& 2.1&[8,16] \\ 
0.0488&0.3937(18)& 1.1&[7,16]&0.3889(22)& 1.7&[7,16]&0.3950(25)& 1.9&[8,16] \\ 
0.0583&0.4295(17)& 1.1&[7,16]&0.4254(21)& 1.6&[7,16]&0.4318(22)& 1.7&[8,16] \\ 
0.0769&0.4939(16)& 0.9&[7,16]&0.4907(19)& 1.4&[7,16]&0.4972(19)& 1.2&[8,16] \\ 
0.0952&0.5515(15)& 0.8&[7,16]&0.5490(18)& 1.2&[7,16]&0.5554(17)& 0.8&[8,16] \\ 
0.1132&0.6045(14)& 0.7&[7,16]&0.6024(18)& 1.1&[7,16]&0.6086(16)& 0.5&[8,16] \\ 
0.1482&0.7006(14)& 0.6&[7,16]&0.6992(17)& 1.0&[7,16]&0.7048(15)& 0.4&[8,16] \\ 
0.1818&0.7875(14)& 0.7&[7,16]&0.7865(17)& 0.9&[7,16]&0.7917(16)& 0.5&[8,16] \\ 
\hline
\end{tabular}
\end{center}
\caption{{}Results for the meson correlators for the CI Dirac
operator at $\beta_1=7.90$ on the $16^3\times32$ lattice.}
\label{tab:ci1}
\end{table}

\begin{table}
\begin{center}
\hspace*{-11mm}
\begin{tabular}{l|l|c|c|l|c|c|l|c|c} \hline\hline
\multicolumn{9}{c}{$16^3\times 32$, $\beta=7.90$, CI} \\ \hline
$am_q$ & $am_\mathrm{V}$ & $\chi^2_\mathrm{df}$ & $t$ & $am_\mathrm{N}$ &
$\chi^2_\mathrm{df}$ & $t$ & $am_\mathrm{N^*}$ & $\chi^2_\mathrm{df}$& $t$ \\
\hline
0.0129& 0.625(44)& 1.4&[6,9]  & 0.790(22)& 0.2&[4,8]  &  & &  \\
0.0134& 0.625(41)& 1.4&[6,9]  & 0.795(21)& 0.2&[4,8]  &  & &  \\
0.0139& 0.625(38)& 1.4&[6,9]  & 0.799(21)& 0.3&[4,8]  &  & &  \\
0.0144& 0.625(36)& 1.3&[6,9]  & 0.803(20)& 0.4&[4,8]  &  & &  \\
0.0149& 0.626(34)& 1.3&[6,9]  & 0.807(20)& 0.5&[4,8]  &  & &  \\
0.0159& 0.628(31)& 1.3&[6,9]  & 0.813(19)& 0.7&[4,8]  &  & &  \\
0.0169& 0.629(29)& 1.2&[6,9]  & 0.819(19)& 0.8&[4,8]  &  & &  \\
0.0178& 0.631(27)& 1.1&[6,9]  & 0.824(19)& 0.9&[4,8]  &  & &  \\
0.0198& 0.635(24)& 1.1&[6,9]  & 0.831(21)& 1.4&[5,8]  &  & &  \\
0.0247& 0.644(19)& 0.6&[6,11] & 0.854(18)& 1.6&[5,8]  &  & &  \\
0.0296& 0.654(15)& 0.4&[6,11] & 0.876(15)& 0.8&[5,12] &  & &  \\
0.0392& 0.673(11)& 0.2&[6,11] & 0.919(13)& 0.7&[5,12] &  & &  \\
0.0488& 0.6909(93)& 0.1&[6,11]& 0.960(12)& 0.5&[5,12] & 1.480(40)& 2.0&[3,6]\\
0.0583& 0.7075(79)& 0.1&[6,11]& 0.999(11)& 0.4&[5,12] & 1.484(34)& 1.4&[3,6]\\
0.0769& 0.7404(61)& 0.2&[6,11]& 1.0720(96)& 0.5&[5,12]& 1.511(29)& 0.7&[3,6]\\
0.0952& 0.7729(52)& 1.7&[6,12]& 1.1396(87)& 0.6&[5,12]& 1.549(27)& 0.4&[3,6]\\
0.1132& 0.8031(48)& 1.9&[7,12]& 1.1983(85)& 0.4&[6,12]& 1.592(26)& 0.4&[3,6]\\
0.1482& 0.8696(38)& 1.8&[7,12]& 1.3171(77)& 0.6&[6,12]& 1.683(24)& 0.5&[3,6]\\
0.1818& 0.9346(33)& 1.5&[7,12]& 1.4278(72)& 0.6&[6,12]& 1.775(23)& 0.6&[3,6] \\
\hline
\end{tabular}
\end{center} 
\caption{{}Results for the vector meson and the nucleon for the CI Dirac
operator at $\beta=7.90$ on the $16^3\times32$ lattice.}
\label{tab:ci2}
\end{table}

\begin{table}
\begin{center}
\hspace*{-11mm}
\begin{tabular}{l|l|c|c|l|c|c|l|c|c} \hline\hline
\multicolumn{9}{c}{$12^3\times 24$, $\beta=7.90$, CI} \\ \hline
$am_q$ & $am_\mathrm{PS}$(P) & $\chi^2_\mathrm{df}$&t &  $am_\mathrm{AA}$(P) & 
 $\chi^2_\mathrm{df}$ &$t$ & $am_\mathrm{PS}$(P-S) & 
 $\chi^2_\mathrm{df}$ & $t$  \\
\hline
0.020& 0.2560(64)& 1.2&[7,12]&0.2688(86)& 0.7&[6,12]&0.2527(66)& 0.8&[6,12]\\
0.030& 0.3107(47)& 0.8&[7,12]&0.3191(67)& 0.6&[6,12]&0.3112(49)& 1.0&[6,12]\\
0.040& 0.3561(40)& 0.6&[7,12]&0.3642(65)& 0.5&[7,12]&0.3510(58)& 0.7&[8,12]\\
0.049& 0.3960(36)& 0.5&[7,12]&0.4023(58)& 0.3&[7,12]&0.3922(52)& 0.9&[8,12]\\
0.058& 0.4321(33)& 0.5&[7,12]&0.4370(52)& 0.2&[7,12]&0.4293(47)& 1.0&[8,12]\\
0.077& 0.4967(30)& 0.4&[7,12]&0.4993(45)& 0.4&[7,12]&0.4954(41)& 1.1&[8,12]\\
0.095& 0.5543(28)& 0.4&[7,12]&0.5552(41)& 0.7&[7,12]&0.5544(36)& 1.0&[8,12]\\
0.113& 0.6073(27)& 0.4&[7,12]&0.6072(38)& 1.0&[7,12]&0.6084(33)& 0.9&[8,12]\\
0.148& 0.7033(25)& 0.4&[7,12]&0.7025(33)& 1.2&[7,12]&0.7059(30)& 1.0&[8,12]\\
0.182& 0.7902(24)& 0.5&[7,12]&0.7892(30)& 1.1&[7,12]&0.7934(29)& 1.2&[8,12]\\ 
\hline
\end{tabular}
\end{center}
\caption{{}Results for the pseudoscalar correlators for the CI Dirac
operator at $\beta=7.90$ on the $12^3\times24$ lattice.}
\end{table}

\begin{table}
\begin{center}
\begin{tabular}{l|l|c|c|l|c|c} \hline\hline
\multicolumn{6}{c}{$12^3\times 24$, $\beta=7.90$, CI} \\ \hline
$am_q$ & $am_\mathrm{V}$ & $\chi^2_\mathrm{df}$ & $t$ & $am_\mathrm{N}$ &
$\chi^2_\mathrm{df}$ & $t$  \\
\hline
0.020&0.696(37)& 1.2&[6,11]&  0.778(53)& 0.1&[5,9]  \\
0.030&0.709(24)& 1.7&[6,11]&  0.851(28)& 0.5&[5,10] \\
0.040&0.716(18)& 2.1&[6,11]&  0.906(21)& 0.8&[5,10] \\
0.049&0.724(14)& 2.2&[6,11]&  0.954(17)& 1.2&[5,10] \\
0.058&0.734(12)& 2.0&[6,11]&  0.996(15)& 0.9&[5,12] \\
0.077&0.7584(89)& 1.6&[6,11]& 1.070(12)& 0.9&[5,12] \\
0.095&0.7865(71)& 1.3&[6,11]& 1.138(11)& 0.9&[5,12] \\
0.113&0.8170(60)& 0.9&[6,12]& 1.207(12)& 1.1&[6,12] \\
0.148&0.8792(47)& 0.7&[6,12]& 1.325(10)& 1.5&[6,12] \\
0.182&0.9413(41)& 0.6&[6,12]& 1.4351(96)& 1.8&[6,12] \\ 
\hline
\end{tabular}
\end{center}
\caption{{}Results for the vector meson and the nucleon for the CI Dirac
operator at $\beta=7.90$ on the $12^3\times24$ lattice.}
\end{table}

\begin{table}
\begin{center}
\hspace*{-11mm}
\begin{tabular}{l|l|c|c|l|c|c|l|c|c} \hline\hline
\multicolumn{9}{c}{$8^3\times 24$, $\beta=7.90$, CI} \\ \hline
$am_q$ & $am_\mathrm{PS}$(P) & $\chi^2_\mathrm{df}$ & $t$ & 
  $am_\mathrm{PS}$(A) & $\chi^2_\mathrm{df}$ & $t$ & 
  $am_\mathrm{PS}$(P-S) &$\chi^2_\mathrm{df}$& $t$ \\
\hline
0.020&0.308(20)& 0.2&[7,12]&  0.282(12)& 0.6&[6,11]&  0.192(23)& 0.0&[6,11]  \\
0.030&0.325(11)& 0.9&[7,12]&  0.329(10)& 1.0&[6,11]&  0.293(11)& 1.9&[6,11]  \\
0.040&0.3651(90)& 0.9&[7,12]& 0.365(10)& 0.7&[7,11]&  0.353(12)& 1.9&[8,11]  \\
0.049&0.4026(80)& 0.8&[7,12]& 0.4034(93)& 0.6&[7,11]& 0.396(10)& 1.9&[8,11]  \\
0.058&0.4373(74)& 0.7&[7,12]& 0.4389(88)& 0.5&[7,11]& 0.4348(93)& 1.9&[8,11] \\
0.077&0.5006(65)& 0.6&[7,12]& 0.5034(80)& 0.6&[7,11]& 0.5020(80)& 1.9&[8,11] \\
0.095&0.5577(59)& 0.6&[7,12]& 0.5614(75)& 0.6&[7,11]& 0.5612(70)& 1.8&[8,11]\\
0.113&0.6104(54)& 0.5&[7,12]& 0.6146(70)& 0.7&[7,11]& 0.6151(64)& 1.7&[8,11]\\
0.148&0.7064(48)& 0.4&[7,12]& 0.7109(62)& 0.7&[7,11]& 0.7125(55)& 1.4&[8,11]\\
0.182&0.7935(43)& 0.4&[7,12]& 0.7977(56)& 0.7&[7,11]& 0.8002(50)& 1.2&[8,11]\\ 
\hline
\end{tabular}
\end{center}
\caption{{}Results for the meson correlators for the CI Dirac
operator at $\beta=7.90$ on the $8^3\times24$ lattice.}
\end{table}

\begin{table}
\begin{center}
\begin{tabular}{l|l|c|c|l|c|c} \hline\hline
\multicolumn{6}{c}{$8^3\times 24$, $\beta=7.90$, CI} \\ \hline
$am_q$ & $am_\mathrm{V}$ & $\chi^2_\mathrm{df}$ & $t$ & $am_\mathrm{N}$ &
$\chi^2_\mathrm{df}$ & $t$  \\
\hline
0.020&0.581(47)& 0.3&[6,9]&   &  &  \\
0.030&0.634(29)& 0.0&[6,9]&   0.88(11)& 0.1&[5,9]   \\
0.040&0.665(24)& 0.0&[6,9]&   0.954(60)& 0.2&[5,10] \\
0.049&0.688(21)& 0.1&[6,9]&   1.003(45)& 0.1&[5,10] \\
0.058&0.708(19)& 0.1&[6,9]&   1.050(37)& 0.1&[5,10] \\
0.077&0.745(15)& 0.1&[6,10]&  1.133(28)& 0.3&[5,10] \\
0.095&0.781(13)& 0.1&[6,10]&  1.206(25)& 0.5&[5,10] \\
0.113&0.816(12)& 0.1&[6,10]&  1.282(27)& 0.6&[6,10] \\
0.148&0.8814(94)& 0.1&[6,12]& 1.399(25)& 0.9&[6,10] \\
0.182&0.9461(82)& 0.2&[6,12]& 1.505(24)& 1.2&[6,10] \\ 
\hline
\end{tabular}
\end{center}
\caption{{}Results for the vector meson and the nucleon for the CI Dirac
operator at $\beta=7.90$ on the $8^3\times24$ lattice.}
\end{table}

\begin{table}
\begin{center}
\hspace*{-11mm}
\begin{tabular}{l|l|c|c|l|c|c|l|c|c} \hline\hline
\multicolumn{9}{c}{$16^3\times 32$, $\beta=8.35$, CI} \\ \hline
$am_q$ & $am_\mathrm{PS}$(P) & $\chi^2_\mathrm{df}$ & $t$ & 
 $am_\mathrm{PS}$(A) & $\chi^2_\mathrm{df}$ & $t$ &
 $am_\mathrm{PS}$(P-S) &$\chi^2_\mathrm{df}$& $t$ \\
\hline
0.010&0.1734(40)& 0.7&[7,16]&0.1757(58)& 1.8&[6,16]&0.1599(74)& 0.9&[8,16]\\
0.015&0.2015(29)& 0.9&[7,16]&0.2043(46)& 1.9&[6,16]&0.1941(61)& 0.8&[8,16]\\
0.020&0.2268(25)& 1.0&[7,16]&0.2294(41)& 2.0&[6,16]&0.2226(52)& 0.8&[8,16]\\
0.030&0.2712(21)& 1.2&[7,16]&0.2731(34)& 1.8&[6,16]&0.2704(41)& 0.8&[8,16]\\
0.039&0.3099(19)& 1.3&[7,16]&0.3113(30)& 1.6&[6,16]&0.3112(34)& 0.9&[8,16]\\
0.049&0.3448(18)& 1.3&[7,16]&0.3460(27)& 1.5&[6,16]&0.3462(32)& 1.1&[9,16]\\
0.058&0.3770(17)& 1.4&[7,16]&0.3780(24)& 1.3&[6,16]&0.3795(28)& 1.3&[9,16]\\
0.077&0.4356(16)& 1.4&[7,16]&0.4363(21)& 1.1&[6,16]&0.4394(23)& 1.5&[9,16]\\
0.095&0.4887(16)& 1.3&[7,16]&0.4893(19)& 1.0&[6,16]&0.4932(20)& 1.6&[9,16]\\
0.113&0.5380(15)& 1.3&[7,16]&0.5385(18)& 1.0&[6,16]&0.5430(19)& 1.6&[9,16]\\
0.148&0.6286(15)& 1.2&[7,16]&0.6290(17)& 1.1&[6,16]&0.6340(17)& 1.6&[9,16]\\
0.182&0.7117(14)& 1.1&[7,16]&0.7120(16)& 1.2&[6,16]&0.7171(17)& 1.6&[9,16]\\ 
\hline
\end{tabular}
\end{center}
\caption{{}Results for the meson correlators for the CI Dirac
operator at $\beta=8.35$ on the $16^3\times32$ lattice.}
\end{table}

\begin{table}
\begin{center}
\begin{tabular}{l|l|c|c|l|c|c} \hline\hline
\multicolumn{6}{c}{$16^3\times 32$, $\beta=8.35$, CI} \\ \hline
$am_q$ & $am_\mathrm{V}$ & $\chi^2_\mathrm{df}$ & $t$ & $am_\mathrm{N}$ &
$\chi^2_\mathrm{df}$ & $t$  \\
\hline
0.010& 0.450(21)& 0.4&[7,12]& 0.636(25)& 2.0&[4,11]\\
0.015& 0.452(15)& 0.4&[7,12]& 0.652(17)& 2.1&[4,11]\\
0.020& 0.460(12)& 0.4&[7,12]& 0.674(14)& 2.0&[4,11]\\
0.030& 0.4813(88)& 0.4&[7,12]&0.708(13)& 1.8&[5,13]\\
0.039& 0.5019(71)& 0.6&[7,12]&0.746(12)& 1.8&[5,13]\\
0.049& 0.5217(60)& 0.9&[7,12]&0.775(11)& 1.9&[6,13]\\
0.058& 0.5410(54)& 1.2&[7,12]&0.810(10)& 2.0&[6,13]\\
0.077& 0.5790(45)& 1.6&[7,12]&0.8699(88)& 1.8&[7,14]\\
0.095& 0.6164(39)& 1.8&[7,12]&0.9326(79)& 1.8&[7,14]\\
0.113& 0.6533(35)& 1.8&[7,12]&0.9931(73)& 1.8&[7,14]\\
0.148& 0.7254(30)& 1.8&[7,12]&1.1094(65)& 1.7&[7,14]\\
0.182& 0.7950(27)& 1.8&[7,12]&1.2202(60)& 1.6&[7,14]\\ 
\hline
\end{tabular}
\end{center}
\caption{{}Results for the vector meson and the nucleon for the CI Dirac
operator at $\beta = 8.35$ on the $16^3\times32$ lattice.}
\end{table}

\begin{table}
\begin{center}
\hspace*{-11mm}
\begin{tabular}{l|l|c|c|l|c|c|l|c|c} \hline\hline
\multicolumn{9}{c}{$12^3\times 24$, $\beta = 8.35$, CI} \\ \hline
$am_q$ & $am_\mathrm{PS}$(P) & $\chi^2_\mathrm{df}$ & $t$ & 
 $am_\mathrm{PS}$(A) & $\chi^2_\mathrm{df}$ & $t$ &
 $am_\mathrm{PS}$(P-S) &$\chi^2_\mathrm{df}$& $t$ \\
\hline
0.010&0.185(14)& 0.9&[7,12]& 0.169(17)& 1.8&[6,12]&  0.120(27)& 2.1&[8,12] \\
0.020&0.2287(95)& 1.6&[7,12]&0.227(11)& 1.6&[6,12]&  0.204(16)& 1.4&[8,12] \\
0.030&0.2708(72)& 1.6&[7,12]&0.2684(89)& 1.2&[6,12]& 0.260(11)& 0.7&[8,12] \\
0.039&0.3089(60)& 1.5&[7,12]&0.3055(76)& 1.0&[6,12]& 0.3050(92)& 0.5&[8,12]\\
0.049&0.3437(53)& 1.4&[7,12]&0.3396(67)& 1.0&[6,12]& 0.3411(89)& 0.5&[9,12]\\
0.058&0.3758(49)& 1.4&[7,12]&0.3713(62)& 1.1&[6,12]& 0.3765(78)& 0.5&[9,12]\\
0.077&0.4344(45)& 1.3&[7,12]&0.4296(55)& 1.4&[6,12]& 0.4392(66)& 0.7&[9,12]\\
0.095&0.4877(43)& 1.3&[7,12]&0.4831(51)& 1.5&[6,12]& 0.4948(59)& 0.9&[9,12]\\
0.113&0.5370(41)& 1.2&[7,12]&0.5330(49)& 1.4&[6,12]& 0.5456(55)& 1.1&[9,12]\\
0.148&0.6275(39)& 1.0&[7,12]&0.6245(45)& 1.1&[6,12]& 0.6377(51)& 1.5&[9,12]\\
0.182&0.7102(37)& 0.8&[7,12]&0.7080(42)& 0.9&[6,12]& 0.7210(49)& 1.7&[9,12]\\
 \hline
\end{tabular}
\end{center}
\caption{{}Results for the meson correlators for the CI Dirac
operator at $\beta = 8.35$ on the $12^3\times24$ lattice.}
\end{table}

\begin{table}
\begin{center}
\begin{tabular}{l|l|c|c|l|c|c} \hline\hline
\multicolumn{6}{c}{$12^3\times 24$, $\beta = 8.35$, CI} \\ \hline
$am_q$ & $am_\mathrm{V}$ & $\chi^2_\mathrm{df}$ & $t$ & $am_\mathrm{N}$ &
$\chi^2_\mathrm{df}$ & $t$  \\
\hline
0.010&0.468(51)& 2.4&[7,12]& 0.84(14)& 1.4&[4,11] \\
0.020&0.502(30)& 2.7&[7,12]& 0.779(47)& 2.0&[4,11]\\
0.030&0.512(20)& 2.6&[7,12]& 0.773(31)& 2.0&[5,12]\\
0.039&0.522(16)& 2.7&[7,12]& 0.798(22)& 1.9&[5,12]\\
0.049&0.535(14)& 2.9&[7,12]& 0.818(20)& 2.0&[6,12]\\
0.058&0.550(13)& 2.9&[7,12]& 0.847(17)& 1.7&[6,12]\\
0.077&0.584(12)& 2.6&[7,12]& 0.916(18)& 1.1&[7,12]\\
0.095&0.619(11)& 2.2&[7,12]& 0.972(16)& 0.8&[7,12]\\
0.113&0.6544(96)& 1.8&[7,12]&1.028(15)& 0.8&[7,12]\\
0.148&0.7239(81)& 1.2&[7,12]&1.138(14)& 0.8&[7,12]\\
0.182&0.7917(70)& 0.8&[7,12]&1.244(13)& 1.0&[7,12]\\ 
\hline
\end{tabular}
\end{center}
\caption{{}Results for the vector meson and the nucleon for the CI Dirac
operator at $\beta=8.35$ on the $12^3\times24$ lattice.}
\end{table}

\begin{table}
\begin{center}
\hspace*{-11mm}
\begin{tabular}{l|l|c|c|l|c|c|l|c|c} \hline\hline
\multicolumn{9}{c}{$16^3\times 32$, $\beta = 8.70$, CI} \\ \hline
$am_q$ & $am_\mathrm{PS}$(P) & $\chi^2_\mathrm{df}$ & $t$ &
 $am_\mathrm{PS}$(A) & $\chi^2_\mathrm{df}$ & $t$ &
 $am_\mathrm{PS}$(P-S) &$\chi^2_\mathrm{df}$& $t$ \\
\hline
0.005&0.141(12)& 0.8&[5,14]& 0.122(11)& 0.8&[5,14]&  0.073(15)& 0.7&[7,12] \\
0.010&0.1607(58)& 0.9&[5,14]&0.1603(82)& 0.9&[5,14]& 0.1446(77)& 0.6&[8,14]\\
0.020&0.2054(40)& 1.2&[7,14]&0.2042(73)& 1.1&[7,14]& 0.2061(52)& 0.8&[8,14]\\
0.030&0.2453(34)& 1.3&[7,14]&0.2431(61)& 1.0&[7,14]& 0.2512(43)& 1.1&[8,14]\\
0.039&0.2807(30)& 1.4&[7,14]&0.2781(52)& 1.1&[7,14]& 0.2891(37)& 1.6&[8,14]\\
0.049&0.3130(28)& 1.6&[7,14]&0.3101(45)& 1.1&[7,14]& 0.3229(34)& 2.3&[8,14]\\
0.058&0.3429(26)& 1.6&[7,14]&0.3400(39)& 1.1&[7,14]& 0.3540(31)& 2.9&[8,14]\\
0.077&0.3978(23)& 1.7&[7,14]&0.3950(32)& 1.1&[7,14]& 0.4104(27)& 3.9&[8,14]\\
0.095&0.4480(21)& 1.6&[7,14]&0.4455(28)& 1.1&[7,14]& 0.4613(24)& 4.2&[8,14]\\
0.113&0.4950(19)& 1.5&[7,14]&0.4928(25)& 1.0&[7,14]& 0.5083(22)& 4.0&[8,14]\\
0.148&0.5821(18)& 1.4&[7,14]&0.5804(22)& 1.0&[7,14]& 0.5947(20)& 3.2&[8,14]\\
0.182&0.6625(17)& 1.3&[7,14]&0.6614(20)& 1.0&[7,14]& 0.6740(18)& 2.5&[8,14] \\
\hline
\end{tabular}
\end{center}
\caption{{}Results for the pseudoscalar correlators for the CI Dirac
operator at $\beta=8.70$ on the $16^3\times32$ lattice.}
\end{table}

\begin{table}
\begin{center}
\begin{tabular}{l|l|c|c|l|c|c} \hline\hline
\multicolumn{6}{c}{$16^3\times 32$, $\beta=8.70$, CI} \\ \hline
$am_q$ & $am_\mathrm{V}$ & $\chi^2_\mathrm{df}$ & $t$ & $am_\mathrm{N}$ &
$\chi^2_\mathrm{df}$ & $t$  \\
\hline
0.005&0.335(28)& 0.1&[9,14]& 0.527(63)& 1.0&[5,12]\\
0.010&0.354(19)& 0.5&[9,14]& 0.540(29)& 0.6&[5,12]\\
0.020&0.366(12)& 1.2&[9,14]& 0.568(23)& 0.8&[6,12]\\
0.030&0.385(10)& 1.8&[9,14]& 0.599(20)& 1.1&[6,12]\\
0.039&0.4054(88)& 2.0&[9,14]&0.643(16)& 1.4&[6,14]\\
0.049&0.4254(77)& 2.0&[9,14]&0.673(15)& 1.5&[6,14]\\
0.058&0.4454(69)& 1.9&[9,14]&0.702(14)& 1.5&[6,14]\\
0.077&0.4851(56)& 1.7&[9,14]&0.761(13)& 1.6&[6,14]\\
0.095&0.5243(48)& 1.5&[9,14]&0.820(12)& 1.6&[6,14]\\
0.113&0.5633(35)& 1.2&[7,14]&0.878(11)& 1.6&[6,14]\\
0.148&0.6383(29)& 1.1&[7,14]&0.9918(92)& 1.7&[6,14]\\
0.182&0.7102(26)& 1.2&[7,14]&1.1013(82)& 1.7&[6,14]\\ 
\hline
\end{tabular}
\end{center}
\caption{{}Results for the vector meson and the nucleon for the CI Dirac
operator at $\beta=8.70$ on the $16^3\times32$ lattice.}
\end{table}

\begin{table}[tbp]
\small
\begin{center}
\begin{tabular}{l|l|l|l} \hline\hline
 \multicolumn{4}{c}{$16^3\times 32$, $\beta=7.9$, CI} \\ 
\hline
$am_q$ & $m_\mathrm{PS}/m_\mathrm{V}$ & $m_\mathrm{N}/m_\mathrm{V}$ & 
$m_{\mathrm{N}^*}/m_\mathrm{V}$ \\
\hline
0.0129&  0.320(25) &   1.263(92) &     \\
0.0134&  0.328(24) &   1.272(86) &     \\
0.0139&  0.335(23) &   1.279(82) &      \\
0.0144&  0.341(22) &   1.284(78) &      \\
0.0149&  0.347(21) &   1.289(75) &      \\
0.0159&  0.358(20) &   1.295(70) &      \\
0.0169&  0.369(19) &   1.300(64) &     \\
0.0178&  0.378(18) &   1.306(60) &     \\
0.0198&  0.397(17) &   1.308(55) &     \\
0.0247&  0.435(14) &   1.325(43) &     \\
0.0296&  0.469(12) &   1.339(36) &      \\
0.0392&  0.526(10) &   1.365(27) &      \\
0.0488&  0.569(8)  &   1.390(22) & 2.14(6) \\
0.0583&  0.607(7)  &   1.412(19) & 2.09(5) \\
0.0769&  0.667(6)  &   1.448(15) & 2.04(4) \\
0.0952&  0.713(5)  &   1.474(13) & 2.00(4) \\
0.1132&  0.752(5)  &   1.492(12) & 1.98(3) \\
0.1482&  0.805(4)  &   1.514(9)  & 1.93(3) \\
0.1818&  0.843(3)  &   1.527(7)  & 1.89(2) \\
 \hline
\end{tabular}
%\vspace{-2mm}
\end{center}
\caption{{}Mass ratios on the $16^3\times 32$ lattice at $\beta=7.9$ 
with $D^\mathrm{CI}$.}
\label{tab:hratios_ci_mq}
\end{table}

\begin{table}[tbp]
\small
\begin{center}
\begin{tabular}{l|c|c} \hline\hline
 \multicolumn{3}{c}{$16^3\times 32$, $\beta=7.9$, CI} \\ 
\hline
 $x$ & $m_\mathrm{V}(x)/m_\mathrm{V}(x_0)$ & 
$m_{\mathrm{N}}(x)/m_\mathrm{V}(x_0)$ \\
\hline
 0.320(25) &0.781(52) &  0.99(3)  \\
 0.328(24) &0.780(48) &  0.99(3)  \\
 0.335(23) &0.780(45) &  1.00(3)  \\
 0.341(22) &0.781(43) &  1.00(3)  \\
 0.347(21) &0.782(41) &  1.01(3)  \\
 0.358(20) &0.784(37) &  1.01(3)  \\
 0.369(19) &0.786(34) &  1.02(3)  \\
 0.378(18) &0.788(31) &  1.03(3)  \\
 0.397(17) &0.793(27) &  1.04(3)  \\
 0.435(14) &0.804(20) &  1.07(2)  \\
 0.469(12) &0.817(16) &  1.09(2)  \\
 0.526(10) &0.841(11) &  1.15(2)  \\
 0.569(8)  &0.863(8)  &  1.20(2)  \\
 0.607(7)  &0.883(8)  &  1.25(2)  \\
 0.667(6)  &0.924(5)  &  1.34(2)  \\
 0.713(5)  &0.965(5)  &  1.42(2)  \\
 0.752(5)  &1.003(4)  &  1.50(2)  \\
 0.805(4)  &1.086(7)  &  1.64(2)  \\
 0.843(3)  &1.167(9)  &  1.78(2)  \\
 \hline                            
\end{tabular}
%\vspace{-2mm}
\end{center}
\caption{Hadron masses in units of $m_V(x_0=0.75)$ on the $16^3 \times 
32$ lattice at $\beta=7.9$ with $D^{\rm CI}$.}
\label{tab:hratios_mq_CI}
\end{table}

\end{appendix}

\end{document}